\documentclass[10pt]{thesis}
\usepackage[pctex32]{graphics}
\usepackage{amssymb}
\renewcommand{\baselinestretch}{2}
\usepackage{hyperref}
\usepackage[enableskew]{youngtab}

\def\be{\begin{equation}}
\def\ee{\end{equation}}
\def\bea{\begin{eqnarray}}
\def\eea{\end{eqnarray}}

\def\IR{\relax{\rm I\kern-.18em R}}
\def\binomial#1#2{\left(\,{\buildrel 
{\raise4pt\hbox{$\displaystyle{#1}$}}\over
{\raise-6pt\hbox{$\displaystyle{#2}$}}}\,\right)}

\def\[{\lfloor{\hskip 0.35pt}\!\!\!\lceil}
\def\]{\rfloor{\hskip 0.35pt}\!\!\!\rceil}

\def\a{\alpha}
\def\b{\beta}

\def\d{\delta}
\def\e{\epsilon}

\def\g{\gamma}
\def\h{\eta}

\def\l{\lambda}

\def\n{\nu}
\def\o{\omega}

\def\q{\theta}
\def\r{\rho}
\def\s{\sigma}

\def\u{\upsilon}

\def\z{\zeta}

\def\F{\Phi}
\def\G{\Gamma}

\def\L{\Lambda}
\def\O{\Omega}
\def\P{\Pi}

\def\S{\Sigma}

\def\cj{{\cal J}}

\def\cl{{\cal L}}
\def\cm{{\cal M}}
\def\cn{{\cal N}}
\def\co{{\cal O}}

\def\cs{{\cal S}}

\def\cu{{\cal U}}

\newcommand{\drawsquare}[2]{\hbox{%
\rule{#2pt}{#1pt}\hskip-#2pt
\rule{#1pt}{#2pt}\hskip-#1pt
\rule[#1pt]{#1pt}{#2pt}}\rule[#1pt]{#2pt}{#2pt}\hskip-#2pt
\rule{#2pt}{#1pt}}

\renewcommand{\Box}{\,\raisebox{-.45pt}{\drawsquare{7}{0.6}}\,}

\def\bo{{\raise.15ex\hbox{\large$\Box$}}}               
\def\pa{\partial}                                       
\def\de{\nabla}                                         
\def\TH{{\raise.2ex\hbox{$\displaystyle \bigodot$}\mskip-4.7mu \llap H
\;}}
\def\face{{\raise.2ex\hbox{$\displaystyle \bigodot$}\mskip-2.2mu \llap
{$\ddot
        \smile$}}}                                      

\def\sl#1{\rlap{\hbox{$\mskip 1 mu /$}}#1}      
\def\Tilde#1{\widetilde{#1}}                    
\def\Hat#1{\widehat{#1}}                        
\def\Bar#1{\overline{#1}}                       
\def\leftrightarrowfill{$\mathsurround=0pt \mathord\leftarrow \mkern-6mu
        \cleaders\hbox{$\mkern-2mu \mathord- \mkern-2mu$}\hfill
        \mkern-6mu \mathord\rightarrow$}
\def\dvec#1{\vbox{\ialign{##\crcr
        \leftrightarrowfill\crcr\noalign{\kern-1pt\nointerlineskip}
        $\hfil\displaystyle{#1}\hfil$\crcr}}}           

\catcode`@=11
\def\un#1{\relax\ifmmode\@@underline#1\else
        $\@@underline{\hbox{#1}}$\relax\fi}
\catcode`@=12

\def\frac#1#2{{\textstyle{#1\over\vphantom2\smash{\raise.20ex
        \hbox{$\scriptstyle{#2}$}}}}}                   
\def\sfrac#1#2{{\vphantom1\smash{\lower.5ex\hbox{\small$#1$}}\over
        \vphantom1\smash{\raise.4ex\hbox{\small$#2$}}}} 
\def\bfrac#1#2{{\vphantom1\smash{\lower.5ex\hbox{$#1$}}\over
        \vphantom1\smash{\raise.3ex\hbox{$#2$}}}}       
\def\afrac#1#2{{\vphantom1\smash{\lower.5ex\hbox{$#1$}}\over#2}}    

\newskip\humongous \humongous=0pt plus 1000pt minus 1000pt

\newif\ifdtup

  \def\pp{{\mathchoice
          {
              \kern 1pt%
              \raise 1pt
              \vbox{\hrule width5pt height0.4pt depth0pt
                    \kern -2pt
                    \hbox{\kern 2.3pt
                          \vrule width0.4pt height6pt depth0pt
                          }
                    \kern -2pt
                    \hrule width5pt height0.4pt depth0pt}%
                    \kern 1pt
           }
            {
              \kern 1pt%
              \raise 1pt
              \vbox{\hrule width4.3pt height0.4pt depth0pt
                    \kern -1.8pt
                    \hbox{\kern 1.95pt
                          \vrule width0.4pt height5.4pt depth0pt
                          }
                    \kern -1.8pt
                    \hrule width4.3pt height0.4pt depth0pt}%
                    \kern 1pt
            }
            {
              \kern 0.5pt%
              \raise 1pt
              \vbox{\hrule width4.0pt height0.3pt depth0pt
                    \kern -1.9pt  
                    \hbox{\kern 1.85pt
                          \vrule width0.3pt height5.7pt depth0pt
                          }
                    \kern -1.9pt
                    \hrule width4.0pt height0.3pt depth0pt}%
                    \kern 0.5pt
            }
            {
              \kern 0.5pt%
              \raise 1pt
              \vbox{\hrule width3.6pt height0.3pt depth0pt
                    \kern -1.5pt
                    \hbox{\kern 1.65pt
                          \vrule width0.3pt height4.5pt depth0pt
                          }
                    \kern -1.5pt
                    \hrule width3.6pt height0.3pt depth0pt}%
                    \kern 0.5pt
            }
        }}

  \def\mm{{\mathchoice
                       {
                             \kern 1pt
               \raise 1pt    \vbox{\hrule width5pt height0.4pt depth0pt
                                  \kern 2pt
                                  \hrule width5pt height0.4pt depth0pt}
                             \kern 1pt}
                       {
                            \kern 1pt
               \raise 1pt \vbox{\hrule width4.3pt height0.4pt depth0pt
                                  \kern 1.8pt
                                  \hrule width4.3pt height0.4pt depth0pt}
                             \kern 1pt}
                       {
                            \kern 0.5pt
               \raise 1pt
                            \vbox{\hrule width4.0pt height0.3pt depth0pt
                                  \kern 1.9pt
                                  \hrule width4.0pt height0.3pt depth0pt}
                            \kern 1pt}
                       {
                           \kern 0.5pt
             \raise 1pt  \vbox{\hrule width3.6pt height0.3pt depth0pt
                                  \kern 1.5pt
                                  \hrule width3.6pt height0.3pt depth0pt}
                           \kern 0.5pt}
                       }}

\def\dslash{\not{\hbox{\kern-2pt $\partial$}}}
\def\Dslash{\not{\hbox{\kern-4pt $D$}}}
\def\pslash{\not{\hbox{\kern-2.3pt $p$}}}
 \newtoks\slashfraction
 \slashfraction={.13}
 \def\slash#1{\setbox0\hbox{$ #1 $}
 \setbox0\hbox to \the\slashfraction\wd0{\hss \box0}/\box0 }
 
\def\plpl{\raise-2pt\hbox{$\raise3pt\hbox{$_+$}\hskip-6.67pt\raise0.0pt
\hbox{$^+$}\hskip 0.01pt$}}
\def\mimi{\raise-2pt\hbox{$\raise3pt\hbox{$_-$}\hskip-6.67pt\raise0.0pt
\hbox{$^-$}\hskip 0.01pt$}}

\begin{document}


\hbox{\ }

\renewcommand{\baselinestretch}{1}
\small \normalsize

\begin{center}
\large{{ABSTRACT}} 

\vspace{3em} 

\end{center}
\hspace{-.15in}
\begin{tabular}{ll}
Title of dissertation:    & {\large  FORMULATION OF FREE HIGHER}\\
&				      {\large  SPIN SUPERSYMMETRIC THEORIES} \\
&				      {\large  IN SUPERSPACE} \\
\ \\
&                          {\large  Joseph Phillips, Doctor of Philosophy, 2004} \\
\ \\
Dissertation directed by: & {\large  Professor S. James Gates, Jr.} \\
&  				{\large	 Department of Physics } \\
\end{tabular}

\vspace{3em}

\renewcommand{\baselinestretch}{2}
\large \normalsize

The N = 1 superfield formalism in four-dimensions is well formulated and understood, yet there remain unsolved problems.  
In this thesis, superfield actions for free massless and massive  higher spin superfield theories are formulated in four dimensions.  
The discussion of massless models is restricted to half integer superhelicity.  These models describe multiplets with helicities $(s,~s-1/2)$ where $s$ is an integer.  The investigation of massive models covers recent work on superspin-3/2 and superspin-1 multiplets.  Superspin-3/2 multiplets contain component fields with spins $(2,~3/2,~3/2,~1)$ and superspin-1 multiplets contain component fields with spins $(3/2,~1,~1,~1/2)$.  The super projector method is used to distinguish supersymmetric subspaces.  Here, this method is used to write general superspace actions.  The underlying geometrical structure of superspace actions is elucidated when they are written in terms of super projectors.  This thesis also discusses the connection between four-dimensional massive theories and five-dimensional massless theories.  This connection is understood in non-supersymmetric field theory but has not been established in superspace.  A future direction of the five-dimensional models would involve finding an anti-de Sitter supergravity background.  In order to construct this model using the knowledge gained from this thesis, an understanding of the Casimir operators of four-dimensional anti-de Sitter superspace would be necessary.  These Casimir operators have not yet appeared in the literature and are presented in this thesis.


\thispagestyle{empty}
\hbox{\ }
\vspace{1in}
\renewcommand{\baselinestretch}{1}
\small\normalsize
\begin{center}

\large{{FORMULATION OF FREE HIGHER SPIN \\
SUPERSYMMETRIC THEORIES IN SUPERSPACE}}\\
\ \\
\ \\
\large{by} \\
\ \\
\large{Joseph Phillips}
\ \\
\ \\
\ \\
\ \\
\normalsize
Dissertation submitted to the Faculty of the Graduate School of the \\
University of Maryland, College Park in partial fulfillment \\
of the requirements for the degree of \\
Doctor of Philosophy \\
2004
\end{center}

\vspace{7.5em}

\noindent Advisory Commmittee: \\

Professor S. James Gates, Jr., Chair/Advisor\\

Professor Melanie Becker\\

Professor Paul Green\\

Professor Rabi Mohapatra\\

Professor Jogesh Pati\\

\thispagestyle{empty}
\hbox{\ }
\vspace{3.5in}
\renewcommand{\baselinestretch}{1.5}
\small\normalsize

\begin{center}
\large{\copyright \hbox{ }Copyright by\\
Joseph Phillips \\
2004}
\end{center}

\pagestyle{plain}
\pagenumbering{roman}
\setcounter{page}{2}
\newpage

\renewcommand{\baselinestretch}{2}
\small\normalsize
\hbox{\ }
 
\begin{center}
\Large{{DEDICATION}} \\[0.5in]
\end{center} 

To my wife, whose love, support and friendship have proven invaluable.

\newpage

\renewcommand{\baselinestretch}{2}
\small\normalsize
\hbox{\ }
 
\begin{center}
\Large{{ACKNOWLEDGMENTS}} \\[0.5in]
\end{center} 

The success that I have enjoyed as a graduate student is due, in part, to the hard work that I have contributed, but more importantly, to the input and support of several Professors.  First and foremost is the guidance of Dr. S. James Gates, Jr.  I would also like to acknowledge the input and technical guidance of Dr. Ioseph L. Buchbinder and Dr. Sergei M. Kuzenko.  I would like to thank Dr. Melanie Becker, Dr. Markus Luty, and Dr. Vincent G. J. Rodgers for proposing interesting lines of research that challenged my understanding.  I would also like to thank Dr. John Schwarz for his hospitality and support during my stay at the California Institute of Technology.  I must also acknowledge Dr. Pierre Ramond for his warm hospitality at the University of Florida where this thesis was written.  Finally, I must thank William Linch, III, for proving time and again that science, like wine, is always better when shared with a good friend.
\newpage
\tableofcontents 
\newpage

\setcounter{page}{1}
\pagenumbering{arabic}

\newpage
\chapter{Introduction}
\label{Introduction}

Supersymmetry plays a significant role in our understanding of physics beyond the standard model.  There are two ways of studying supersymmetric theories, standard field theory methods and superspace methods.  Standard field theory methods, also known as component methods, can be employed to write theories with supersymmetric multiplets.  In these formulations the supersymmetry algebra must be checked by explicit calculation.  The second way is to work in superspace, where supersymmetry is manifest.  The difference between the two approaches is similar to the difference between using the 4-vector or the 3-vector formulations of electromagnetism.  Although, both approaches are equivalent, certain problems can be solved more efficiently in one formalism versus the other.

This thesis focuses on using the superspace approach to construct higher spin, $s>1$, actions.  There are two reasons why superspace is the more efficient approach to writing actions for higher spin theories.  {\it First, the superspace formulation produces actions that are comparatively shorter than component formulations.}  This benefit will be invaluable as the number of auxiliary fields grows in higher spin theories.  Higher spin theories contain both propagating fields and auxiliary fields, and are therefore more complicated then regular theories even at the classical level.  Supersymmetric higher spin multiplets contain the usual higher spin propagating and auxiliary fields associated with the fermionic and bosonic sectors, as well as new auxiliary fields that are necessary for a supersymmetric description.  This proliferation of fields means that higher spin supersymmetric component actions will contain a large number of terms and in general will be unwieldy.  The superspace description replaces these large actions by comparatively short expressions written in terms of functions of the superspace coordinates, called superfields.  {\it The second reason why superspace is more efficient is that superfields solve the question of auxiliary fields at least for one supersymmetry, $N=1$, in four dimensions.}  A general understanding of the structure and uniqueness of the auxiliary field sector of supersymmetric theories remains a fundamental problem\cite{Gates2002}.  The construction of supersymmetric theories in the component approach involves solving the auxiliary field problem as well as getting the correct dynamics.  Superfields naturally contain all auxiliary fields that are necessary to close the supersymmetry algebra.  Thus, finding the correct dynamics is the only problem that needs to be solved in the superspace approach.  

Higher spin quantum field theories are non-renormalizable, but higher spin theories can still be defined classically as representations of the Poincar\'e algebra.  In the same sense, higher spin supersymmetric theories can be defined as representations of the Poincar\'e superalgebra.  There are two reasons to work in this general direction.  A purely academic reason is that the construction of higher spin supersymmetric field theories remains an unsolved problem in field theory.  A more popular reason would be to give some insight into superstring theory.  Superstring theory predicts an infinite tower of massive higher spin interacting supersymmetric quantum states.  The results of this thesis can shed light on the complexity of such a theory.  For example, one of the off-shell superspin-$3/2$ theories discussed in this thesis contains four vectors, three scalars, one 2-form, and one traceless rank two tensor(the massive graviton), and it seems likely that this list would nearly double at the superspin-$5/2$ level.  Even using superfields, the task of writing actions for superspin$> 3/2$ has proven intractable at this point in time.  The level of difficulty and complexity that arises for these classical actions with one higher spin multiplet must be considered trivial when compared to superstring theory.  It has been conjectured that the off-shell formulation of eleven-dimensional supergravity contains 32,768 bosonic and 32,768 fermionic  degrees of freedom\cite{Gates2002}, and this theory only represents the low energy limit of string theory.  It seems that any knowledge of higher spin supersymmetric models will be helpful in the future.

High spin theories have a long history that started in 1939, when M. Fierz and W. Pauli wrote the action that corresponds to a massive spin-2 representation of the Poincar\'e algebra.  30 years later, actions for free arbitrary higher spin massive\cite{Singh1974} and massless\cite{Fronsdal1978} theories were written.  Since then, a large amount of work has been done on interacting higher spin theories and higher spin theories in curved spacetime backgrounds \cite{Aragone1979} - \cite{Berkovits1997}.  Massless higher spin theories contain off-shell fields with spin $s$ and $s-2$, whereas, massive theories contain off-shell fields with spins $s,~s-2,~s-3,\dots,~0$.  Thus, massless theories require one auxiliary field while massive theories require $s-1$ auxiliary fields.  The supersymmetric massless arbitrary spin theories were found using gauge invariance as a guide\cite{Kuzenko1993}.\footnote{It is interesting to note that $N=0$ massive theories were found in 1974, 4 years before massless theories.}  In chapter \ref{Half-Integer Massless Arbitrary Superspin Actions}, the work of \cite{Kuzenko1993} on massless half integer arbitrary high spin theories is reformulated in terms of superspace projection operators.  Research in supersymmetric massive higher spin theories has only recently begun \cite{Buchbinder2002a, Buchbinder2002b}.  The main focus of chapter \ref{Massive Superspin Actions} is to review and extend this work.

It is also well known that massless actions in five dimensions can be mapped into massive actions in four dimensions.  Although five-dimensional theories using $N=1$ superfields have been relatively popular recently, \cite{Marcus1983} - \cite{Paccetti2004}, there has been no explanation of a correspondence to massive four-dimensional theories.  The authors of \cite{Gregoire2004} came closest to this connection by deriving a massive theory that is related to the massless five-dimensional theory in \cite{Linch2003}, but they did not understand how the two models were related.  Chapter \ref{Dimensional Reduction and Oxidation in N=1 Superspace}, explores this relationship by first dimensionally reducing the five-dimensional theory from \cite{Linch2003}.   The reduced theory is then shown to be  equivalent to the four-dimensional massive theory of \cite{Gregoire2004}.  Two new massless five-dimensional theories are postulated by applying the reverse of dimensional reduction to the new massive superspin-3/2 models derived in chapter \ref{Massive Superspin Actions}.  The creation of higher dimensional models from existing lower dimensional models is usually termed dimensional oxidation.

The main results of this thesis are contained in a set of six superspace actions.  There are four new versions of free massive models.  Three new versions of massive superspin-$3/2$ models are given in equations (\ref{Newresult1}), (\ref{newminmass}), and (\ref{newnewminmass}).  The last two theories are relatively important, since they complete the list of massive extensions of the three minimal linearized supergravities.  One new version of superspin-$1$ is given in equation (\ref{ss1final}).  Two new versions of massless five-dimensional linearized supergravity models are given in equations (\ref{5dnewminfinaction}) and (\ref{final5DNNMIN}).  These two theories are created by dimensional oxidization starting directly from (\ref{newminmass}) and (\ref{newnewminmass}).  A few other results are worth mentioning.  First, the gauge variation of the most general local action for arbitrary higher spin superfields without compensating superfields is given in equation (\ref{gaugetransH}).   Another important result is the list of field strengths for these theories, equations (\ref{fshs1})-(\ref{fshs5}).  Finally, at the end of the last chapter the Casimir operators for four-dimensional AdS superspace are given.

This thesis begins with a mathematic review of general superspace concepts in chapter \ref{Mathematical Background}.  More specific introductory material is presented at the beginning of each subsequent chapter.  The review material is largely based on introductory texts \cite{Buchbinder1995, Galperin2001, Gates1, Gates1983} and is meant to guide the interested reader through the minimal subset of material needed to understand the technical chapters.
\newpage
\chapter{Mathematical Background}
\label{Mathematical Background}

This chapter introduces all mathematical concepts that will be necessary for derivations in the following chapters.  Notations and conventions are taken mainly from \cite{Buchbinder1995} and are given in passing as part of the discussion.  Superspace is motivated from general field theory concepts and the Poincar\'e superalgebra is derived based on this analysis.  Much attention is given to the representation theory of the Poincar\'e superalgebra.  In particular, the on-shell equations that determine massive and massless representations of superfields are derived.  Superspace integration theory is reviewed and some simple superspace actions are discussed.

\section{Poincare Superalgebra}
\label{Poincare Superalgebra}
This section is dedicated to introducing the concept of superspace.  A discussion of superspace must begin with spinor representation theory.  Spinors are representations of the Lorentz algebra and their representations differ greatly depending on the spacetime dimension, unlike bosons.  The Lorentz algebra is:
\bea
\label{Lorentz}
[\cm_{ab},\cm_{cd}]=\h_{ad}\cm_{bc}-\h_{ac}\cm_{bd}
+\h_{bc}\cm_{ad}-\h_{bd}\cm_{ac}~~,~~
\eea
where $\cm_{ab}$ is the generator of boosts and rotations and $\h_{ab}$ is the spacetime metric.  The most obvious representation of this algebra is given by the differential operator $\cm_{ab}(x):=(x_a\pa_b-x_b\pa_a)$.  $\cm_{ab}(x)$ is a representation of $\cm_{ab}$ that acts on the space of fields.  A representation of (\ref{Lorentz}) that acts on the space of bosonic tensors is $(\cm_{ab})_c{}^d:=(\h_{ca}\d_b{}^d-\h_{cb}\d_a{}^d)$.  Here matrix multiplication over the vector valued indices c and d is assumed.  $(\cm_{ab})_c{}^d$ is an adjoint matrix representation of the Lorentz algebra.  The spinor representation is also realized as a matrix and is easily constructed by using the Dirac or ``gamma'' matrices which are defined through the Dirac algebra:
\bea
\label{Dirac}
\{\g_a,\g_b\}=-2\h_{ab}I~~.~~
\eea
It is a short exercise to show that $(\cm_{ab})_\a{}^\b:={1\over4}[\g_a,\g_b]_\a{}^\b$ also satisfies the Lorentz algebra.  The objects that are transformed by $(\cm_{ab})_\a{}^\b$ under Lorentz rotations are spinors.  Note that the sign in front of the metric is convention.  Changing this sign does not change the fact that $(\cm_{ab})_\a{}^\b$ is a representation of the Lorentz algebra.  

The construction of an explicit matrix realization of (\ref{Dirac}) is necessary to work with spinors.  There are two important properties of the Dirac matrices that have serious consequences for spinors in various dimensions.  The first property is the irreducible matrix dimension, which is the smallest spinor representation.  In general, the Dirac algebra is solved by matrices that have dimension $d=2^{\lfloor{D\over2}\rfloor}$.  The floor operation in the exponent pairs spacetime dimensions together.  This paring leads to the general feature that in even spacetime dimensions the $2^{\lfloor{D\over2}\rfloor}$-dimensional matrices are reducible.  In even dimensions there always exists a matrix, $\g^{D+1}$, that anti-commutes with the set of Dirac matrices and therefore can be used to construct a projection operator and reduce the representation to $\frac 122^{\lfloor{D\over2}\rfloor}$.  The reality of the $2^{\lfloor{D\over2}\rfloor}$-dimensional matrices is the second important property and is determined by the signature of the metric.   A complex spinor has twice as many degrees of freedom as a real or Majorana spinor.  Thus, there is the opportunity to reduce the number of real degrees of freedom of the spinor to $\frac 122^{\lfloor{D\over2}\rfloor}$, if the projection operator is real.  This actually occurs in $4n+2$ dimensions and explains the different types of $N=2$ supersymmetry in two and ten dimensions.

The irreducible spinor representation in four dimensions can be constructed using the 2x2 Pauli matrices as a basis:
\bea
\nonumber
\s_1=\left(\begin{array}{cc}
0 & 1 \\
1 & 0\ \\
\end{array}\right)~~,~~
\s_2=\left(\begin{array}{cc}
0 & -i \\
i & 0\ \\
\end{array}\right)~~,~~
\s_3=\left(\begin{array}{cc}
1 & 0 \\
0 & -1\ \\
\end{array}\right)~~,~~
\eea
\bea
\rightarrow ~~~~~\s_i\s_j=\d_{ij}I+i\e_{ijk}\s_k~~.~~
\eea
Matrices that are larger than two dimensions can be constructed  by using the outer product notation $\s_i\otimes\s_j$.  The outer product notation can be realized as placing $\s_j$ in every entry of $\s_i$, for example:
\bea
\s_3\otimes\s_2=\left(\begin{array}{cc}
\s_2 & 0 \\
0 & -\s_2 \\
\end{array}\right)~~.~~
\eea
The outer product structure is respected under multiplication:
\bea
(\s_2\otimes\s_3) \cdot (\s_1\otimes\s_2)=\s_2\s_1\otimes\s_3\s_2=
-\s_3\otimes\s_1~~.~~
\eea
Taking the metric, $\h_{ab}$, to be mostly plus:
\bea
\h_{ab}=\left(\begin{array}{cccc}
-1 & 0 & 0 & 0\\
0 & +1 & 0 & 0\\
0 & 0 & +1 & 0\\
0 & 0 & 0 & +1\\
\end{array}\right)~~,~~
\eea
the Dirac algebra is satisfied by four-dimensional matrices:
\bea
(\g_a)_\a{}^\b &=& (\s_1\otimes I, i\s_2\otimes\s^1,i\s_2\otimes\s_2,i\s_2\otimes\s_3 )~~~~
{\rm Weyl~Basis}~~,~~\cr
(\g_a)_\a{}^\b &=& (-i\s_2\otimes\s_1, \s_2\otimes\s_2, \s_1\otimes I, \s_3\otimes I)~~~~
{\rm Real~Basis}~~.~~
\eea
Since the Weyl basis is block diagonal it is obvious that the four-dimensional Dirac matrices are reducible.  On the other hand, the real basis means that Majorana spinors can also be defined.  The only question is if the projection respects the reality.  The projection matrices can be constructed using the matrix $\g^5 = -\frac i{4!}\e^{abcd}\g_{a}\g_b\g_c\g_{d}$ which anti-commutes with $(\g_a)$ and squares to $+1$.  The projection matrices are:
\bea
P_\pm:={1\over2}(I\pm\g^5)~~,~~
\eea
In the real basis $\g^5$ is purely imaginary, so $P_+{}^*=P_-$, Thus, conjugation does not commute with the projection and the irreducible spinor in $D=4$ is a two-component complex spinor or a four component Majorana spinor.  The two component spinor representation is the easiest with which to calculate and will be used throughout this thesis.  Complex conjugation maps between the two chiralities.  Projecting and then conjugating the Majorana spinor shows how the complex irreducible representation coincides with the Majorana spinor:
\bea
(P_+\Psi)^*=P_+{}*\Psi^*=P_-\Psi~~.~~
\eea
Thus, the Majorana spinor is a four component object that holds a 2-component irreducible spinor and its conjugate.

In order to set up the standard formalism of superspace in 
$D=4$, some more notation is necessary.  The non-zero projections of the Dirac matrices are defined as the sigma matrices of \cite{Buchbinder1995}:
\bea
P_+ \g^a P_-=: (\s^a)_{\a\dot\a}~~,~~P_- \g^a P_+=: (\tilde\s^a)^{\dot\a \a}~~.~~
\eea
On the surface the dotted index can be thought of as keeping track of the orthogonality of the projected subspaces $P_+P_-=0$.  
It is natural to contract spinors to make invariant objects and since spinors should anti-commute the spinor metric should be anti-symmetric.  In the two-dimensional spinor space the Levi-Cevita tensor is the only anti-symmetric tensor.  It is defined under the conventions of \cite{Buchbinder1995} as:
\bea
\e_{\a\b}=-\e_{\b\a}~~,~~\e_{12}=-1~~,~~\cr
\e^{\a\b}=-\e^{\b\a}~~,~~\e^{12}=1~~,~~\cr
\e_{\a\b}\e^{\g\d}=-\d_\a{}^\g\d_\b{}^\d+\d_\b{}^\g\d_\a{}^\d~~,~~
\eea
Raising and lowering of spinor indices is defined by:
\bea
\Psi^\a=\e^{\a\b}\Psi_\b~~,~~\chi_\a=\e_{\a\b}\chi^\b~~.~~
\eea
These definitions imply:
\bea
\e^{\d\a}\e^{\g\b}\e_{\a\b}=-\e^{\d\g}~~.~~
\eea
The opposite chirality Levi-Cevita tensor $\e_{\dot\a\dot\b}$ is defined through conjugation , i.e $(\e_{\a\b})^*=\e_{\dot\a\dot\b}$.  Conjugation of products of spinors inverts the order,  $(\chi_\a\psi_\b)^*$ $=$ $\bar\psi_{\dot\b}\bar\chi_{\dot\a}$.  The projected Dirac algebra takes the form:
\bea
(\s_a\tilde\s_b+\s_b\tilde\s_a)_\a{}^\b=-2\h_{ab}\d_\a{}^\b
~~,~~\cr
(\tilde\s_a\s_b+\tilde\s_b\s_a)^{\dot\a}{}_{\dot\b}=-2\h_{ab}\d^{\dot\a}{}_{\dot\b}~~,~~
\eea
and the Fierz identities become:
\bea
\label{sigmafierz}
Tr(\s_a\tilde\s_b)=(\s_a)_{\a\dot\a}(\tilde\s_b)^{\dot\a\a}=-2\h_{ab}~~,~~~~~~~~~~~~~\cr
(\s^a)_{\a\dot\a}(\tilde\s_a)^{\dot\b\b}
=-2\d_\a{}^\b\d^{\dot\b}{}_{\dot\a}~~,~~~~~~~~~~~\cr
\s_a\tilde\s_b\s_c=(\h_{ac}\s_b-\h_{bc}\s_a-\h_{ab}\s_c)
+i\e_{abcd}\s^d~~,~~\cr
\tilde\s_a\s_b\tilde\s_c=(\h_{ac}\tilde\s_b-\h_{bc}\tilde\s_a-\h_{ab}\tilde\s_c)
-i\e_{abcd}\tilde\s^d~~,~~
\eea
The sigma matrices are also related by raising indices:
\bea
\label{related}
\e^{\a\b}\e^{\dot\a\dot\b}(\s_{\rm a})_{\b\dot\b}
=(\tilde\s_{\rm a})^{\dot\b\b}~~.~~
\eea
All of these identities and relations will be used without comment in the rest of this thesis.

In superspace it is convenient to convert all vector indices into spinor indices using the sigma matrices.  This is accomplished by contracting all vector indices with sigma matrices and separating out the irreducible spin-tensor representations.  The vector is just:
\bea
V^{\dot\a\a}:=(\tilde\s_b)^{\dot\a\a}V^b~~.~~
\eea
So a pair of undotted and dotted indices represents a vector and it is natural to use the notation $\un a:=\a\dot\a$.  A rank two tensor is a better example.  A rank two tensor has the following decomposition in the vector space:
\bea
\label{TensorDecomp}
T_{ab}=\Tilde T_{ab}+X_{ab}+{1\over4}\h_{ab}T~~,~~
\eea
here $\Tilde T_{ab}$ is symmetric and traceless, $X_{ab}$ is antisymmetric, and $T$ is the trace of $T_{ab}$.  Contracting with sigma matrices $T_{ab}$ becomes:
\bea
\label{spindecomp}
T_{\a\dot\a\b\dot\b}={1\over4}(T_{(\a\b)(\dot\a\dot\b)}+\e_{\a\b}\e_{\dot\a\dot\b}T^\g{}_\g{}^{\dot\g}{}_{\dot\g}+\e_{\a\b}T^\g{}_\g{}_{(\dot\a\dot\b)}
+\e_{\dot\a\dot\b}T_{(\a\b)}{}^{\dot\g}{}_{\dot\g})~~.~~
\eea
The symmetrization and anti-symmetrization notation $(\a\b\g\cdots)$ and $[ab\cdots]$ does not contain any factorials, therefore, $(\a\b)=\a\b+\b\a$ and $[ab]=ab-ba$.  The explicit writing of factorials is the only difference between the notation in this thesis and \cite{Buchbinder1995}.  The Levi-Cevita tensors in (\ref{spindecomp}) appear through the identity:
\bea
\label{contract}
X_{\a\b}-X_{\b\a}=\e_{\a\b}X^\g{}_\g~~.~~
\eea
The ability to replace pairs of anti-symmetric spinor indices with Levi-Cevita symbols is one reason why $D=4$ superspace is so successful.  In higher dimensions, anti-symmetric pairs of indices carry larger spacetime tensor representations, and, along with the larger spinor dimension, make superspace formulations in higher dimensions quite complicated.  Substitution of (\ref{TensorDecomp}) into (\ref{spindecomp}) shows the peculiar equivalence of spinor tensors to vector tensors: 
\bea
T_{(\a\b)(\dot\a\dot\b)}=(\s^a)_{(\a(\dot\a}(\s^b)_{\b)\dot\b)}T_{ab}
=(\s^a)_{(\a(\dot\a}(\s^b)_{\b)\dot\b)}\Tilde T_{ab}~~,~~\cr
T^\g{}_\g{}^{\dot\g}{}_{\dot\g}=Tr(\tilde\s^a\s^b)T_{ab}=-2T~~~~~~,~~\cr
T_{(\a\b)}{}^{\dot\g}{}_{\dot\g}=
\e^{\dot\b\dot\a}(\s^a)_{(\a\dot\a}(\s^b)_{\b)\dot\b}T_{ab}
=\e^{\dot\b\dot\a}(\s^a)_{(\a\dot\a}(\s^b)_{\b)\dot\b}X_{ab}~~.~~
\eea
So an anti-symmetric rank two vector tensor is equivalent to a symmetric rank two spin tensor.  Another tensor that will show up in this thesis is the Weyl tensor, which is the propagating degree of freedom for a spin-2 particle.  The Weyl tensor is traceless and has the following symmetries:
\bea
C_{abcd}=C_{cdab}=-C_{bacd}=-C_{abdc}~~,~~C_{[abc]}{}^d=0~~,~~
\eea
and can be written as a completely symmetric rank four spin tensor:
\bea
\label{spinweyl}
C_{\un a\un b\un c\un d}=
\e_{\dot\a\dot\b}\e_{\dot\g\dot\d}C_{\a\b\g\d}
+\e_{\a\b}\e_{\g\d}\Bar C_{\dot\a\dot\b\dot\g\dot\d}~~.~~
\eea
A nontrivial check of this is that the number of components  of each tensor are equal.  The rank four vector tensor, $C_{abcd}$ has ten components:
\bea
{\tiny\yng(2,2)}\ominus{\tiny\yng(1,1)}=20-10=10~~,~~
\eea
whereas the rank four tensor, $C_{\a\b\g\d}$ has five complex components:
\bea
{\tiny\yng(4)}=5~~.~~
\eea
In later chapters, it will be necessary to introduce tensors with large numbers of symmetric indices.  These indices will be denoted by placing the number of symmetric indices in parenthesis.  In this notation the Weyl tensor becomes $C_{\a_1\a_2\a_3\a_4}$$=C_{\a(4)}$.
The spin tensor notation begins to put bosons and fermions on the same footing by using the same language to describe these fields, but supersymmetry requires that both types of fields are treated equally.

Supersymmetry is a symmetry between bosons and fermions.  The most basic of these objects are the complex scalar field, $\varphi$, and the Weyl spinor $\psi_\a$.  In order to write some transformation that facilitates this symmetry a spinor parameter  $\z^\a$ must be introduced.  The simplest transformation of $\varphi$ is:
\bea
\d_\z \varphi =  \z^\a\psi_\a~~.~~
\eea
Since propagating fermions have mass dimension $\frac 32$, the only possibility for the supersymmetry variation of $\psi_\a$ is:
\bea
\d_\z \psi_\a = -2i\bar\z^{\dot\a}\pa_{\un a}\varphi~~.~~
\eea
The commutator of two of these transformations acting on $\varphi$ is:
\bea
\label{delphi}
[\d_{\z_1}, \d_{\z_2}]\varphi
=-2i(\z^\a_2\bar\z^{\dot\a}_1
-\z^\a_1\bar\z^{\dot\a}_2)\pa_{\un a}\varphi~~.~~
\eea
These simple supersymmetry transformations seem to close onto the partial derivative or momentum generator, but the commutator acting on the spinor reveals a problem:
\bea
[\d_{\z_1}, \d_{\z_2}]\psi_\b
=-2i(\z^\a_2\bar\z^{\dot\a}_1-\z^\a_1\bar\z^{\dot\a}_2)
\pa_{\dot\a\b}\psi_\a~~.~~
\eea
This is almost in same form as (\ref{delphi}), except that the free index $\b$ is on the partial derivative and not on $\psi_\a$.  A short manipulation puts the commutator acting on $\psi_\b$ into the form:
\bea
\label{psicom}
[\d_{\z_1}, \d_{\z_2}]\psi_\b
=-2i(\z^\a_2\bar\z^{\dot\a}_1-\z^\a_1\bar\z^{\dot\a}_2)
\pa_{\un a}\psi_{\b}
-2i(\z_{2\b}\bar\z^{\dot\a}_1-\z_{1\b}\bar\z^{\dot\a}_2)
\pa_{\dot\a}{}^\g\psi_{\g}~~.~~
\eea
The first term is now of the form (\ref{delphi}), but the second term looks like the equation of motion for $\bar\psi_{\dot\a}$.  This is an extremely fundamental result in supersymmetry.  It has nothing to do with the simplicity of this derivation.  If $\psi_\a$ is on-shell, e.g. it obeys the equation of motion $\pa_{\dot\a}{}^\g\psi_\g=0$, then the simple transformations close properly on all of the fields in this example.  The supersymmetry algebra closes on-shell.

There are some problems with a definition of supersymmetry that closes on-shell.  Quantum fields are integrated over arbitrary values of momentum, not just values that correspond to propagating states.  So supersymmetric quantum field theories need to be defined off-shell.  Another problem is that the equations of motion are different depending on the theory.  This means that the supersymmetry transformations must be engineered for each theory.  In other words, a supersymmetry that is implemented on-shell is model dependent.  This is not a problem for any one specific model, but if the goal is to understand the fundamentals of supersymmetry, a model independent formulation is necessary.

One way to make the $\varphi$, $\psi_\a$ system model independent is to add another scalar field, $F$.  $F$ is a field with mass dimension 2, and it shows up in the transformation of the spinor:
\bea
\d_\z \psi_\a = -2i\bar\z^{\dot\a}\pa_{\un a}\varphi + \z_\a F~~.~~
\eea
The addition of $F$ does not change (\ref{delphi}).  The transformation of $F$ can be discerned from the commutator acting on $\psi_\a$:
\bea
[\d_{\z_1}, \d_{\z_2}]\psi_\b
=-2i(\z^\a_2\bar\z^{\dot\a}_1-\z^\a_1\bar\z^{\dot\a}_2)
\pa_{\dot\a\a}\psi_{\b}
-2i(\z_{2\b}\bar\z^{\dot\a}_1-\z_{1\b}\bar\z^{\dot\a}_2)
\pa_{\dot\a}{}^\g\psi_{\g}
+\z_{2\b} \d_1F-\z_{1\b} \d_2F~~,~~
\eea
\bea
\Rightarrow~~\d F = 2i\bar\z^{\dot\a}\pa_{\dot\a}{}^\g\psi_\g
~~.~~
\eea
This algebra closes properly on F:
\bea
[\d_{\z_1}, \d_{\z_2}]F
=-2i(\bar\z^{\dot\a}_2\z^\a_1-\bar\z^{\dot\a}_1\z^\a_2)\pa_{\un a} F
~~.~~
\eea
The auxiliary field $F$ has allowed a description of supersymmetry that closes off-shell and is model independent.  The supersymmetry algebra consistently closes into the partial derivative on all fields in the multiplet.  The abstract algebra takes the form:
\bea
\label{abstract}
[\d_{\z_1}, \d_{\z_2}]=-2i(\bar\z^{\dot\a}_2\z^\a_1-\bar\z^{\dot\a}_1\z^\a_2)\pa_{\un a}~~.~~
\eea
This algebra would be more useful if it was written in terms of abstract generators.  The partial derivative can be replaced by the momentum generator $P_{\un a}=-i\pa_{\un a}$.  The variation, $\d_\z$, can be replaced by a spinor valued generator:
\bea
\label{deltaq}
\d_\z=:i\z^\a Q_\a+i\bar\z_{\dot\a}\Bar Q^{\dot\a}~~,~~
\eea
so that $\d_\z$ preserves reality and $(Q_\a)^*=-\Bar Q_{\dot\a}$.
The algebra for the generator $Q_\a$ can be deduced by substituting (\ref{deltaq}) into (\ref{abstract}):
\bea
[\d_{\z_1}, \d_{\z_2}]
=\z^\a_1\z^\b_2 \{Q_\a, Q_\b\}
+\bar\z_{1\dot\a}\bar\z_{2\dot\b}\{\Bar Q^{\dot\a},\Bar Q^{\dot\b}\}
+(\bar\z_2{}^{\dot\a}\z^\a_1
-\bar\z_1{}^{\dot\a} \z^\a_2) \{Q_\a,\Bar Q_{\dot\a}\}~~,~~
\eea
and comparing this to (\ref{abstract}).  The full Poincar\'e superalgebra is:
\bea
\nonumber
\{Q_\a,\Bar Q_{\dot\a}\}=2P_{\un a}~~,~~
\eea
\bea
\nonumber
\{Q_\a,Q_\b\}=0~~,~~\{\Bar Q_{\dot\a},\Bar Q_{\dot\b}\}=0~~,~~
[\cj_{\a\b},\Bar\cj_{\dot\g\dot\d}]=0~~,~~
\eea
\bea
\nonumber
[\cj_{\a\b},\cj_{\g\d}]=\frac i2\Big\{
\e_{\a\g}\cj_{\b\g}+\e_{\a\d}\cj_{\b\g}
+\e_{\b\g}\cj_{\a\d}+\e_{\b\d}\cj_{\a\g}
\Big\}~~,~~
\eea
\bea
\nonumber
[\Bar\cj_{\dot\a\dot\b},\Bar\cj_{\dot\g\dot\d}]=\frac i2\Big\{
\e_{\dot\a\dot\g}\Bar\cj_{\dot\b\dot\g}
+\e_{\dot\a\dot\d}\Bar\cj_{\dot\b\dot\g}
+\e_{\dot\b\dot\g}\Bar\cj_{\dot\a\dot\d}
+\e_{\dot\b\dot\d}\Bar\cj_{\dot\a\dot\g}
\Big\}~~,~~
\eea
\bea
\nonumber
[\cj_{\a\b},P_{\un c}]=\frac i2(\e_{\g\a}P_{\b\dot\g}+\e_{\g\b}P_{\a\dot\g})~~,~~
[\bar\cj_{\dot\a\dot\b},P_{\un c}]=\frac i2(\e_{\dot\g\dot\a}P_{\g\dot\b}+\e_{\dot\g\dot\b}P_{\g\dot\a})~~,~~
\eea
\bea
\nonumber
[\cj_{\a\b},Q_\g]=\frac i2(\e_{\g\a}Q_\b+\e_{\g\b}Q_\a)
~~,~~[\cj_{\a\b},\Bar Q_{\dot\g}]=0~~,~~
\eea
\bea
[\Bar\cj_{\dot\a\dot\b},\Bar Q_{\dot\g}]
=\frac i2(\e_{\dot\g\dot\a}\Bar Q_{\dot\b}
+\e_{\dot\g\dot\b}\Bar Q_{\dot\a})
~~,~~[\Bar\cj_{\dot\a\dot\b},Q_\g]=0~~.~~
\eea
$P_{\un a}$ is the usual translation generator and $\cj_{ab}$ has been written in spin tensor notation through:
\bea
\cj_{\a\b}=-\frac 18(\s^{[a}\tilde\s^{b]})_{\a\b}\cj_{ab}
~~,~~
\eea
and $Q_\a$ and $\Bar Q_{\dot\a}$ are the supersymmetry generators.
\section{On-shell Spin States of Irreducible Representations}
\label{On-shell Spin States of Irreducible Representations}
This section focuses on determining the spin states that make up irreducible on-shell massless and massive representations of the Poincar\'e superalgebra.  An easy way to do this is to choose a Lorentz frame and use fermionic harmonic oscillator representations.  The derivation presented here is taken from \cite{Galperin2001}, where the representation theory for arbitrary numbers of supersymmetry is discussed.  For massless representations, $P^a=(p,0,0,p)$, and the algebra becomes:
\bea
\{Q_\a, \Bar Q_{\dot\a}\}=2(I+\s_3)_{\a\dot\a}p~~.~~
\eea
The only non-zero relation is $ \{Q_1, \Bar Q_{\dot1}\}=4p$.  The vacuum, $|\l>$ with helicity $\l$, is annihilated by all generators accept $\Bar Q_{\dot1}$.  $\Bar Q_{\dot1}$ anti-commutes with itself, so its square is zero and the states are:
\bea
|\l>~~,~~\Bar Q_{\dot1}|\l>~~.~~
\eea
Thus, an on-shell irreducible massless representation of supersymmetry contains two states.  One of helicity $\l$ and another of helicity $\l-\frac 12$.  The helicity of $\Bar Q_{\dot1}$ can be determined by noting that on-shell:
\bea
\{\pa_{\a\dot\b}\Bar Q^{\dot\b}, \pa_{\b\dot\a}Q^\b\}=+2i\pa_{\un a}\Box=0~~,~~\cr
\Rightarrow~~\pa_{\a\dot\b}\Bar Q^{\dot\b}=0~~,~~
\eea
which is the correct supplementary condition for a helicity $-\frac 12$ field.

The massive representations are determined using $p^a=(m,0,0,0)$ with which the algebra becomes:
\bea
\{Q_\a, \Bar Q_{\dot\a}\}=2(I)_{\a\dot\a}m~~.~~
\eea
The vacuum, $|s>$ with spin $s$, is annihilated by $Q_\a$.  The spin states are:
\bea
|s>~~,~~\Bar Q_{\dot\a}|s>~~,~~\Bar Q^{\dot\a}\Bar Q_{\dot\a}|s>
~~.~~
\eea
$|s>$ can be raised only twice since $Q_\a Q_\b Q_\g=0$.  The first excited state can be reduced with respect to $s$.  If $s=0$, this state has spin $\frac 12$.  If $s\not=0$, the irreducible tensor combinations are $s\pm\frac 12$, corresponding to the addition of spin angular momentum.  Thus, an on-shell irreducible massive representation of supersymmetry contains four states with spins-$(s, s\pm\frac 12, s)$ for $s\not=0$ and $(0,\frac 12,0)$ for $s=0$.

\section{Superspace}
\label{Superspace}
The Poincar\'e superalgebra has a field representation over a space that contains the usual spacetime coordinates.  The anti-commuting nature of the supersymmetry generators cannot be accounted for by the usual bosonic momentum and position coordinates.  It is necessary to use anti-commuting momentum and position coordinates.  This is done by introducing Grassmann variables, $\q_\a$, and their derivatives, $\pa_\a$ such that:
\bea
\label{defgrass}
\pa_\a\q^\b=\d_\a{}^\b~~,~~
\pa^\a\q_\b=\d_\b{}^\a~~,~~
\q_\a\q_\b=-\q_\b\q_\a~~,~~\pa_\a\pa_\b=-\pa_\b\pa_\a~~,~~\cr
\bar\pa_{\dot\a}\bar\q^{\dot\b}=\d^{\dot\b}{}_{\dot\a}~~,~~
\bar\pa^{\dot\a}\bar\q_{\dot\b}=\d_{\dot\b}{}^{\dot\a}~~,~~
\bar\q_{\dot\a}\bar\q_{\dot\b}=-\bar\q_{\dot\b}\bar\q_{\dot\a}
~~,~~
\bar\pa_{\dot\a}\bar\pa_{\dot\b}=-\bar\pa_{\dot\b}\bar\pa_{\dot\a}~~,~~
\eea
where the contravariant and covariant partial derivatives are related by $\pa^\a=-\e^{\a\b}\pa_\b$ and $\bar\pa_{\dot\a}=-\e_{\dot\a\dot\b}\bar\pa^{\dot\b}$.  The spacetime variables $x^{\un a}$ and the anti-commuting variables $\q_\a$ and $\bar\q_{\dot\a}$ define superspace.  Superfields are functions of the Grassmann and bosonic coordinates.  Since products of more than two anti-commuting coordinates vanish, i.e. $\q_\a\q_\b\q_\g=0$, a superfield has a finite taylor expansion in the Grassmann coordinates:
\bea
\label{taylor}
V(x,\q,\bar\q)=A(x)+\q^\a\l_\a(x)+\bar\q_{\dot\a}\bar\chi^{\dot\a}(x)
+\q^\a\q_\a E(x)
+\bar\q_{\dot\a}\bar\q^{\dot\a}F(x)\cr
+\q^\a\bar\q^{\dot\a}V_{\un a}(x)
+\q^\a\bar\q_{\dot\a}\bar\q^{\dot\a}\psi_\a(x)
+\bar\q_{\dot\a}\q^\a\q_\a\bar\h^{\dot\a}(x)
+\q^\a\q_\a\bar\q_{\dot\a}\bar\q^{\dot\a} D(x)~~.~~
\eea
There is one subtlety in the question of the conjugation of the Grassmann derivatives.  To see this equate $\bar\pa_{\dot\a}\Bar V$ to $(\pa_\a V)^*$:
\bea
\bar\pa_{\dot\a}\Bar V=-(-1)^{\e(V)}\bar\l_{\dot\a}+\co(\q,\bar\q)
=(\pa_\a V)^*=(\l_\a+\co(\q,\bar\q))^*=\bar\l_{\dot\a}+\co(\q,\bar\q)~~,~~\cr
\Rightarrow~~(\pa_\a V)^*=-(-)^{\e(V)}\bar\pa_{\dot\a}\Bar V~~.~~
\eea
In this expression $\e(V)$ equals 0 if $V$ is bosonic and 1 if $V$ is fermionic.  With the coordinates and partial derivatives in hand, superfield representations of the Poincar\'e superalgebra can be constructed.  By dimensional analysis $Q_\a$ is proportional to $\pa_\a$ and $\bar\q^{\dot\a}\pa_{\un a}$.  A combination of these terms that satisfies the superalgebra and has the property $(Q_\a)^*=-\Bar Q_{\dot\a}$ is:
\bea
\label{qfield}
Q_\a = i\pa_\a +\bar\q^{\dot\a}\pa_{\un a}~~,~~
\Bar Q_{\dot\a}=-i\bar\pa_{\dot\a}-\q^\a\pa_{\un a}~~.~~
\eea
The boost and rotation generator takes the form:
\bea
\cj_{\a\b}=-\frac i4x_{(\a}{}^{\dot\b}\pa_{\b)\dot\b}
+\frac i2\q_{(\a}\pa_{\b)}-i\cm_{\a\b}~~,~~\cr
\Bar\cj_{\dot\a\dot\b}=-\frac i4x_{(\dot\a}{}^{\b}\pa_{\dot\b)\b}
+\frac i2\bar\q_{(\dot\a}\bar\pa_{\dot\b)}-i\Bar\cm_{\dot\a\dot\b}~~,~~
\eea
where it is understood that the Lorentz generator $\cm_{\a\b}$ acts only on external superfield indices.  There is no restriction on the external index structure of superfields.

Superfields naturally contain multiplets of fermionic and bosonic fields.  Since irreducible representations of the Poincar\'e superalgebra contain at most two bosons and two fermions, a general superfield is not an on-shell irreducible representation.  Another observation about superfields is that the mass dimension of the component fields changes at each level of the Taylor series expansion.  This fact places supersymmetric auxiliary fields, i.e. $F$, naturally into the superfield description of supersymmetry.

\section{Covariant Spinor Derivatives}
\label{Covariant Spinor Derivatives}

The expansion of a superfield in powers of $\q_\a$ and $\bar\q_{\dot\a}$ reveals too many regular fields to describe irreducible supersymmetric multiplets.  There must be some consistent way to project irreducible multiplets out of superfields.  If there exists an operator that anti-commutes with the supersymmetry generator $Q_\a$, it could be used to label the irreducible representations contained in a general superfield.  Such an operator exists and can be found by taking the other linearly independent combination of the terms that make up $Q_\a$:
\bea
D_\a := \pa_\a+i\bar\q^{\dot\a}\pa_{\un a}
~~,~~
\Bar D_{\dot\a}:=-\bar\pa_{\dot\a}-i\q^\a\pa_{\un a}~~.~~
\eea
It is easy to check that the $D$'s have the following properties:
\bea
\label{dcom}
\{ D_\a, \Bar D_{\dot\a}\}=-2i\pa_{\un a}~~,~~
\eea
\bea
\label{dsym}
\{ D_\a, D_\b\}=0~~,~~\{ \Bar D_{\dot\a}, \Bar D_{\dot\b}\}=0~~,~~
\eea
\bea
\{ D_\a, Q_\b\}=0~~,~~\{D_\a, \Bar Q_{\dot\a}\}=0~~.~~
\eea
The duplication of the $Q$ algebra by the $D$ algebra seems relatively trivial.  It is important to make the distinction that the $Q$'s exist as abstract supersymmetry generators, where as, the $D$'s are a byproduct of the field representation and only exist in superspace.  Thus, the $Q$'s must be understood as fundamental objects and the $D$'s as differential operators.  If the $D$ operators are taken to be abstract generators then the abstract algebra of $D$'s  and $Q$'s would be an $N=2$ supersymmetry algebra.  When passing to a field representation from the $N=2$ algebra, another set of $N=2$ $D$'s would appear in the superfield space.

The supersymmetry transformation of $D_\a$ acting on some superfield is:
\bea
\label{dtrans}
[\d_\z, D_\a V(x, \q, \bar\q)]&=&D_\a\d_\z V(x,\q,\bar\q)\cr
&=&D_\a (i\z^\a Q_\a+i\bar\z_{\dot\a}\Bar Q^{\dot\a})V(x,\q,\bar\q)\cr
&=&(i\z^\a Q_\a+i\bar\z_{\dot\a}\Bar Q^{\dot\a})D_\a V(x,\q,\bar\q)~~.~~
\eea
The object $D_\a V(x,\q,\bar\q)$ transforms properly, or covariantly, under supersymmetry and is therefore a superfield.  This means that any object constructed out of some combination of covariant spinor derivatives, partial derivatives and superfields is again a superfield.  

Before the covariant spinor derivatives can be put to use, some notation and identities are in order.  The notation for the square of the covariant spinor derivative is:
\bea
D^2:=D^\a D_\a~~,~~\Bar D^2:= \Bar D_{\dot\a}\Bar D^{\dot\a}~~,~~
\eea
which can be used with (\ref{dsym}) to find:
\bea
D_\a D_\b = \frac 12\e_{\a\b}D^2~~,~~
\Bar D_{\dot\a}\Bar D_{\dot\b}=-\frac 12\e_{\dot\a\dot\b}\Bar D^2
~~.~~
\eea
The following list of the most useful identities is derived using (\ref{dcom}):
\bea
\nonumber
[D^2,\Bar D_{\dot\a}] = -4i\pa_{\un a}D^\a~~,~~
[\Bar D^2, D_{\a}] = 4i\pa_{\un a}\Bar D^{\dot\a}
\eea
\bea
\nonumber
\{ D^2 , \Bar D^2 \}-2D^\a\Bar D^2 D_\a=16\Box~~,~~
\eea
\bea
\nonumber
D^2\Bar D^2D^2=16\Box D^2~~,~~
\Bar D^2 D^2 \Bar D^2=16\Box \Bar D^2
\eea
\bea
\nonumber
[D^2,\Bar D^2]=-4i\pa^{\un a}[D_\a,\Bar D_{\dot\a}]~~,~~
\eea
\bea
D^\a\Bar D^{\dot\a}D_\a=-\frac 12\{D^2,\Bar D^{\dot\a}\}~~.~~
\label{dalg}
\eea
For future reference, the superspace conjugation rules are:
\bea
(D_\a V)^*=(-1)^{\e(V)}\Bar D_{\dot\a}V^*~~~~,~~~~
(D^2V)^*=\Bar D^2 V^*~~.~~
\eea 
Also, the order of all fermions is  inverted under conjugation:
\bea
(\psi_\a\chi_\b\bar\q_{\dot\a})^*=\q_\a\bar\chi_{\dot\b}\bar\psi_{\dot\a}~~.~~
\eea

The covariant spinor derivative can be used to covariantly project out the component fields in the superfield expansion (\ref{taylor}).  $D_\a$ and $\Bar D_{\dot\a}$ are approximately $\pa_\a$ and $\bar\pa_{\dot\a}$, and can be used strip off the $\q$'s in the expansion.  Once a derivative is applied, $\q$ and $\bar\q$ must be set to zero.  Setting $\q=\bar\q=0$ will be denoted by a vertical bar $|$.  The projection of (\ref{taylor}) is:
\bea
\nonumber
V|=A~~,~~D_\a V|=\l_\a~~,~~\Bar D_{\dot\a}V|=\bar\chi_{\dot\a}~~,~~
\eea
\bea
\nonumber
D^2 V|=-4E~~,~~\Bar D^2 V|=-4F~~,~~
[D_\a,\Bar D_{\dot\a}]V|=2V_{\un a}~~,~~
\eea
\bea
D_\a\Bar D^2V|=-4\s_\a~~,~~ 
\Bar D_{\dot\a}D^2V| =-4\bar\r_{\dot\a} ~~,~~
D^2\Bar D^2V|=32G~~.~~
\eea
The last three components are not the same as in the original expansion (\ref{taylor}).  This arises because the $D$'s are not purely spinor derivatives.  The last three components are related by field redefinitions to those in the expansion.  This can be seen by noting:
\bea
\nonumber
D_\a V|=\pa_\a V|~~,~~D^2V|=-\pa^\a\pa_\a V|~~,~~
\eea
\bea
\nonumber
\Bar D_{\dot\a} V|=-\bar\pa_{\dot\a}V|~~,~~
\Bar D^2V|=-\bar\pa_{\dot\a}\bar\pa^{\dot\a}V|~~,~~
\eea
\bea
\nonumber
[D_\a, \Bar D_{\dot\a}]V|=-[\pa_\a,\bar\pa_{\dot\a}]V|~~,~~
\eea
\bea
\label{dlimit}
D_\a\Bar D^2 V|=-\pa_\a\bar\pa_{\dot\a}\bar\pa^{\dot\a}V|
-2i\pa_{\un a}\bar\pa^{\dot\a}V|~~.~~
\eea
The simplicity of this operation should not be taken for granted.  Defining components in any particular theory is an art form.  The last three components can be defined in several different ways just by themselves.  Also, derivatives, or masses can be used on the lower components in order to use them to shift the higher components.  Further, in massive theories, inverse masses can be used to shift the lower components by the higher components.

Since the fermionic coordinates are complex, it pays to use some simple complex analysis on the superfields.  A well defined operation is to use superfields that depend only on one of the complex coordinates, say $\q_\a$.  Unfortunately, this type of superfield does not remain purely a function of $\q$ after a supersymmetry transformation:
\bea
\d_\z V(x,\q,0)\not= X(x,\q,0)
\eea
A supersymmetrically invariant way to define a holomorphic superfield is $\Bar D_{\dot\a}\Phi=0$.  The superfield $\Phi$ will satisfy this constraint after a supersymmetry transformation, (\ref{dtrans}).  Superfields vanishing under $\Bar D_{\dot\a}$ are called chiral and superfields vanishing under $D_\a$ are called anti-chiral.  The meaning of the nomenclature becomes clear at the component level:
\bea
\label{chicomp}
\Phi|=\varphi~~,~~D_\a \Phi|=\psi_\a~~,~~-\frac 14D^2\Phi|=F~~,~~
\eea
since a chiral field has a spinor of only one chirality.  In the particle physics community, theories involving only one type of chirality are called chiral.  TypeIIB supergravity or TypeIIB superstrings are chiral theories.

The existence of chiral superfields in four dimensions is rather significant.  Constraints in geometrical descriptions of supergravity can be understood as chirality preserving.  Furthermore, the renormalization properties of super Yang-Mills theories are governed by the fact that the superpotential is a holomorphic function of chiral fields.

\section{On-shell Massive Irreducible Representations}
\label{On-shell Massive Irreducible Representations}
Constraints for massive representations can be found rather easily by noting that $\Box$ is invertible on-shell since $\Box=m^2$.  In particular, the third line in (\ref{dalg}) implies the partition of unity:
\bea
\label{divone}
I = {1\over{16m^2}}\Bar D^2D^2 +{1\over{16m^2}}D^2 \Bar D^2
-{1\over{8m^2}}D^\a\Bar D^2D_\a~~.~~
\eea
A small amount of effort reveals that these three terms are orthogonal and square to themselves.  Thus, they satisfy the projector algebra, $P_iP_j=\d_{ij}P_j$.  This means that massive representations can be decomposed with respect to these operators.  The first two projectors are the identity on chiral and anti-chiral fields, respectively.  This means, that a chiral field is already fully reduced with respect to massive representation theory.  A chiral superfield with spin s has component fields $(s,s\pm\frac12,s)$.  The third projector is a more complicated.  It selects out what are called linear superfields.  Linear superfields vanish when acted upon by $D^2$ and $\Bar D^2$.  Linearity ensures that the third projector is unity on these representations, which follows from (\ref{divone}).  The complication can be seen from the component expansion of the linear spinor superfield:
\bea
\Psi_\a|=\l_\a~~~~D_\a\Psi_\b|=A_{\a\b}
~~~~\Bar D_{\dot\a}\Psi_\a|=B_{\un a}
~~~~[D_\a, \Bar D_{\dot\a}]\Psi_\b|=\psi_{\un a\b}~~.~~
\eea
Decomposition of the Lorentz indices gives a scalar from the trace of $A_{\a\b}$ and another spinor from the trace of $\psi_{\un a\b}$.  Note that this is not the case for the scalar linear superfield.  This means that the simple linear constraint is not good enough to deal with superfields having external indices.  There are two consistent choices of constraints that imply linearity.  $D^\a\Psi_\a=0$ or $D_{(\a}\Psi_{\b)}=0$ both imply $D^2\Psi_\a=0$.  If the constraint is $D^\a\Psi_\a=0$, the components are:
\bea
\Psi_\a|=\l_\a~~,~~D_{(\a}\Psi_{\b)}|=A_{\a\b}
~~,~~\Bar D_{\dot\a}\Psi_\a|=B_{\un a}
~~,~~[D_{(\a}, \Bar D_{\dot\a}]\Psi_{\b)}|=\psi_{\un a\b}~~.~~
\eea
This is the superspin-1 representation with spin content ($\frac12, 1, 1, \frac32$).  The alternative constraint gives the components:
\bea
\Psi_\a|=\l_\a~~,~~D^\a\Psi_\a|=A
~~,~~\Bar D_{\dot\a}\Psi_\a|=B_{\un a}
~~,~~[D^\a, \Bar D_{\dot\a}]\Psi_\a|=\psi_{\dot\a}~~,~~
\eea
which has spins ($0,\frac 12,\frac 12,1$) and is therefore a superspin-$\frac 12$ representation.

The constraints $D^\a\Psi_\a=0$ and $D_{(\a}\Psi_{\b)}=0$ can be derived from the linear projector by multiplying a general superfield with $s$ indices, $\Psi_{\a(s)}$, by the projector and decomposing some indices:
\bea
-{1\over{8m^2}}D^\b\Bar D^2D_\b\Psi_{\a(s)}
=-{1\over{8m^2}}\Big[{1\over (s+1)!}D^\b\Bar D^2D_{(\b}\Psi_{\a(s))}
-{s\over (s+1)!}D_{(\a_s}\Bar D^2D^\b\Psi_{\a(s-1))\b}
\Big]~~.~~
\eea
This equation shows how there are two subspaces within the linear subspace for superfields with all dotted indices.  The subspaces for which $D^\a\Psi_{\a_s\a(s-1)}=0$ and $D_{(\a}\Psi_{\b(s))}=0$ are called transversal linear and longitudinal linear, respectively.  This derivation only reveals the subspaces associated with linear superfields with indices of only one chirality.  The constraints that govern irreducible representations can always be determined by finding the eigenvalues of the superspin operator:
\bea
\label{casimir}
C=m^4\Big\{Y(Y+1)I+(-{1\over{8m^2}})(\frac 34+{\bf B})
D^\a\Bar D^2D_\a
\Big\}~~,~~
\eea
where $Y$ is one-half of the total number of external spinor indices and $\bf B$ has the following properties:
\bea
{\bf B}^2=-{1\over{8m^2}}Y(Y+1)D^\a\Bar D^2D_\a-{\bf B}~~,~~\cr
-8m^2{\bf B}=D^\a\Bar D^2D_\a~{\bf B}=
{\bf B}~D^\a\Bar D^2D_\a~~,~~\cr
{\bf B}={1\over 4m^2}(\cm_{\a\b}P^\b{}_{\dot\a}
-\Bar\cm_{\dot\a\dot\b}P_\a{}^{\dot\b})[D^\a, \Bar D^{\dot\a}]~~.~~
\eea
It is also assumed in (\ref{casimir}) that the superfield obeys the usual non supersymmetric supplementary condition $\pa^{\un a}V_{\un a\dots}=0$ if it has at least one pair of dotted and undotted indices.  For example, the constraints that select the superspin-$\frac 32$ representation contained in the real vector superfield $H_{\un a}$ can be obtained by acting with the superspin operator.  Assuming $H_{\un a}$ to be linear, since the chiral part is superspin-1, the casimir operator acting on $H_{\un a}$ becomes:
\bea
CH_{\un a}=m^4\Big\{\frac {11}4+{\bf B}
\Big\}H_{\un a}
=m^4(\frac {11}4+1)H_{\un a}
-{1\over 8m^2}\Big\{D_\a\Bar D^2D^\g H_{\g\dot\a}+c.c.\Big\}~~.~~
\eea
Further constraining $H_{\un a}$ to be transversal, $D^\a H_{\un a}=0$, the superspin eigenvalue is:
\bea
CH_{\un a}=m^4\frac {15}4H_{\un a}
=m^4\frac 32(\frac 32+1)H_{\un a}~~.~~
\eea
Therefore, $H_{\un a}$ satisfies a massive irreducible representation of the Poincar\'e superalgebra if it is constrained to be transverse linear and obeys the massive D'Alembertian:
\bea
D^\a H_{\un a}=0~~,~~(\Box-m^2)H_{\un a}=0~~.~~
\eea
One of the main goals of this thesis is to construct actions that reproduce these superspin-$\frac 32$ constraints as equations of motion.
\section{On-shell Massless Irreducible Representations}
\label{On-shell Massless Irreducible Representations}
Massless representations are complicated by the fact that $\Box$ is no longer invertible.  This means that there is no partition of unity and also no projectors.  The constraints that are implied by $\Box=0$ must be determined completely from algebraic considerations.  The following derivation will yield a set of constraints that must be satisfied by a massless representation.  Contracting two partial derivatives on the basic covariant spinor derivative equation, (\ref{dcom}), leads to:
\bea
\{\pa_{\a\dot\b}\Bar D^{\dot\b}, \pa_{\b\dot\a}D^\b\}=+2i\pa_{\un a}\Box=0~~.~~
\eea
This means that $\pa_{\a\dot\b}\Bar D^{\dot\b}=0$ on massless physical states.  The D-algebra identities (\ref{dalg}) show that $\pa_{\a\dot\b}\Bar D^{\dot\b}=$$-\frac i4[\Bar D^2, D_\a]$.  $D^2=0$ on physical states as well.  This can be seen by analyzing the following operator:
\bea
-2i\pa_{\un a}D^2=\{D_\a,\Bar D_{\dot\a}\}D^2=D_\a\Bar D_{\dot\a}D^2
=D_\a[\Bar D_{\dot\a},D^2]=0~~.~~
\eea
Since a massless particle should have energy, i.e. $P_0\not=0$, then this means $D^2=\Bar D^2=0$ for massless states.  The list of constraints that determine irreducibility in superspace are:
\bea
\nonumber
D^2=0~~,~~
\Bar D^2=0~~,~~
\eea
\bea
\label{maslescon}
\pa_{\a\dot\b}\Bar D^{\dot\b}=0~~,~~
 \pa_{\b\dot\a}D^\b=0~~.~~
\eea
These constraints and the usual constraints for massless representations of the Poincar\'e algebra:
\bea
\label{suplement}
\Box=0~~,~~
\pa^{\un c}V_{\g\a(A-1)\dot\a(B)}=0~~,~~
\pa^{\un c}V_{\a(A)\dot\g\dot\a(B-1)}=0~~,~~
\eea
completely define the massless irreducible representations of the Poincar\'e superalgebra.  The last two equations in (\ref{suplement}) exist as long as $A\geq1$ and $B\geq1$, respectively. They are called supplementary conditions and are necessary to diagonalize the Lorentz spin Casimir operator.  The constraints can be solved for superfields of any index structure.  This section closes with three examples of massless representations.

Consider first the chiral scalar superfield, $\Phi$.  Being chiral, this field already satisfies $\Bar D^2\Phi=\pa_{\a\dot\b}\Bar D^{\dot\b}\Phi=0$.  Since $4i\pa_{\b\dot\a}D^\b\Phi=\Bar D_{\dot\a}D^2\Phi$, $\Phi$ must also satisfy $D^2\Phi=0$.  Assuming $\Phi$ has satisfied these conditions it has the component structure:
\bea
\Phi|=\varphi~~~~D_\a\Phi|=\psi_\a
\eea
which are the helicity states 0 and $\frac 12$.

Next take the chiral spinor $W_\a$.  Being chiral $W_\a$ solves the same constraints as $\Phi$, but $W_\a$ must also satisfy the supplementary condition, a long with the linearity constraint:
\bea
\pa^{\g\dot\g}W_\g=0~~,~~D^2W_\a=0~~.~~
\eea
If $W_\a$ is transverse linear, these two constraints are satisfied.  The component structure of $W_\a$ takes the form:
\bea
W_\a|=\l_\a~~,~~D_{(\a}W_{\b)}|=f_{\a\b}~~,~~
\eea
containing the helicity states $\frac 12$ and $1$.  This is the supersymmetric Maxwell multiplet (remember that $f_{\a\b}\sim F_{ab}$).  This multiplet is the starting point for supersymmetric Yang-Mills theory.

The final example is the chiral rank 3 spin tensor $W_{\a\b\g}$.  The same reasoning applies to this field as in the $W_\a$ case.  The components are:
\bea
W_{\a\b\g}|=f_{\a\b\g}~~,~~D_{(\a}W_{\b\g\d)}|=C_{\a\b\g\d}~~,~~
\eea
with helicity states $\frac 32$ and 2, corresponding to the gravitino and graviton.  Here $f_{\a\b\g}$ is the curl of the gravitino and $C_{\a\b\g\d}$ is the Weyl tensor in spinor notation, see (\ref{spinweyl}).

\section{Superspace Actions}
\label{Superspace Actions}
The previous sections have introduced superfields and the equations that govern their physics.   If supersymmetry were purely a classical phenomenon equations of motion would suffice to study supersymmetry.  If supersymmetry exists, it is at high energy and is quantum mechanical.  This motivates the construction of action principles that can be used to quantize superfields.  This section introduces integration over anti-commuting variables and promotes the usual spacetime measure to a superspace measure.  The dynamics of several actions are analyzed.  The derivation of component actions from superspace actions is also discussed.  

In general, integration can be defined over anti-commuting variables.  The differentials, $d\q_\a$ and $d\bar\q_{\dot\a}$, are placed in the integration measure in the only Lorentz invariant possibility, $d^2\q=\frac 14\e^{\a\b}d\q_\a d\q_\b$ and $d^2\bar\q=\frac 14\e_{\dot\a\dot\b}d\bar\q^{\dot\a}d\bar\q^{\dot\b}$.   An integral should be invariant under the change of variables $\q_\a=\q_\a^\prime+\e_\a$.  Integrating over the general superfield from (\ref{taylor}):
\bea
\label{int1}
\int d^2\q d^2\bar\q V(\q, \bar\q)=\int d^2\q^\prime d^2\bar\q^\prime \Big(\q^\prime{}^2\bar\q^\prime{}^2 D
+\q^\prime{}^2\bar\q_{\dot\a}{}^\prime(2\bar\e^{\dot\a}D
+\bar\h^{\dot\a})+\cdots\Big)~~,~~
\eea
shows that only the highest component of $V$ is invariant under the change of variables.  Thus, the lower components of $V$ should have no bearing on this integral since they can be shifted to arbitrary values.  Next, look at Gauss's Law:
\bea
\int d^2\q d^2\bar\q \pa^2\bar\pa^2V(\q)=V(\q)|_\pa=0~~,~~
\eea
which is zero because there is no boundary for the anti-commuting space.  Explicitly, plugging the expansion for $V(\q)$ into this expression yields:
\bea
\int d^2\q d^2\bar\q \pa^2\bar\pa^2V(\q)
=16D\int d^2\q d^2\bar\q=0~~,~~
\eea
showing that the volume of the anti-commuting space is zero.  The same result holds for integration over one Grassmann variable, $\int d\z=0$.  These results mean that an integral vanishes unless all of the differentials are multiplying their corresponding  coordinates.  This removes the arbitrary shifting problem for the lower components in (\ref{int1}), since these terms vanish.  The integral of $V$ becomes:
\bea
\int d^2\q d^2\bar\q V(\q, \bar\q)
=D\int d^2\q d^2\bar \q (\q^2\bar\q^2)~~.~~
\eea
This last integral is conventionally set to $1$.  This convention makes the action of integration formally equal to the action of the derivative, e.g.: 
\bea
\int d^2\q\q^2=\frac 14\e^{\a\b}\int d\q_\a d\q_\b\q^2=1~~\sim~~\frac 14\pa^\a\pa_\a \q^2=1~~.~~
\eea
There is a hidden sign coming from the conventions of (\ref{defgrass}).  The full superspace integral is then defined as:
\bea
\int d^8z=\int d^4xd^2\q d^2\bar\q=
\frac 1{16}\int d^4x \pa^\a\pa_\a\bar\pa_{\dot\a}\bar\pa^{\dot\a}
~~.~~
\eea
From (\ref{dlimit}), the limit as $\q$ goes to zero of a $D$ operator is equivalent to replacing all $D$'s by $\pa$ operators up to spacetime derivatives.  Thus, the superspace integration measure can be written as:
\bea
\label{fullmeasure}
\int d^8z V= \frac 1{16}\int d^4x (D^2\Bar D^2 V)|
= \frac 1{16}\int d^4x (\Bar D^2D^2 V)|~~.~~
\eea
It is also consistent to have purely chiral measures:
\bea
\int d^6z = \int d^4x d^2\q~~,~~
\int d^6\bar z = \int d^4x d^2\bar\q~~,~~
\eea
which act on chiral and anti-chiral Lagrangians, respectively.  This possibility follows from (\ref{fullmeasure}) after integrating over half of $d^8z$:
\bea
\int d^8z \cl = -\frac 14\int d^6z \Bar D^2\cl~~.~~
\eea
The chiral Lagrangian $-\frac 14\Bar D^2\cl$ is integrated over the chiral measure.  This may seem trivial, but it allows the introduction of purely chiral Lagrangians that have no interpretation as $\Bar D^2\cl$.

The functional variation of a general superfield follows from general functional analysis:
\bea
\nonumber
\d V(z^\prime)=\int d^8z\d V(z)\d^8(z-z^\prime)
=\int d^8z\d V(z){\d V(z^\prime)\over\d V(z)}~~,~~
\eea
\bea
\label{genvar}
\Rightarrow~~~~{\d V(z^\prime)\over\d V(z)}=\d^8(z-z^\prime)~~.~~
\eea
where $\d^8(z)=\d^4(x)\q^2\bar\q^2$ is the superspace delta function.  Chiral superfields have different variations:
\bea
\nonumber
\d \Phi(z^\prime)=\int d^8z\d \Phi(z)\d^8(z-z^\prime)
=\int d^6z\d \Phi(z)(-\frac 14\Bar D^2\d^8(z-z^\prime))
=\int d^6z\d \Phi(z){\d \Phi(z^\prime)\over\d \Phi(z)}
~~,~~
\eea
\bea
\label{chivar}
{\d \Phi(z^\prime)\over\d \Phi(z)}=-\frac 14\Bar D^2\d^8(z-z^\prime)=\d_+(z,z^\prime)~~,~~
\eea
where $\d_+(z,z^\prime)$ is the chiral delta function.  External indices on superfields are dealt with in the same fashion as external indices on fields.  The superspace measures and super functional variation are all that is needed to construct actions in superspace.  This section ends with some simple examples that give equations of motion that are equivalent to the constraints derived in sections \ref{On-shell Massive Irreducible Representations} and \ref{On-shell Massless Irreducible Representations}.

If a the lowest component of a chiral superfield is to represent a propagating scalar field, then it must have mass dimension 1.  The superspace measure $d^8z$ has mass dimension $-2$; $-4$ from $d^4x$ and $+2$ from $d^2\q d^2\bar\q$.  This means that there is a unique quadratic action for the chiral scalar superfield:
\bea
\cs[\Phi, \Bar \Phi]=\int d^8z \Phi\Bar\Phi~~.~~
\eea
Using (\ref{chivar}), the equation of motion for $\Phi$ is:
\bea
{\d \cs[\Phi, \Bar \Phi]\over \d\Phi}=-\frac 14\int d^8z \Bar D^2\d^8(z-z^\prime)\Bar\Phi=-\frac 14\Bar D^2\Bar \Phi=0~~,~~
\eea
which is exactly what is needed for a chiral field to satisfy the massless constraints (\ref{maslescon}).  Since the mass dimension of the chiral measure is -3, a mass parameter, $m$, affords the opportunity to write another term in the action:
\bea
\cs_m[\Phi, \Bar \Phi]=\int d^8z \Phi\Bar\Phi
+\frac 12m\int d^6z \Phi\Phi+\frac 12m\int d^6\bar z\Bar\Phi\Bar\Phi~~.~~
\eea
Here the equation of motion for $\Phi$ is:
\bea
{\d \cs_m[\Phi, \Bar \Phi]\over \d\Phi}=-\frac 14\Bar D^2\Bar \Phi
-\frac 14m\int d^6z\Bar D^2\d^8(z-z^\prime)\Phi\cr
=-\frac 14\Bar D^2\Bar \Phi
+m\int d^8z\d^8(z-z^\prime)\Phi~~~~\,~\cr
=-\frac 14\Bar D^2\Bar \Phi
+m\Phi=0~~.~~~~~~~~~~~~~~~~~~\,
\eea
Multiplying this equation by $D^2$ and substituting it into the conjugate equation leads to:
\bea
(\Box-m^2)\Bar\Phi=0~~.~~
\eea
Thus, the action $\cs_m[\Phi, \Bar \Phi]$ represents a massive superspin-$0$ irreducible representation.  The component action of this theory is obtained by replacing the measure as in equation (\ref{fullmeasure}) and using the component definitions (\ref{chicomp}):
\bea
\label{phicomp1}
\cs_m[\Phi, \Bar \Phi]=+\frac 1{16}\int d^4xD^2\Bar D^2 (\Phi\Bar\Phi)|
-\frac 18m\int d^4xD^2 (\Phi\Phi)|
-\frac 18m\int d^4x\Bar D^2(\Bar\Phi\Bar\Phi)|\cr
=+\frac 1{16}\int d^4x
(D^2\Phi\Bar D^2 \Bar\Phi+\Phi D^2\Bar D^2 \Bar\Phi
+2D^\a\Phi D_\a\Bar D^2\Bar\Phi)|~~~~~~~~~~~~~~~~~\cr
-\frac 14m\int d^4x(\Phi D^2 \Phi+D^\a\Phi D_\a\Phi)|
-\frac 14m\int d^4x(\Bar\Phi \Bar D^2 \Bar \Phi
+\Bar D_{\dot\a}\Bar\Phi \Bar D^{\dot\a}\Bar\Phi)|\cr
=\int d^4x\Big\{\varphi \Box \bar\varphi
-\frac i2\psi^\a \pa_{\un a} \Bar \psi^{\dot\a}
-\frac 14m\psi^\a\psi_\a
-\frac 14m\bar\psi_{\dot\a}\bar\psi^{\dot\a}
+F\Bar F
+m\varphi F
+m\bar\varphi \Bar F\Big\}
\eea
With $m=0$, this action corresponds to a massless scalar and spinor.  The equation of motion for $F$ is $\Bar F=0$.  With $m\not=0$ the equation of motion for $F$ is $\Bar F+m\varphi=0$.  Upon substitution the action becomes:
\bea
\label{phiint}
=\int d^4x\Big\{\varphi (\Box-m^2 )\bar\varphi
-\frac i2\psi^\a \pa_{\un a} \Bar \psi^{\dot\a}
-\frac 14m\psi^\a\psi_\a
-\frac 14m\bar\psi_{\dot\a}\bar\psi^{\dot\a}
\Big\}
\eea
The auxiliary fields have been integrated out of this action.  The action (\ref{phicomp1}) has off-shell supersymmetry and obeys the abstract Poincar\'e superalgebra, where as, (\ref{phiint}) has on-shell supersymmetry and obeys the algebra given by (\ref{delphi}) and (\ref{psicom}).  The most striking observation about superspace actions is that a superspace action containing 3 terms is equivalent to the off-shell action containing 7 terms.  Superfields are at least an efficient way of packaging supersymmetric theories and arguably the only way to discuss more complicated supersymmetric models.

The real scalar superfield, $V=\Bar V$, has at the highest spin component a real vector.  The gauge invariance of this massless vector field can be incorporated at the superfield level:
\bea
\d V = -\frac i2(\L-\Bar \L)~~,~~\Bar D_{\dot\a}\L=0~~.~~
\eea
Projecting the vector component of this equation gives:
\bea
\d V_{\un a}=\frac 12\d[D_\a,\Bar D_{\dot\a}]V|
=-\frac i4[D_\a,\Bar D_{\dot\a}](\L-\Bar\L)|\cr
=+\frac 12\pa_{\un a}(\L+\Bar\L)|=\pa_{\un a}\a~~,~~~~~~
\eea
which is the correct gauge transformation for the vector field.  Since $V_{\un a}(x)$ should have mass dimension $+1$, $V(z)$ must have dimension $0$.  There are two possible terms that can be written for a real scalar superfield of mass dimension $0$; $VD^\a\Bar D^2D_\a V$ and $V\{D^2, \Bar D^2\}V$.  The latter is not gauge invariant so the action is:
\bea
\label{simplemax}
\cs[V]=\frac 18\int d^8z V D^\a\Bar D^2D_\a V
\eea
The equation of motion is $\frac 14D^\a\Bar D^2 D_\a V=0$.  If a chiral spinor superfield is defined as $W_\a=-\frac 14\Bar D^2 D_\a V$, then the equation of motion is the appropriate constraint equation for a massless irreducible representation.  Before calculating the component action, it pays to analyze the gauge transformations of the first few components of $V$:
\bea
\d V|=-\frac i2(\L-\Bar \L)|~~,~~\d D_\a V|= -\frac i2D_\a\L|~~,~~
\d D^2 V|=-\frac i2D^2\L|~~.~~
\eea
These components transform directly into components of $\L$ that are not the gauge parameter $\a$ and are therefore purely algebraic.  An algebraic gauge transformation means that there exists a gauge in which these components are zero.  Algebraic gauge choices of this nature are called Wess-Zumino gauges.  Wess-Zumino gauges occur whenever there is gauge freedom in a superfield theory.  Taking the last two components as:
\bea
\l_\a=-\frac 14\Bar D^2D_\a V|~~,~~D=\frac 1{32}\{D^2,\Bar D^2\}V|~~,~~
\eea
the component action becomes:
\bea
\cs[V] = \int d^4x\Big\{-\frac 14F^{ab}F_{ab}
-i\l^\a\pa_{\un a}\bar\l^{\dot\a}+2D^2
\Big\}~~.~~
\eea
A massive theory can be obtained by simply adding a mass term:
\bea
\cs_m[V]=\int d^8z\Big\{ \frac 18V D^\a\Bar D^2D_\a V+m^2V^2\Big\}~~.~~
\eea
The equation of motion now reads:
\bea
\frac 14D^\a\Bar D^2D_\a V+2m^2V=0~~.~~
\eea
Taking $D^2$ on this equation yields $D^2V=0$ and the equation of motion becomes:
\bea
(-2\Box +2m^2)V=0~~,~~
\eea
thus, forming an irreducible massive superspin-$\frac 12$ representation.  Going to components in this theory is harder since the mass term ruins the gauge invariance and therefore there is no Wess-Zumino gauge.  With the following definitions:
\bea
\nonumber
\frac 1mB=V|~~,~~\frac 1m\chi_\a=D_\a V|~~,~~
\frac 1mG=-\frac 14D^2 V|~~,~~
\eea
\bea
\nonumber
V_{\un a}=\frac 12[D_\a,\Bar D_{\dot\a}]V|~~,~~
\l_\a=-\frac 14\Bar D^2V|~~,~~
\eea
\bea
D=\frac 1{16}D^\a \Bar D^2D_\a V|~~,~~
\eea
and a little determination the component action is:
\bea
\nonumber
\cs_m[V]=\int d^4x\Big\{-\frac 14F^{ab}F_{ab}-\frac 12m^2 V^aV_a
-\frac 12\pa^aB\pa_aB-\frac 12m^2B^2
\eea
\bea
-i\l^\a\pa_{\un a}\bar\l^{\dot\a}-i\chi^\a\pa_{\un a}\bar\chi^{\dot\a}
-m(\l^\a\chi_\a+\bar\l_{\dot\a}\bar\chi^{\dot\a})
+2\Bar GG+2(D+\frac 12B)^2
\Big\}~~.~~
\eea
The first two terms are the Proca Lagrangian for a massive vector field, and there is a Dirac mass term for the spinors.  This action contains eleven terms and the superfield action only contained two.

This section ends with a description of linearized supergravity.  Finding massive extensions of this theory is one of the main goals of this thesis.  Linearized supergravity is described by a real vector field $H_{\un a}=\Bar H_{\un a}$ and a chiral scalar $\s$.  The action takes the form:
\bea
\nonumber
\cs_{SUGRA}[H_{\un a}, \s]=\int d^8z\Big\{\frac 18H^aD^\b\Bar D^2D_\b H_a
-3\s\bar\s+\frac 1{48}([D_\a,\Bar D_{\dot\a}]H^{\un a})^2
\eea
\bea
\label{linsugra}
-(\pa_a H^a)^2+2i(\s-\bar\s)\pa_a H^a
\Big\}~~,~~
\eea
and has the following gauge invariance:
\bea
\label{sugrgauge}
\d H_{\un a}=\Bar D_{\dot\a}L_\a-D_\a\Bar L_{\dot\a}~~,~~
\d \s =-\frac 1{12}\Bar D^2D^\a L_\a~~.~~
\eea
The curl of the equation of motion for $H^{\un a}$ is:
\bea
-i\pa_{(\a}^{\dot\g}\Big({\d\cs_{SUGRA}[H_{\un a}, \s]\over \d H^{\b)\dot\g}}\Big)
=D^\g W_{\g\a\b}=0~~,~~
\eea
with
\bea
W_{\a\b\g}:=\frac i{3!8}\Bar D^2\pa_{(\a}^{\dot\g}D_\b H_{\g)\dot\g}~~.~~
\eea
This is the correct equation for $W_{\a\b\g}$ to propagate.  A few words should be said about supergravity.  The superfields $H_a$ and $\s$ are called prepotentials.  These superfields are prepotentials in the sense that they are potentials relative to the usual potentials that are geometrically defined to describe supergravity.  For instance, the potential for gravity is the metric field, $g_{ab}$.  If gravity could be written by replacing $g_{ab}$ with $\pa_{(a} B_{b)}$, $B_a$ would be the prepotential.  Although this seems preposterous, it actually occurs in superspace.  Further, the chiral superfield $\s$ is called a compensating superfield.  Without $\s$, the gauge invariance (\ref{sugrgauge}) would correspond to conformal supergravity.  The lowest component of $\s$ shares the scale invariance of the trace of the metric in $H_a$.  There is a Wess-Zumino gauge in which the scale transformation parameter is used to set $\s|=0$.  This removes the scale transformation that could be used on the trace of the metric.

This concludes the mathematical introduction to flat superspace.  The Poincar\'e superalgebra was derived and superfield representations were discussed.  Superspace integration theory was defined and used to construct basic actions that lead to the constraint equations that determine massless and massive irreducible representations of the Poincar\'e superalgebra.

\newpage
\chapter{Half-Integer Massless Arbitrary Superspin Actions}
\label{Half-Integer Massless Arbitrary Superspin Actions}

This chapter is devoted to deriving classical actions for free massless superfields with half-integer superspin.  Massless half-integer superspin-$Y$ multiplets have component fields with helicities $Y+\frac 12$ and $Y$.  Superprojectors will be used to construct the actions.  Using the language of superprojectors is beneficial in two ways.  First, the action takes on a geometrical interpretation, based on the various irreducible subspaces of the real vector superfield $H_{\un a}$.  Second, the orthogonality of the superprojectors, simplifies many of the manipulations in the next chapter on massive theories.  Section \ref{Geometric Actions} explains how to construct the general quadratic action for the real tensor superfield $H_{\un a(s)}$ and then express this action in terms of superprojectors.  Section \ref{Massless Actions}, develops the massless free theories by imposing gauge invariance and introducing compensating superfields.
\section{Geometric Actions}
\label{Geometric Actions}
This section begins with a derivation of a general action which is local and quadratic in a tensor valued real superfield $H_{\un a(s)}(x, \q,\bar\q)$(i.e. $H_{\un a(s)}$$=H_{\un a_1\cdots \un a_s}$).  $H_{\un a(s)}$ is irreducible on all $2s$ spinor indices.  This superfield contains highest component spin of $s+1$.  Once the general action is established, it is rewritten in terms of the super projection of the superfield with one vector index, $H_{\un a}$ .  Although the action describes the dynamics of the superfield $H_{\un a(s)}$ with $s$ vector indices, it will be shown that application of superprojectors to a single vector index of the superfield is sufficient to describe an appropriate dynamical set of equations.  Furthermore, using the $3s+2$ set of superprojectors specific for each $H_{\un a(s)}$, would hardly allow a general description for all spins.

\subsection{General Action}
\label{General Action}

In order to make a general statement about higher spin theories, the types of terms that can exist in a superspace action must be enumerated.  The mass dimension of the superfield $H_{\un a(s)}$ is set to zero, which gives the highest spin component mass dimension $1$.  This means that the mass dimension of the superspace integration measure can only be canceled by four covariant spinor derivatives.  Thus, the general action is fourth order in spinor covariant derivatives.  At first this seems daunting, since there are a priori $\sim2^4$ terms plus various Lorentz contractions.  It is also rather complicated to prove the linear independence of some set of terms using only the covariant derivative algebra.  It seems necessary to find some simple way of organizing all possible terms and finding an acceptable set of linearly independent terms.  The following considerations show that all possible fourth order differential operators can be encompassed by a set of seven Lorentz irreducible operators.  This set is further reduced by requiring that the action is real and does not violate parity.  The final action for $H_{\un a(s)}$ contains five linearly independent terms.

There are six possible fourth order operators.  Separated into Lorentz irreducible pieces they are:
\bea
\label{4thops1}
D_\a D_\b\Bar D_{\dot\a}\Bar D_{\dot\b}=
-\frac 14\e_{\a\b}\e_{\dot\a\dot\b}D^2\Bar D^2
~~,~~
\eea
\bea
\label{4thops2}
\Bar D_{\dot\a} \Bar D_{\dot\b} D_\a D_\b=
-\frac 14\e_{\a\b}\e_{\dot\a\dot\b}\Bar D^2 D^2
~~,~~
\eea
\bea
\label{4thops3}
D_\a\Bar D_{\dot\a} \Bar D_{\dot\b} D_\b=
-\frac 14\e_{\dot\a\dot\b}D_{(\a}\Bar D^2 D_{\b)}
-\frac 14\e_{\a\b}\e_{\dot\a\dot\b}D^\g\Bar D^2 D_\g
~~,~~
\eea
\bea
\label{4thops4}
\Bar D_{\dot\a} D_\a D_\b\Bar D_{\dot\b}=
+\frac 14\e_{\a\b}\Bar D_{(\dot\a}D^2 \Bar D_{\dot\b)}
-\frac 14\e_{\a\b}\e_{\dot\a\dot\b}D^\g\Bar D^2 D_\g
~~,~~
\eea
\bea
\nonumber
D_\a \Bar D_{\dot\a}D_\b \Bar D_{\dot\b}=
\frac 14D_{(\a}\Bar D_{(\dot\a}D_{\b)}\Bar D_{\dot\b)}
-\frac 18\e_{\a\b}\Bar D_{(\dot\a}D^2\Bar D_{\dot\b)}
+\frac 18\e_{\dot\a\dot\b} D_{(\a}\Bar D^2 D_{\b)}
\eea
\bea
\label{4thops5}
+\frac 18\e_{\a\b}\e_{\dot\a\dot\b}(D^\g\Bar D^2D_\g+D^2\Bar D^2)~~,~~
\eea
\bea
\nonumber
\Bar D_{\dot\a}D_\a \Bar D_{\dot\b}D_\b =
\frac 14\Bar D_{(\dot\a}D_{(\a}\Bar D_{\dot\b)}D_{\b)}
-\frac 18\e_{\a\b}\Bar D_{(\dot\a}D^2\Bar D_{\dot\b)}
+\frac 18\e_{\dot\a\dot\b} D_{(\a}\Bar D^2 D_{\b)}
\eea
\bea
\label{4thops5}
+\frac 18\e_{\a\b}\e_{\dot\a\dot\b}(D^\g\Bar D^2D_\g+\Bar D^2D^2)~~.~~
\eea
Note that $(D^\a\Bar D^2 D_\a)^*$$=\Bar D_{\dot\a}D^2\Bar D^{\dot\a}$.  Thus, there are only seven linearly independent fourth order Lorentz irreducible operators, namely; $D^\g\Bar D^2D_\g$, $D^2\Bar D^2$, $\Bar D^2D^2$, $D_{(\a}\Bar D^2 D_{\b)}$, $\Bar D_{(\dot\a}D^2\Bar D_{\dot\b)}$, $\Bar D_{(\dot\a}D_{(\a}\Bar D_{\dot\b)}D_{\b)}$, $D_{(\a}\Bar D_{(\dot\a}D_{\b)}\Bar D_{\dot\b)}$.  The next step is to construct the general action and determine if this list is diminished because of parity violations or equivalence under integration by parts.  The action is assumed to be real and quadratic in the real pseudo tensor valued superfield $H_{\un a(s)}$.  The action is also assumed not to violate parity.  A few comments on the parity operator, $P$, are in order.  Parity changes all chiralities, thus $P D_\a=\Bar D_{\dot\a}$ and $PD^2=-\Bar D^2$.  Furthermore, since $P[D_\a,\Bar D_{\dot\a}]=-[D_\a,\Bar D_{\dot\a}]$, $H_{\un a(s)}$ must be understood to be a pseudo tensor so that the physical component field, $g_{\un a\un b(s)}:=[D_\a,\Bar D_{\dot\a}]H_{\un b(s)}|$, is a tensor.  Taking into account real and imaginary coefficients, the operators in the previous paragraph form an action with one real coefficient, $w$, and three imaginary coefficients; $x,y,z$:
\bea
\nonumber
\cs_{Gen}[H_{\un a(s)}]=\int d^8z \Big\{
H^{\un a(s)}\Big(wD^\g\Bar D^2 D_\g 
+xD^2\Bar D^2
+x^*\Bar D^2 D^2\Big)H_{\un a(s)}
\eea
\bea
\nonumber
+H^{\un a\un c(s-1)}\Big(
yD_{(\a}\Bar D_{(\dot\a}D_{\b)}\Bar D_{\dot\b)}
+y^*\Bar D_{(\dot\a}D_{(\a}\Bar D_{\dot\b)}D_{\b)}
\Big)H_{\un b\un c(s-1)}
\eea
\bea
+H^{\un a\un c(s-1)}\Big(
zD_{(\a}\Bar D^2 D_{\b)}
H^\b_{\dot\a\un c(s-1)}
-z^*\Bar D_{(\dot\a} D^2 \Bar D_{\dot\b)}
H^{\dot\b}_{\a\un c(s-1)}\Big)
\Big\}~~.~~
\eea
The first term is standard even with $s=0$.  The imaginary parts of $x,~y,$ and $z$ all violate parity, thus these coefficients are real.    Even at the level of integration by parts, $x$ and $y$ are constrained to be real.  The action becomes:
\bea
\label{ruffgenact}
\nonumber
\cs_{Gen}[H_{\un a(s)}]=
\int d^8z \Big\{H^{\un a(s)}\Big(wD^\g\Bar D^2 D_\g 
+x\{D^2,\Bar D^2\}\Big)H_{\un a(s)}
\eea
\bea
\nonumber
+2yH^{\un a\un c(s-1)}
D_{(\a}\Bar D_{(\dot\a}D_{\b)}\Bar D_{\dot\b)}
H_{\b\dot\b\un c(s-1)}
\eea
\bea
+H^{\un a\un c(s-1)}\Big(
+zD_{(\a}\Bar D^2 D_{\b)}
H^\b_{\dot\a\un c(s-1)}
-z\Bar D_{(\dot\a} D^2 \Bar D_{\dot\b)}
H^{\dot\b}_{\a\un c(s-1)}\Big)
\Big\}~~.~~
\eea
This is the most general quadratic superfield action for a pseudo tensor $H_{\un a(s)}$.  Although it contains only four real coefficients, it is in a rather useless form.  This action can be polished further by writing it in terms of the more canonical fourth order operators.

For this action to be more useful from the perspective of the literature, it is necessary to show where the canonical fourth order operators $\pa_{\un a}\pa_{\un b}$ and $[D_\a,\Bar D_{\dot\a}][D_\b,\Bar D_{\dot\b}]$ fit into this description of the  action.  Rewriting the action with the canonical operators is just like changing basis vectors in a vector space.  The canonical operators can be written in terms of the operators in (\ref{ruffgenact}):
\bea
H^{\un a\un c(s-1)}\pa_{\un a}\pa_{\un b}H^{\un b}{}_{ \un c(s-1)}
=H^{\un a\un c(s-1)}\Big[-\frac 18D_{(\a}\Bar D_{(\dot\a}D_{\b)}\Bar D_{\dot\b)}
-\frac 12\e_{\a\b}\e_{\dot\a\dot\b}\Box\Big]H^{\un b}{}_{ \un c(s-1)}
~~,~~
\eea
\bea
\nonumber
H^{\un a\un c(s-1)}[D_\a,\Bar D_{\dot\a}]
[D_\b,\Bar D_{\dot\b}]H^{\un b}{}_{ \un c(s-1)}=
H^{\un a\un c(s-1)}\Big[+\frac 12D_{(\a}\Bar D_{(\dot\a}D_{\b)}\Bar D_{\dot\b)}
+2\e_{\a\b}\e_{\dot\a\dot\b}\Box
\eea
\bea
+\frac 12\e_{\dot\a\dot\b}D_{(\a}\Bar D^2D_{\b)}
-\frac 12\e_{\a\b}\Bar D_{(\dot\a}D^2\Bar D_{\dot\b)}
+\e_{\a\b}\e_{\dot\a\dot\b}D^\g\Bar D^2D_\g\Big]H^{\un b}{}_{ \un c(s-1)}~~.~~
\eea
Noting that all $\{D^2, \Bar D^2\}$ terms can be replaced via $\{D^2, \Bar D^2\}=2D^\a\Bar D^2D_\a+16\Box$, the basis can be chosen so that the canonical operators, $\pa_{\un a}\pa_{\un b}$ and $[D_\a,\Bar D_{\dot\a}][D_\b,\Bar D_{\dot\b}]$, can be substituted for $xD_{(\a}\Bar D_{(\dot\a}D_{\b)}\Bar D_{\dot\b)}$, and $zD_{(\a}\Bar D^2 D_{\b)}
H^\b_{\dot\a\un c(s-1)}
-z\Bar D_{(\dot\a} D^2 \Bar D_{\dot\b)}
H^{\dot\b}_{\a\un c(s-1)}$.  The action then takes the canonical form:
\bea
\nonumber
\cs_{Gen}[H_{\un a(s)}]=\int d^8z \Big\{\a_1 H^{\un a(s)}D^\g\Bar D^2 D_\g H_{\un a(s)}
+\a_2H^{\un a(s)}\Box H_{\un a(s)}
\eea
\bea
\label{cleangenact}
+\a_3H^{\un a\un c(s-1)}\pa_{\un a}\pa^{\un b}
H_{\un b\un c(s-1)}
+\a_4H^{\un a\un c(s-1)}
[D_\a, \Bar D_{\dot\a}][D^\b,\Bar D^{\dot\b}]
H_{\un b\un c(s-1)}
\Big\}~~.~~
\eea
(\ref{cleangenact}) is the general action quadratic in $H_{\un a(s)}$.  It is local and parity preserving.  This action contains only four real coefficients.  One of these coefficients can be absorbed into the normalization of $H_{\un a(s)}$ and will be chosen canonically in later sections.
\subsection{Superprojectors}
\label{Superprojectors}

Although the action $\cs_{Gen}$ is in a standard form, it would be more useful to construct the higher spin actions, with the knowledge of which terms contained the highest component spins.  The off-shell projectors mentioned in equation (\ref{divone}) provide a geometric solution to elucidating the structure of this action.  Once the action is written in terms of superprojectors, the coefficients that specify the highest spin part of the action can be determined.  This is exactly the procedure used in \cite{Gates2003a} to show how the new massless theory derived in \cite{Buchbinder2002a} fits in with the known forms of linearized supergravity.

To begin, the decomposition of general tensor valued superfields must be obtained.  Acting on a superfield with a set of one type of spinor index $V_{\a(s)}$ with the now off-shell identity, (\ref{divone}), leads to:
\bea
IV_{\a(s)} = (\frac 1{16}\Box^{-1}\Bar D^2D^2 
+\frac 1{16}\Box^{-1}D^2 \Bar D^2
-\frac 18\Box^{-1}D^\b\Bar D^2D_\b )V_{\a(s)}~~.
\eea
The third term can be reorganized by noting the tableaux relation:
\bea
{\tiny\yng(1)}~\otimes~{\tiny\yng(5)}
~=~{\tiny\yng(5,1)}~\oplus~{\tiny\yng(6)}~~,
\eea
which can be written in terms of the tensor $V_{\a(s)}$ and $D_\b$ as:
\bea
D_\b V_{\a(s)}=\frac 1{(s+1)!}D_{(\b}V_{\a(s))}
+\frac s{(s+1)!}\e_{\b(\a_s}D^\g V_{\a(s-1))\g}~~.
\eea
Using this result, the identity becomes:
\bea
\nonumber
IV_{\a(s)} = \frac 1{16}\Box^{-1}\Bar D^2D^2V_{\a(s)} 
+\frac 1{16}\Box^{-1}D^2 \Bar D^2V_{\a(s)}
\eea
\bea
-\frac 1{8(s+1)!}\Box^{-1}D^\b\Bar D^2D_{(\b} V_{\a(s))}
+\frac s{8(s+1)!}\Box^{-1}D_{(\a_s}\Bar D^2D^\g V_{\a(s-1))\g}~~.~~
\eea
Viewing this as an expansion of $V{}_{\a}$ in terms of a set of chiral superfields, the first two terms have superspin $\frac s2$ and the last two terms have superspin $\frac{s + 1}2$ and $\frac{s - 1}2$.  This decomposition can be applied to superfields of any index type by noting that the following operator is invertible off-shell:
\bea
\triangle_{\dot\a}{}^\a=\Box^{-\frac 12}\pa_{\dot\a}{}^\a~~~~
\triangle_{\dot\a}{}^\b\triangle_\b{}^{\dot\b}
=\d^{\dot\b}{}_{\dot\a}~~.
\eea
This operator allows the conversion of an index of one type of chirality into the other.  Thus, the superprojector decomposition of the vector valued superfield $H_{\un a}$, can be obtained by converting one spinor index on $H_{\un a}$ and proceeding with the decomposition of the object with two undotted indices.  The invertible conversion of $H_{\un a}$ is defined as:
\bea
\nonumber
\triangle_\b{}^{\dot\a}H_{\un a}=\frac 12V_{\a\b}-\frac 12\e_{\a\b}V~~,~~
\eea
\bea
\label{HconvertV}
\Rightarrow~~~~
V_{\a\b}=\triangle_{(\b}{}^{\dot\a}H_{\a)\dot\a}
~~,~~V=\triangle^{\un a}H_{\un a}~~.~~
\eea
Since there are two irreducible spin tensors, $V_{\a\b}$ and $V$, the super projection must be evaluated on both of these pieces:
\bea
\nonumber
I H_{\un a}=\frac 12\triangle_{\dot\a}{}^\b IV_{\a\b}
-\frac 12\triangle_{\un a}IV
\eea
\bea
\nonumber
=\frac 12\triangle_{\dot\a}{}^\b \Big(
\frac 1{16}\Box^{-1}\Bar D^2D^2V_{\a\b} 
+\frac 1{16}\Box^{-1}D^2 \Bar D^2V_{\a\b}
\eea
\bea
\nonumber
-\frac 1{8\cdot3!}\Box^{-1}D^\g\Bar D^2D_{(\g} V_{\a\b)}
+\frac 2{8\cdot3!}\Box^{-1}D_{(\a}\Bar D^2D^\g V_{\b)\g}\Big)
\eea
\bea
-\frac 12\triangle_{\un a}\Big(\frac 1{16}\Box^{-1}\Bar D^2D^2 
+\frac 1{16}\Box^{-1}D^2 \Bar D^2
-\frac 18\Box^{-1}D^\b\Bar D^2D_\b \Big)V~~,~~
\eea
Upon converting back to $H_{\un a}$ using (\ref{HconvertV}), the super projection becomes:
\bea
\nonumber
IH_{\un a}=\frac 1{2}\Box^{-2}\pa_{\dot\a}{}^\b \Big\{
\frac 1{16}\{\Bar D^2,D^2\}\pa_{(\a}{}^{\dot\b}H_{\b)\dot\b}
\eea
\bea
\nonumber
-\frac 1{4\cdot3!}D^\g\Bar D^2D_{(\g} 
\pa_{\a}{}^{\dot\b}H_{\b)\dot\b}
+\frac 1{4\cdot3!}D_{(\a}\Bar D^2D^\g 
\Big(\pa_{\b)}{}^{\dot\b}H_{\g\dot\b}
+\pa_{\g}{}^{\dot\b}H_{\b)\dot\b}\Big)\Big\}
\eea
\bea
-\frac 12\pa_{\un a}\Box^{-2}\Big(\frac 1{16}\{\Bar D^2,D^2\} 
-\frac 18D^\b\Bar D^2D_\b \Big)\pa^{\un d}H_{\un d}
~~.~~
\eea
Taking $H_{\un a}$ to be real, five orthogonal projection can be extracted from this equation.  There are three that are transverse, or have no divergence:
\bea
(\P^T_{1}){}_{\un a}{}^{\un b}H_{\un b}:=\frac 1{32}\Box^{-2}\pa_{\dot\a}{}^\b 
\{\Bar D^2,D^2\}\pa_{(\a}{}^{\dot\b}H_{\b)\dot\b}~~,
\eea
\bea
(\P^T_{1/2}){}_{\un a}{}^{\un b}H_{\un b}:=\frac 1{8\cdot3!}\Box^{-2}\pa_{\dot\a}{}^\b
D_{(\a}\Bar D^2D^\g 
(\pa_{\b)}{}^{\dot\b}H_{\g\dot\b}
+\pa_{|\g|}{}^{\dot\b}H_{\b)\dot\b})~~,
\eea
\bea
\label{trans}
(\P^T_{3/2}){}_{\un a}{}^{\un b}H_{\un b}:=-\frac 1{8\cdot3!}\Box^{-2}\pa_{\dot\a}{}^\b
D^\g\Bar D^2D_{(\g} 
\pa_{\a}{}^{\dot\b}H_{\b)\dot\b}~~,~~
\eea
and two that are longitudinal:
\bea
(\P^L_{0}){}_{\un a}{}^{\un b}H_{\un b}:=-\frac 1{32}\pa_{\un a}\Box^{-2}\{\Bar D^2,D^2\}\pa^{\un c}H_{\un c}~~,
\eea
\bea
\label{long}
(\P^L_{1/2}){}_{\un a}{}^{\un b}H_{\un b}:=\frac 1{16}\pa_{\un a}\Box^{-2}D^\b\Bar D^2D_\b\pa^{\un c}H_{\un c}~~.~~~
\eea
This process can be continued to construct the projections of all high spin superfields.  Surprisingly, these five projectors are all that is needed to make a thorough investigation of massless and massive half-integer superspin theories.

Using the expressions for the superprojectors, the general action, (\ref{cleangenact}), can be rewritten.  The following calculations can be performed by inserting the identity and gathering the non-zero terms.  The first term is the sum of the linear projectors:
\bea
\a_1H^{\un a\un b(s-1)}D^\g\Bar D^2 D_\g H_{\un a\un b(s-1)}=
-8\a_1H^{\un a\un c(s-1)}\Box(\P^T_{3/2}+\P^T_{1/2}+\P^L_{1/2}){}_{\un a}{}^{\un b}H_{\un b\un c(s-1)}~~.~
\eea
The second term is the identity off-shell and is therefore the sum of all five projectors:
\bea
\a_2H^{\un a}\Box H_{\un a(s)}=\a_2 H^{\un a\un c(s-1)}
\Box(\P^T_{3/2}+\P^T_{1}+\P^T_{1/2}+\P^L_{1/2}+\P^L_0)_{\un a}{}^{\un b}H_{\un b\un c(s-1)}~~.~
\eea
The third term is only partial derivatives, so it must be a sum of longitudinal superprojectors:
\bea
\a_3H^{\un a\un c(s-1)}\pa_{\un a}\pa^{\un b}IH_{\un b\un c(s-1)}
=-2\a_3H^{\un a\un c(s-1)}(\P^L_{1/2}+\P^L_0)_{\un a}{}^{\un b}H_{\un b\un c(s-1)}~~.~
\eea
The final term, the double commutator, is formidable.  The chiral transverse superprojector, $\P^T_1$, trivially vanishes under this operator.  The longitudinal linear superprojector, $\P^L_{1/2}$, vanishes since $[D_\a, \Bar D_{\dot\a}]\pa^{\un a}\propto [D^2, \Bar D^2]$ multiplies a linear operator.  Further, the transverse projector, $\P^T_{3/2}$ vanishes because of the clash between the triple symmetrization and the antisymmetry of $D^\a D^\b$.  The other chiral superprojector is also trivial:
\bea
\a_4H^{\un a\un d(s-1)}[D_\a, \Bar D_{\dot\a}]
[D^\b, \Bar D^{\dot\b}](\P^L_0)_{\un b}{}^{\un c}H_{\un c\un d(s-1)}
=8\a_4H^{\un a\un c(s-1)}
(\P^L_0)_{\un a}{}^{\un b}H_{\un b\un c(s-1)}~~.~
\eea
The only non-trivial calculation involves $\P^T_{1/2}$:
\bea
\nonumber
\a_4H^{\un a\un d(s-1)}[D_\a,\Bar D_{\dot\a}]
[D^\b, \Bar D^{\dot\b}](\P^T_{1/2})_{\un b}{}^{\un c}H_{\un c\un d(s-1)}
\eea
\bea
\nonumber
=-12i\a_4H^{\un a\un e(s-1)}\Box[D_\a,\Bar D_{\dot\a}]
(\Box^{-2}\frac 1{8\cdot3!})D^\b\Bar D^2D^\d
\pa_{(\b}{}^{\dot\g}H_{\d)\dot\g\un e(s-1)}~~,
\eea
\bea
=-24\a_4H^{\un a\un c(s-1)}\Box (\P^T_{1/2})_{\un a}{}^{\un b}
H_{\un b\un c(s-1)}~~.~~
\eea
The second line above is the middle line in an $\sim8$ line calculation and is shown just to give some direction to the interested reader.  The final answer for the super projection of the double commutator is:
\bea
\a_4H^{\un a\un c(s-1)}[D_\a,\Bar D_{\dot\a}]
[D^\b, \Bar D^{\dot\b}]H_{\un b\un c(s-1)}
=\a_4H^{\un a\un c(s-1)}(8\P^L_0-24\P^T_{1/2})_{\un a}{}^{\un b}H_{\un b\un c(s-1)}~~.~
\eea
Having decomposed all four terms with superprojectors, the super projected action can be written:
\bea
\nonumber
\cs_{Gen}[H_{\un a(s)}]=\int d^8zH^{\un a\un c(s-1)}\Box\Big(
[-8\a_1+\a_2]\P^T_{3/2}
+\a_2\P^T_1
+[-8\a_1+\a_2-24\a_4]\P^T_{1/2}
\eea
\bea
\label{finalaction}
+[-8\a_1+\a_2-2\a_3]\P^L_{1/2}
+[\a_2-2\a_3+8\a_4]\P^L_0
\Big)_{\un a}{}^{\un b}H_{\un b\un c(s-1)}~~.~~
\eea
It is interesting to note a few things about the form of this action.  First, there are four unknown coefficients, one of which can be absorbed into $H_{\un a(s)}$.  Second, a little linear algebra shows that it is impossible to set the coefficients such that there is only one superprojector.  This is a consequence of the construction, since it starts by using only local operators and one superprojector by itself is not local.  Another method for deriving this action would be to sum all superprojectors with arbitrary coefficients and demand locality.  It seems that this method would involve intense calculations and not use the orthogonality of the superprojectors efficiently.

It is canonical to set $\a_1=-\frac 1{16}$ which can be thought of as absorbing $\a_1$ into the definition of $H_a$.  Further, in what follows, $\a_2=0$, so that the $\P_{1}^T$ superprojector is absent.  The use of the superprojector, $\P_{1}^T$, in a gauge theory implies the existence of an unconstrained 2-form superfield.  Although a theory of this nature may have some utility, it certainly does not describe an irreducible representation of supersymmetry.
\section{Massless Actions}
\label{Massless Actions}
To generate free massless superfield representations starting with the general action, it is necessary to impose gauge invariance.  The highest component of the superfield should transform like the linearized version of gravity with more indices.  So the gauge transformation is just the obvious extension of linearized supergravity:
\bea
\d H_{\un a(s)}=\frac 1{s!}\Bar D_{(\dot\a_s}L_{|\a(s)|\dot\a(s-1))}
-\frac 1{s!}D_{(\a_s}\Bar L_{\a(s-1))\dot\a(s)}~~.
\eea
This gauge transformation is precisely that which leaves invariant the highest superspin projection of $H_{\un a(s)}$:
\bea
\label{arbitraryhighspin}
\P^{T}_{s+\frac 12}H_{\un a(s)}\sim \Box^{-(s+1)}
\pa_{\dot\a_1}{}^{\b_1}\cdots\pa_{\dot\a_s}{}^{\b_s}
D^\g\Bar D^2D_{(\g}
\pa_{\b_1}^{~\dot\b_1}\cdots\pa_{\b_s}^{~\dot\b_s}
H_{\a(s))\dot\b(s)}~~.~~
\eea
The propagating superfield-strength is buried in this projection:
\bea
W_{\a(2s+1)}:={i\over 8(2s+1)!}\Bar D^2D_{(\a_{s+1}}
\pa_{\a_{s+2}}{}^{\dot\b_1}\cdots\pa_{\a_{2s+1}}{}^{\dot\b_s}
H_{\a(s))\dot\b(s)}~~.~~
\eea
The action (\ref{finalaction}) is by no means gauge invariant.  In fact, every term has non-zero gauge transformation.  This is not the case for supergravity, when $s=1$, since (\ref{arbitraryhighspin}) is a term in the action.  This will become clear in the following analysis.

The gauge transformations of each term in the action are:
\bea
\nonumber
\d(H^{\un a\un b(s-1)}\Box\P^{L}_{0}H_{\un a\un b(s-1)})=
\frac {i}{4s}\pa_{\un a}H^{\un a\un b(s-1)}\Bar D^2D^\g L_{\g\un b(s-1)}
\eea
\bea
\label{fries}
-\frac i{4s}(s-1)\pa_{\un a}H^{\un a\un b(s-1)}
\Box^{-1}\Bar D^2D^\d\pa_{\d\dot\b_{s-1}}\pa^{\un c}L_{\un b(s-2)\b_{s-1}\un c}
+{\rm c.c.}~~,
\eea
\bea
\nonumber
\d (H^{\un a\un b(s-1)}\Box\P^{L}_{1/2}H_{\un a\un b(s-1)})=
-\frac {i}{4s}\pa_{\un a}H^{\un a\un b(s-1)}D^\g\Bar D^2L_{\g\un b(s-1)}
\eea
\bea
\label{cheesy}
+\frac i{4s}(s-1)\pa_{\un a}H^{\un a\un b(s-1)}\Box^{-1}
D^\d\Bar D^2\pa_{\d\dot\b_{s-1}}\pa^{\un c}L_{\un c\un b(s-2)\b_{s-1}}
+{\rm c.c.}~~,~~
\eea
\bea
\nonumber
\d( H^{\un a\un b(s-1)}\Box\P^{T}_{1/2}H_{\un a\un b(s-1)})=
\frac 1{24s}(2s+1)[D_\a, \Bar D_{\dot\a}]
H^{\un a\un b(s-1)}D^\g\Bar D^2L_{\g\un b(s-1)}
\eea
\bea
\label{hot}
-i\frac {(s-1)}{12s}\pa_{\un a}
H^{\un a\un b(s-1)}\Box^{-1}
D^\d\Bar D^2\pa_{\d\dot\b_{s-1}}\pa^{\un c}L_{\un b(s-2)\b_{s-1}\un c}
+{\rm c.c.}~~,~~
\eea
\bea
\nonumber
\d(H^{\un a\un b(s-1)}\Box\P^{T}_{3/2}H_{\un a\un b(s-1)})=
\frac {(s-1)}{3!s}
\Big\{\frac 12[D_\a, \Bar D_{\dot\a}]-i3\pa_{\un a}\Big\}
H^{\un a\un b(s-1)}D^\g\Bar D^2L_{\g\un b(s-1)}
\eea
\bea
\label{sweetapple}
-\frac {i(s-1)}{3!s}\pa_{\un a}H^{\un a\un b(s-1)}\Box^{-1}
D^\g\Bar D^2\pa_{\g\dot\b_{s-1}}\pa^{\un d}L_{\dot\b(s-2)\b(s-1)\un d}
+{\rm c.c.}~~.~~
\eea
The equation:
\bea
T_{(\a}\G_{\b(p))}=p!T_\a\G_{\b(p)}+pT_{(\b_1}\G_{\b_2\cdots\b_p)\a}~~,
\eea
is helpful to untangle some indices.  Setting $s=1$ in equations (\ref{fries}-\ref{sweetapple}), reveals a peculiar structure.  With $s=1$, these equations state that the gauge variations of non-local objects are actually local.  It should be noted that for arbitrary $s$, the combinations of these terms that exist in the action (\ref{finalaction}) will lead to an overall local gauge transformation of the action.

To create a gauge invariant action it is necessary to introduce compensating superfields whose gauge transformations cancel these contributions.  The compensating superfields will be enumerated in the following section.  There are only two known formulations for higher spin theories\cite{Buchbinder1995, Kuzenko1993}.  These theories are stated based on gauge invariance and there has not been a procedure that proves that the two formulations are the only possibilities.  It will be interesting to see if the superprojector method can clarify this structure or produce any new structures.
\subsection{Compensating Superfields}
\label{Compensating Superfields}
The compensating superfields are responsible for canceling the gauge variation of the general action (\ref{finalaction}) which is:
\bea
\nonumber
\int d^8z\Big\{
-\frac is(8\a_4-2\a_3)\pa_{\un a}H^{\un a\un b(s-1)}\Big[
(s-1)i\Bar D_{\dot\b_{s-1}}\pa^{\un c}L_{\dot\b(s-2)\b(s-1)\un c}
-\frac 14\Bar D^2D^\g L_{\g\un b(s-1)}\Big]
\eea
\bea
\label{gaugetransH}
+\Big[(\frac 1{2s}\a_3-\frac 18)i\pa_{\un a}
+\Big(\frac 1{16}-\frac {(2s+1)}s\a_4\Big)
[D_\a, \Bar D_{\dot\a}]
\Big]H^{\un a\un b(s-1)}
D^\g \Bar D^2 L_{\g\un b(s-1)}
+{\rm c.c.}\Big\}~~.~~
\eea
This result is quite nice.  Written in this fashion, the gauge variation contains both longitudinal and transversal linear terms coupled to derivatives of $H_{\un a(s)}$.  There is one chiral term, but the gauge variation of a chiral superfield will not be able to cancel the linear variations.  This suggests the addition of complex linear compensating superfields.  Further, proceeding by using the most general definitions of these compensating fields, will guarantee that all possibilities are enumerated.  

The first compensating field is called longitudinal complex linear and couples to the divergence of $H_{\un a(s)}$.  The second is called transversally complex linear and couples through the commutator and partial derivative combination.  Thus, two compensating superfields are introduced; transversally linear $\G_{\un a(s-1)}$ and longitudinally linear $G_{\un a(s-1)}$.  These complex linear superfields are constrained:
\bea
\label{constrainttrans}
\Bar D^{\dot\b}\G_{\un a(s-2)\b\dot\b}=0~~,~~
\eea
\bea
\label{constraintlong}
\Bar D_{(\dot\b}G_{\dot\a(s-1))\a(s-1)}=0~~.~~
\eea
In accordance with their definitions, both superfields lie in the kernel of $\Bar D^2$ but not $D^2$, hence the distinction complex.  The compensating superfields can have the following gauge transformations consistent with (\ref{constrainttrans}) and (\ref{constraintlong}):
\bea
\d \G_{\un a(s-1)}=
a\Bar D^{\dot\g} D^2\Bar L_{\dot\g\un a(s-1)}
+b\Bar D^2D^\b L_{\b\un a(s-1)}~~,~~
\eea
\bea
\d G_{\un a(s-1)}=
\frac c{(s-1)!}i\pa^{\un d}\Bar D_{(\dot\a_{s-1}} L_{\dot\a(s-2))\a(s-1)\un d}
+d\Bar D^2D^\b L_{\b\un a(s-1)}~~.~~
\eea
These gauge transformations are exactly what is needed to construct the full gauge invariant action.  There are two steps to completing the action.  First, the coupling terms are written such that they cancel the terms in (\ref{gaugetransH}).  Second, the kinetic terms for the compensating fields must be determined.  Throughout this process it will be necessary to fix the free coefficients $\a_3,~\a_4,~a,~b,~c$, and $d$.  

The coupling terms take the form:
\bea
\nonumber
\cs_{Coupling}=\int d^8z\Big\{
+\frac is(8\a_4-2\a_3)\pa_{\un a}H^{\un a\un b(s-1)}
G_{\un b(s-1)}
\eea
\bea
-4\Big[(\frac 1{2s}\a_3-\frac 18)i\pa_{\un a}
+\Big(\frac 1{16}-\frac {(2s+1)}s\a_4\Big)
[D_\a, \Bar D_{\dot\a}]
\Big]H^{\un a\un b(s-1)}
\Bar\G_{\un b(s-1)}
+{\rm c.c.}
\Big\}~~.~~
\eea
This coupling implies that $c=(s-1)$ and $a=-\frac 14$, which is the canonical normalization.  Further, in order to remove the chiral part of the gauge transformation the coefficients must obey the following equation:
\bea
\frac 1{4s}\Big(8\a_4-2\a_3\Big)\Big(4d+1\Big)
-b\Big(\frac 2s\a_3-\frac {8(2s+1)}s\a_4\Big)=0~~,~~
\eea
which is really two equations, since irreducibility requires that only one compensator be used.  Removing $G_{\un a}$ by setting $8\a_4-2\a_3=0$ implies that $b=0$.  Keeping $G_{\un a(s-1)}$ and removing $\G_{\un a(s-1)}$ by setting $\a_3=\frac s4$ and $16(2s+1)\a_4=s$ implies that $d=-\frac 14$.

All that is left is to find suitable kinetic terms.  The additional terms take the form:
\bea
\nonumber
\cs_{Comp}=\int d^8z\Big\{
eG^{b(s-1)}\Bar G_{\un b(s-1)}
+fG^{b(s-1)}G_{\un b(s-1)}
+f^*\Bar G^{b(s-1)}\Bar G_{\un b(s-1)}
\eea
\bea
+g\G^{b(s-1)}\Bar \G_{\un b(s-1)}
+h\G^{b(s-1)} \G_{\un b(s-1)}
+h^*\Bar \G^{b(s-1)}\Bar \G_{\un b(s-1)}
\Big\}~~.~~
\eea
Looking first at the theory containing only $G_{\un a(s-1)}$, the gauge transformation of the coupling and kinetic terms is
\bea
\nonumber
G^{\un b(s-1)}\Big\{
\Big[\frac 1{4(2s+1)}
+\frac e4\Big]
D^2 \Bar D^{\dot\a}
\Bar L_{\un b(s-1)\dot\a}
+\Big[\frac {(s-1)}{(2s+1)}
+e(s-1)\Big]
i\pa^{\un a}
D_{\b_{s-1}}\Bar L_{\b(s-2)\dot\b(s-1)\un a}
\eea
\bea
\Big[-\frac {(s^2-1)}{s(2s+1)}
+2f(s-1)\Big]
i\pa^{\un a}
\Bar D_{\dot\b_{s-1}}L_{\b(s-1)\dot\b(s-2)\un a}
+{\rm c.c.}\Big\}~~,~~
\eea
thus $e=-(2s+1)^{-1}$ and $f=+\frac {(s+1)}{2s(2s+1)}$.  All coefficients have been set and there is a unique description with the longitudinally linear compensating superfield.  Moving now to the action involving the transversally linear compensator.  The antichiral parts of $D\Bar D\d H$ or $\Bar D D\d H$ multiplying $\Bar \G$ have zero volume in superspace, but the chiral parts do not and can not be canceled by the gauge variations of the $\G$ kinetic terms.  This is connected to having set the coefficient $b=0$.  So the chiral parts of $D\Bar D\d H$ and $\Bar D D\d H$ must vanish separately.  This can be checked by multiplying by the chiral projector:
\bea
\nonumber
\Bar D^2 D^2D^\a\Bar D^{\dot\a}\d H_{\un a\un b(s-1)}=0~~,~~
\eea
\bea
\Bar D^2 D^2\Bar D^{\dot\a}D^\a\d H_{\un a\un b(s-1)}\not=0~~.~~
\eea
Thus, $\a_3$ must be used to remove the second term above, hence $\a_3=0$.  The gauge transformation of the coupling and kinetic terms is:
\bea
\int d^8z\Bar\G^{\un b(s-1)}\Big\{
\Big(\frac {(s+1)}{4s}+\frac {h^*}2\Big)
D^\a\Bar D^2L_{\un b(s-1)\a}
-\frac 14\Big(1+g\Big)
\Bar D^{\dot\a}D^2\Bar  L_{\un b(s-1)\dot\a}
\Big\}+{\rm c.c.}~~,~~
\eea
so the coefficients are $g=-1$ and $h=-\frac {(s+1)}{2s}$.  All coefficients have been set and there is a unique transversal linear theory.

In summary, two higher spin massless actions have been constructed.  The first contains a longitudinal linear compensating superfield and has the following action:
\bea
\nonumber
\cs_{Long}[H_{\un a(s)}, G_{\un b(s-1)}]=\int d^8z\Big\{H^{\un a\un c(s-1)}\Box\Big(
[\frac 12]\P^T_{3/2}
+[\frac 12-\frac {3s}{2(2s+1)}]\P^T_{1/2}
\eea
\bea
\nonumber
+[\frac 12-\frac s2]\P^L_{1/2}
-\frac {s^2}{(2s+1)}\P^L_0
\Big)_{\un a}{}^{\un b}H_{\un b\un c(s-1)}
-\frac {is}{(2s+1)}\pa_{\un a}H^{\un a\un b(s-1)}
\Big(
G_{\un b(s-1)}
-\Bar G_{\un b(s-1)}
\Big)
\eea
\bea
-\frac 1{2(2s+1)}\Big(G^{\un b(s-1)}\Bar G_{\un b(s-1)}
-\frac {(s+1)}{s}G^{\un b(s-1)}G_{\un b(s-1)}
+{\rm c.c.}
\Big)
\Big\}~~,~~
\label{longaction}
\eea
and the second contains a transversally linear compensating superfield with the following action:
\bea
\nonumber
\cs_{Trans}[H_{\un a(s)}, \G_{\un b(s-1)}]=\int d^8z
\Big\{
H^{\un a\un c(s-1)}\Box\frac 12\Big(
\P^T_{3/2}
+\P^T_{1/2}
+\P^L_{1/2}
\Big)_{\un a}{}^{\un b}H_{\un b\un c(s-1)}
\eea
\bea\nonumber
-\frac 12
H^{\un a\un b(s-1)}
\Big(
D_\a \Bar D_{\dot\a}\G_{\un b(s-1)}
-\Bar D_{\dot\a}D_\a \Bar\G_{\un b(s-1)}
\Big)
\eea
\bea
-\frac 12\Big(
\G^{\un b(s-1)}\Bar \G_{\un b(s-1)}
+\frac {(s+1)}{s}\G^{\un b(s-1)} \G_{\un b(s-1)}
+{\rm c.c.}
\Big)
\Big\}~~.~~
\label{transaction}
\eea
These actions are invariant under the following gauge transformations:
\bea
\d \G_{\un a(s-1)}=
-\frac 14\Bar D^{\dot\g} D^2\Bar L_{\dot\g\un a(s-1)}
~~,~~
\eea
\bea
\d G_{\un a(s-1)}=
\frac {(s-1)}{(s-1)!}i\pa^{\un d}\Bar D_{(\dot\a_{s-1}} L_{\dot\a(s-2))\a(s-1)\un d}
-\frac 14\Bar D^2D^\b L_{\b\un a(s-1)}~~,~~
\eea
and are known to be equivalent through a dual auxiliary action super functional \cite{Buchbinder1995}.  The derivation in this chapter reproduces the known results for massless higher spin theories.  It would be interesting to perform a more general analysis in which the variation (\ref{gaugetransH}) is canceled by the simultaneous addition of both $\G_{\un b(s-1)}$ and $G_{\un b(s-1)}$.  It seems that the theories of this nature would be higher spin theories coupled to some lower spin matter.

There is a way to analyze the on-shell properties of these theories by working solely in superspace.  First, superfield-strengths are written and Bianchi identities that relate derivatives of the superfield-strengths are found.  Then the theory is placed on-shell.  The non-vanishing superfield-strengths should form irreducible massless representations of the Poincar\'e superalgebra.  The superfield-strengths are:
\bea
\label{fshs1}
W_{\a(2s+1)}:={i\over 8(2s+1)!}\Bar D^2D_{(\a_{s+1}}
\pa_{\a_{s+2}}{}^{\dot\b_1}\cdots\pa_{\a_{2s+1}}{}^{\dot\b_s}
H_{\a(s))\dot\b(s)}~~,~~
\eea
\bea
\nonumber
G_{\un a\un c(s-1)}^L:={1\over (s!)^2}\Big\{
\Box(
\P^T_{3/2}
+[1-\frac {3s}{(2s+1)}]\P^T_{1/2}
+[1-s]\P^L_{1/2}
-\frac {2s^2}{(2s+1)}\P^L_0
)_{(\a(\dot\a}{}^{\un b}H_{|\un b|\g(s-1))\dot\g(s-1))}
\eea
\bea
+\frac {is}{(2s+1)}\pa_{(\a(\dot\a}
\Big(G_{\g(s-1))\dot\g(s-1))}-\Bar G_{\g(s-1))\dot\g(s-1))}\Big)
\Big\}~~,~~
\eea
\bea
\nonumber
T_{\b(s-1)\dot\b(s-2)}^L:={is(s-1)!\over (2s+1)}
\Bar D^{\dot\b_{s-1}}\pa^{\un a}H_{\un a\un b(s-1)}
-{(s-1)!\over(2s+1)}\Bar D^{\dot\b_{s-1}}\Bar G_{\un b(s-1)}
\eea
\bea
+{(s+1)(s-1)!\over s(2s+1)}\Bar D^{\dot\b_{s-1}}G_{\un b(s-1)}
~~,~~
\eea
\bea
\nonumber
G_{\un a\un c(s-1)}^T:={1\over (s!)^2}\Big\{
\Box(\Pi_{3/2}^T+\Pi_{1/2}^T+\Pi_{1/2}^L)_{(\a(\dot\a}
{}^{\un b}H_{|\un b|\g(s-1))\dot\g(s-1))}
\eea
\bea
-\frac 12\Big(
D_{(\a} \Bar D_{(\dot\a}\G_{\g(s-1))\dot\g(s-1))}
-\Bar D_{(\dot\a}D_{(\a} \Bar\G_{\g(s-1))\dot\g(s-1))}
\Big)
\Big\}~~,~~
\eea
\bea
\Bar T_{\dot\a\un b(s-1)}^T:=
-\frac 14\Bar D^2D^\a
H_{\un a\un b(s-1)}
+{1\over s!}\Bar D_{(\dot\a}\Bar \G_{\dot\b(s-1))\b(s-1)}
\label{fshs5}
+{(s+1)\over s\cdot s!}\Bar D_{(\dot\a}\G_{\dot\b(s-1))\b(s-1)}
\Big\}
~~,~~
\eea
where $T$ and $L$ superscripts pertain to the transverse and longitudinal formulations.  All superfield-strengths except $W_{\a(2s+1)}$ are proportional to equations of motion and therefore vanish on-shell.  Thus, it would only be necessary to find one Bianchi identity that relates $D^\b W_{\b\a(2s)}$ to derivatives of the other superfield-strengths to prove that this is the propagating superfield-strength.  If both $G$ and $\G$ were used to write a gauge invariant action there would probably be a new superfield-strength which contains both of these fields.  If this new superfield-strength is not proportional to the equations of motion, it would constitute a propagating degree of freedom.

This concludes the superprojector derivation of arbitrary higher superspin massless models.  The superprojector analysis has given insight into the general structure of these models.  In particular, the transversal formulation, (\ref{transaction}), contains only linear superprojectors, while the longitudinal model, (\ref{longaction}), contains all linear superprojectors plus the longitudinal chiral superprojector.  A list of superfield-strengths has been presented but has not been analyzed on-shell.  It seems that the superfield-strengths, except for $W_{\a(2s+1)}$, have not appeared in the literature and constitute a new result.

%

%

%

\newpage
\chapter{Massive Higher Superspin Actions}
\label{Massive Superspin Actions}

This chapter is dedicated to reviewing the known massive higher superspin models and to deriving new models.  Massive superspin-$Y$ multiplets have spins $Y-\frac 12, Y, Y$ and $Y+\frac 12$.  Sections \ref{Massive Graviton} reviews the massive non-supersymmetric spin-2 theory developed in 1939\cite{Fierz1939}.  Section \ref{Superspin-3/2 Actions}, describes the known superspin-$\frac 32$ massive superfield theories and uses the superprojectors to derive three new versions of the superspin-$\frac 32$ theory.  Section \ref{Massive Gravitino} reviews the non-supersymmetric massive gravitino.  Section \ref{Superspin-1 Actions} ends this chapter with a review of the known superspin-1 action and a derivation of another version of this model.

\section{Massive Half-Integer Superspin Actions}
\label{Massive Half-Integer Superspin Actions}

This section discusses models with massive gravitons as their highest spin component field.  Four-dimensional massive models of this kind can be used to represent the non-zero modes of higher dimensional supergravity theories.  This makes massive superspin-$\frac 32$ models particularly useful for doing phenomenology in extra dimensions.  This section will review the general method of constructing higher spin massive models.  Superprojectors will be used to construct superspace actions and several new models will be derived.
\subsection{Massive Graviton}
\label{Massive Graviton}
The basic concepts used to construct integer spin massive theories with no supersymmetry will be reviewed by working through the spin-2 model.  Although supersymmetric theories contain both fermionic and bosonic fields, in superspace half-integer superspin theories are described using bosonic valued superfields.  Thus, the methods developed for bosonic field theories are directly applicable at the superfield level to half-integer superspin superfield theories.  

Consider the completely symmetric and traceless bosonic field with $s$ vector indices, $X_{a(s)}$.  This field will be an on-shell massive representation of the Poincar\'e algebra if the following equations are satisfied:
\bea
\label{nosusyonshell}
\pa^a X_{ab(s-1)}=0~~,~~(\Box-m^2)X_{a(s)}=0~~.~~
\eea
This section describes a procedure used to derive action formulations that lead to these equations as a consequence of the equations of motion.  First, the general action is written using arbitrary coefficients.  Then, the equations of motion are analyzed and the coefficients are determined such that (\ref{nosusyonshell}) are satisfied.  In the case that these steps are insufficient to generate the action, a new auxiliary field is added to the theory and the two steps are repeated.  The auxiliary field should carry no propagating degrees of freedom and thus vanish on-shell.  The cases of spin-$1$ and spin-$2$ are enough to illustrate this procedure.

The massive photon is described by a vector field $A_a$.  The general action has three terms.  One coefficient can always be fixed by re-scaling one field.  The mass term on the propagating field will be canonically set to $-\frac 12$ to ensure a factor of $-m^2$ in the field equation.  The general action is:
\bea
\label{arbitproca}
\cs_{\rm Proca}[A_a]=\int d^4x\Big\{\frac 12\a_1A^a\Box A_a
+\frac 12\a_2A^a\pa_a\pa_b A^b
-\frac 12m^2A^aA_a
\Big\}~~,~~
\eea
where $\a_1$ and $\a_2$ are real coefficients.  In general, it is assumed that the fields are well behaved at infinity.  So integration by parts can be performed without keeping track of the boundary terms.  The Euler-Lagrange equations become:
\bea
\label{eulerlagrange}
{ \delta\cs\over \delta\Phi_k}
-\partial^a{\delta\cs\over\delta\partial^a\Phi_k}
={\delta\cs \over \delta\Phi_k}=0~~,~~
\eea
since any terms proportional to $\partial^a\Phi_k$ can be integrated by parts to become terms proportional to $\Phi_k$.
For the massive spin-$1$ action (\ref{arbitproca}) the field equation is:
\bea
\label{eomsp1}
{\d \cs_{\rm Proca}\over{\d A^a}}=\a_1 \Box A_a +\a_2 \pa_a\pa^bA_b-m^2A_a=0~~.
\eea
The divergence of the field equation is:
\bea
\label{diveomsp1}
\partial^a{\d \cs_{\rm Proca}\over{\d A^a}}=
(\a_1+\a_2)\Box\pa^aA_a-m^2\pa^aA_a=0~~,~~
\eea
which implies that $A_a$ has no divergence if $\a_1=-\a_2$.  With this choice, (\ref{eomsp1}) becomes:
\bea
\a_1 \Box A_a -m^2A_a=0~~,~~
\eea
and by taking $\a_1=1$, both conditions for $A_a$ to be a spin-1 representation have been fulfilled.  The off-shell Lagrangian with fixed coefficients is:
\bea
\nonumber
\cs_{Proca}=\frac 12\int d^4x\Big\{A^a\Box A_a-A^a\pa_a\pa_b A^b-m^2A^aA_a\Big\}
\eea
\bea
=\int d^4x\Big\{-\frac 14F^{ab}F_{ab}-\frac 12m^2A^aA_a
\Big\}~~.~~
\eea
This example showcases the general procedure nicely.  There are two distinct mechanisms that are working together to yield the desired effect.  First, some terms that are linearly independent in the field equation (\ref{eomsp1}), become equivalent in the divergence of the field equation (\ref{diveomsp1}).  Second, the mass term allows first order derivatives of the field to appear in the divergence of the field equation (\ref{diveomsp1}).  Setting the coefficients such that the linearly independent terms cancel under the divergence, allows the mass term to set an on-shell first-order differential constraint.  Higher spin theories work the same way, accept for the addition of auxiliary fields.

To illustrate how to use the auxiliary field, it is sufficient to examine the massive spin-2 field, or massive graviton.  An irreducible massive spin-2 field is described by a symmetric traceless rank two tensor, $H_{ab}$.  The general action is:
\bea
\cs_{s=2}[H_{ab}] =\int d^4x\Big\{\frac 12\b_1H^{ab}\Box H_{ab}
+\frac 12\b_2H^{ab}\pa_a\pa^cH_{bc}
-\frac 12m^2H^{ab}H_{ab}
\Big\}~~,~~
\eea
where $\b_1$ and $\b_2$ are real coefficients.  The equation of motion comes from the variation with respect to a symmetric traceless tensor, for which the appropriate functional variation is:
\bea
{\d H^{ab}(x)\over \d H^{cd}(y)}=(\frac 12\d_{(c}{}^a\d_{d)}{}^b
-\frac 14\h^{ab}\h_{cd})\d^4(x-y)~~.~~
\eea
The field equation is then:
\bea
{\d\cs_{s=2}[H_{ab}]\over{\d H^{ab}}}=\b_1\Box H_{ab}
+\b_2\frac 12 \pa_{(a}\pa^cH_{b)c}
-\b_2\frac 14\h_{ab}\pa^c\pa^dH_{cd}
- m^2 H_{ab}=0~~.~~
\eea
Two successive divergences of the field equation are:
\bea
\label{onedivsp2}
\partial^b{\delta \cs_{s=2}[H^{ab}]\over\delta H^{ab}}=
(\b_1+\frac 12\b_2)\Box\pa^bH_{ba}
+\frac 14\b_2\pa_{a}\pa^b\pa^cH_{bc}
-m^2 \pa^bH_{ab}=0~~,~~
\eea
and
\bea
\partial^a\partial^b{\delta \cs_{s=2}[H_{ab}]\over\delta H^{ab}}=
(\b_1+\frac 34\b_2)\Box\pa^a\pa^bH_{ba}
- m^2 \pa^a\pa^bH_{ab}=0~~.~~
\eea
There is no way to set $\b_1$ and $\b_2$ such that the divergence of $H_{ab}$ vanishes, without setting both coefficients to zero.  Although the two kinetic terms have become equivalent after two divergences of the field equation, which allows the possibility of $\pa^a\pa^bH_{ab}=0$, there is no way to set the coefficients to remove the divergence of $H_{ab}$ to yield a propagating spin-2 field.  Some other mechanism must be used.

The key to implementing the auxiliary field, is to observe that if the second divergence of $H_{ab}$ were zero, then the scenario would resemble the massive photon.  To this end, a scalar auxiliary field, $\varphi$, is introduced and coupled directly to the second divergence of $H_{ab}$.  The general action for both fields is: 
\bea
\cs_{\rm Fierz-Pauli}[H_{ab},\varphi]=\cs_{s=2}[H_{ab}]+\int d^4x\Big\{\varphi\pa^a\pa^bH_{ab}+\frac 12\b_3\varphi\Box\varphi
+\frac 12\b_4m^2\varphi\varphi
\Big\}~~,
\eea
where the coefficient of the coupling term has been absorbed into the definition of $\varphi$.  $\b_3$ and $\b_4$ are real coefficients.  The field equations are now:
\bea
\nonumber
{\d\cs_{\rm Fierz-Pauli}[H^{ab},\varphi]\over\d H^{ab}}=\b_1\Box H_{ab}
+\b_2\frac 12 \pa_{(a}\pa^cH_{b)c}
-\b_2\frac 14\h_{ab}\pa^c\pa^dH_{cd}
\eea
\bea
\label{spin2eom}
- m^2 H_{ab}
+\pa_a\pa_b\varphi
-\frac 14\h_{ab}\Box\varphi=0~~,~~
\eea
\bea
\label{aux1}
{\d\cs_{\rm Fierz-Pauli}[H^{ab},\varphi]\over\d\varphi}=
\pa^a\pa^bH_{ab}+\b_3\Box\varphi
+\b_4m^2\varphi=0~~.~~
\eea
The coefficients are first set to eliminate $\varphi$, so that it truly is auxiliary.  Solving for $\pa^a\pa^bH_{ab}$ in (\ref{aux1}) and substituting into the second divergence of  (\ref{spin2eom}) leads to:
\bea
0=\Big[-\b_3(\b_1+\frac 34\b_2)+\frac 34\Big]\Box^2\varphi
+\Big[\b_3-\b_4(\b_1+\frac 34\b_2)
\Big]m^2\Box\varphi
+\b_4m^4\varphi~~,
\eea
Setting the coefficients $\b_3=\frac 34(\b_1+\frac 34\b_2)^{-1}$ and $\b_4=\b_3(\b_1+\frac 34\b_2)^{-1}$ leaves only $\b_4m^4\varphi=0$, forcing $\varphi$ into auxiliary status.   For consistency, (\ref{aux1}) must still be satisfied, so $\pa^a\pa^bH_{ab}=0$.  Now, if $\b_2=-2\b_1$, (\ref{onedivsp2}) implies that $\pa^aH_{ab}=0$, and the equation of motion now reads $(\b_1\Box-m^2)H_{ab}=0$, so $\b_1=1$.  The full action is:
\bea
\nonumber
\cs_{\rm Fierz-Pauli}[H^{ab}, \varphi]=\int d^4x\Big\{\frac 12H^{ab}\Box H_{ab}
-H^{ab}\pa_a\pa^cH_{bc}
-m^2H^{ab}H_{ab}
\eea
\bea
+\varphi\pa^a\pa^bH_{ab}-\frac 34\varphi\Box\varphi
+3m^2\varphi\varphi
\Big\}~~.~~
\eea
It is a simple exercise to define a symmetric rank two tensor $g_{ab}:=H_{ab}+b\h_{ab}\varphi$, and show that for some choice of real coefficient $b$, the Fierz-Pauli action can be written completely in terms of $g_{ab}$.

The procedure of writing the general Lagrangian, fixing coefficients and guessing auxiliary fields gets mapped directly into superspace.  The only difference is the added complication of the covariant spinor derivatives.  In the next section, superprojectors are used to simplify calculations.  The superprojectors come in handy especially when superfield equations are substituted into one another.  The orthogonality of the superprojectors simplifies these calculations.
\subsection{Superspin-3/2 Actions}
\label{Superspin-3/2 Actions}
This section reviews how the superspace projectors can be used to simplify the analysis of the superspin-$\frac 32$ cases that are known in the literature.  Also, a new version of superspin-$\frac 32$ is derived that couples the propagating superfield $V_{\un a}$ to the auxiliary superfield $V^\prime$ using the commutator of spinor derivatives instead of the partial derivative.  Because of this coupling, the auxiliary superfield $V^\prime$ must have positive charge under parity.  Two other superspin-$\frac 32$ versions are derived.  These two theories correspond to the massive analog of new-minimal linearized supergravity and the recently discovered ``new"-new-minimal linearized supergravity of \cite{Buchbinder2002a}.

The general action is taken from chapter \ref{Half-Integer Massless Arbitrary Superspin Actions} with $\a_2=0$, plus a mass term:
\bea
\nonumber
\cs_{Y=\frac 32}[V_a]=
\int d^8z \Big\{
\a_1 V^{\un a}D^\b\Bar D^2D_\b V_{\un a}
+\a_3 V^{\un a}\pa_{\un a}\pa^{\un b}V_{\un b}
\eea
\bea
\label{rawaction}
+\a_4V^{\un a}[D_\a,\Bar D_{\dot\a}][D_{\b},\Bar D_{\dot\b}]V^{\un b}
-\frac 12{\rm m}^2V^{\un a}V_{\un a}\Big\}~~.~~
\eea
In terms of superprojectors this becomes:
\bea
\nonumber
\cs_{Y=\frac 32}[V_a]=
\int d^8z\Big\{ V^{\un a}\Box\Big(
[-2\a_3+8\a_4]\P_{0}^L
+[-8\a_1-2\a_3]\P_{1/2}^L
\eea
\bea
\label{projectedaction}
+[-8\a_1-24\a_4]\P_{1/2}^T
-8\a_1\P_{3/2}^T
\Big)V_{\un a}-\frac 12{\rm m}^2V^{\un a}V_{\un a}\Big\}~~.~~
\eea
As in the case of the spin-2 field, this action can not generate the  constraints:
\bea
\label{32irrepsp}
(\Box-m^2)V_a=0~~,~~
D^\a V_{\un a}=0~~,~~
\Bar D^{\dot\a} V_{\un a}=0~~,~~
\pa^{\un a}V_{\un a}=0~~,~~
\eea
which make $V_a$ an irreducible superspin-$\frac 32$ representation of the Poincar\'e superalgebra.  It turns out that there are several ways to couple an auxiliary superfield to $V_a$ and get the correct dynamics.  The original models of \cite{Buchbinder2002a} use $mV\pa^aV_a$, with $V=\Bar V$.  Using another real superfield $V^\prime$ the coupling can be $mV^\prime[D^\a, \Bar D^{\dot\a}]V_{\un a}$.  Since $PV_a=-V_a$, this coupling would mean that $V^\prime$ has positive parity, whereas, $V$ had negative parity.  The $m$ in the coupling can also be replaced by covariant spinor derivatives.  Using $m$ $\rightarrow$ $D^2$ and a real auxiliary superfield $P$ the coupling is $PD^2\pa^{\un a}V_{\un a}$.  
This coupling was considered in \cite{Gregoire2004} and corresponds to a massive extension of linearized old-minimal supergravity, and is related to the five-dimensional theory in \cite{Linch2003}.  There are two more versions of linearized supergravity\cite{Gates2003a}, so it is fair to assume that there are massive theories associated with each of them.  A chiral spinor auxiliary superfield $\chi_\a$ and the coupling $D^\b\chi_\b[D^\a, \Bar D^{\dot\a}]V_{\un a}$ leads to a massive extension of new-minimal linearized supergravity.  Another chiral spinor $\l_\a$ and the coupling $D^\b\l_\b\pa^{\un a}V_{\un a}$ leads to the massive extension of ``new"-new-minimal linearized supergravity.  All of these possibilities will be considered in what follows.

Starting with the $V^\prime$ theory the auxiliary action takes the form:
\bea
\cs_{Aux}[V_a, V^\prime]=\int d^8z\Big\{m\tilde\g V^\prime[D^\a,\Bar D^{\dot\a}]V_{\un a}+\frac 12V^\prime\Big(\d_1\Box 
+\d_2\{D^2,\Bar D^2\}+\d_3m^2\Big)V^\prime
\Big\}~~,~~
\eea
where $\tilde\g$, $\d_1$, $\d_2$, and $\d_3$ are real coefficients.  The equations of motion are:
\bea
\nonumber
{\d\over\d V^a}\Big(\cs_{Y=\frac 32}[V_a]+\cs_{Aux}[V_a, V^\prime]\Big)=2\Box\Big[(-2\a_3+8\a_4)\P_{0}^L
+(-8\a_1-2\a_3)\P_{1/2}^L
\eea
\bea
\label{primeVeom}
-8\a_1\P_{3/2}^T
+(-8\a_1-24\a_4)\P_{1/2}^T
\Big]V_{\un a}-m^2V_{\un a}
+\tilde\g m[D_\a, \Bar D_{\dot\a}]V=0~~,~~
\eea
\bea
\label{primeVprimeeom}
{\d\over\d V^\prime}\Big(\cs_{Y=\frac 32}[V_a]+\cs_{Aux}[V_a, V^\prime]\Big)
=\tilde\g m[D^\a,\Bar D^{\dot\a}]V_{\un a}
+(\d_1\Box 
+\d_2\{D^2,\Bar D^2\}
+\d_3m^2)V^\prime=0~~.
\eea
Contracting the commutator of covariant derivatives on (\ref{primeVeom}) allows the substitution of (\ref{primeVprimeeom}) and elimination of $V^\prime$.  Only two of the super projections of $V_{\un a}$ are nonzero under this operation:
\bea
[D^\a,\Bar D^{\dot\a}]\P_{0}^LV_{\un a}=
\frac 1{16}\Box^{-1}\{D^2,\Bar D^2\}[D^\a,\Bar D^{\dot\a}]V_{\un a}~~,~~\cr
[D^\a,\Bar D^{\dot\a}]\P_{1/2}^TV_{\un a}=
-\frac i4\Box^{-1}D^\a\Bar D^2D^\d\pa_{(\a}^{~~\dot\b}V_{\d)\dot\b}
~~.~~
\eea
The second relation does not allow the substitution of (\ref{primeVprimeeom}), thus it must be removed:
\bea
\a_4=-\frac 13\a_1~~.~~
\eea
With these coefficients set, the contraction of $[D^\a,\Bar D^{\dot\a}]$ on (\ref{primeVeom}) leads to the following relations:
\bea
\label{deltas}
\d_1-24\tilde\g^2=0~~,~~\cr
(-2\a_3+8\a_4)(\d_1+16\d_2)=0~~,~~\cr
-\d_2-2\tilde\g^2+\frac 18(-2\a_3+8\a_4)\d_3=0~~,~~
\eea
so that $V^\prime=0$.  The vanishing of the auxiliary superfield implies $[D^\a,\Bar D^{\dot\a}]V_{\un a}=0$.  With this constraint it is possible to generate the equations (\ref{32irrepsp}), by analyzing the contraction of  $D^\a$ on (\ref{primeVeom}).  Of the only non-zero superprojectors, $\P_{3/2}^T$ vanishes under this operator and:
\bea
D^\a\P_{1/2}^LV_{\un a}
=-\frac i8\Box^{-1}D^2\Bar D_{\dot\a}\pa_{\un b}V^{\un b}~~,~~
\eea
which must be removed, thus:
\bea
\a_3=-4\a_1~~.~~
\eea
This clears up the ambiguity in (\ref{deltas}) and implies that $D^\a V_{\un a}=0$.  Obtaining the Klein-Gordon equation reveals that $\a_1=-\frac 1{16}$.  After absorbing $\tilde\g$ into $V^\prime$, the action becomes:
\bea
\nonumber
\cs_{Y=\frac 32}[V_a, V^\prime]=
\int d^8z\Big\{ V^{\un a}\Box\Big(
-\frac 13\P_{0}^L
+\frac 12\P_{3/2}^T
\Big)V_{\un a}-\frac 12m^2V^{\un a}V_{\un a}
\eea
\bea
\label{psudomass}
\label{Newresult1}
+mV^\prime[D^\a,\Bar D^{\dot\a}]V_{\un a}+V^\prime(12\Box 
-\frac 34\{D^2,\Bar D^2\}-6m^2)V^\prime\Big\}~~.~~
\eea

The two other actions with massive couplings come from the auxiliary action:
\bea
\cs_{Aux}[V_a, V]=\int d^8z\Big\{m\g V\pa^{\un a}V_{\un a}+\frac 12V\Big(\d_1\Box 
+\d_2\{D^2,\Bar D^2\}+\d_3\rm{m}^2\Big)V
\Big\}~~,~~
\eea
where $\gamma$, $\d_1$, $d_2$ and $\d_3$ are real coefficients.  The $\d$ coefficients in this example should not be confused with the $\d$ coefficients in the $V^\prime$ example.  The equations of motion are:
\bea
\nonumber
{\d\over\d V^a}\Big(\cs_{Y=\frac 32}[V_a]+\cs_{Aux}[V_a, V]\Big)=2\Box\Big[(-2\a_3+8\a_4)\P_{0}^L
+(-8\a_1-2\a_3)\P_{1/2}^L
\eea
\bea
\label{old32Veom}
-8\a_1\P_{3/2}^T
+(-8\a_1-24\a_4)\P_{1/2}^T
\Big]V_{\un a}-{\rm m}^2V_{\un a}
-\g{\rm m}\pa_{\un a}V=0~~,~~
\eea
\bea
\label{old32Vauxeom}
{\d\over\d V}\Big(\cs_{Y=\frac 32}[V_a]+\cs_{Aux}[V_a, V]\Big)=\g{\rm m}\pa^{\un a}V_{\un a}+\Big(\d_1\Box 
+\d_2\{D^2,\Bar D^2\}+\d_3{\rm m}^2\Big)V=0~~.~~
\eea
The divergence of (\ref{old32Veom}) allows the substitution of (\ref{old32Vauxeom}).  Only the longitudinal superprojections are non-zero under the divergence:
\bea
\pa^{\un a}\P^L_{0}V_{\un a}={\frac 1{16}}\Box^{-1}\{ D^2, \Bar D^2 \}\pa_{\un c}V^{\un c}
\eea
\bea
\pa^{\un a}\P^L_{1/2}V_{\un a}=
-{\frac 18}\Box^{-1}D^\d\Bar D^2D_\d \pa_{\un c}V^{\un c}
\eea
Both of these relations allow the substitution of (\ref{old32Vauxeom}), and do not need to be removed at this point.  Setting $V=0$ in the divergence of (\ref{old32Veom}) leads to:
\bea
\label{choice}
\d_1(8\a_1+2\a_3)=0~~,~~\cr
(-2\a_3+8\a_4)(\d_1+16\d_2)=0~~,~~\cr
-2\d_3(-2\a_3+8\a_4)+16\d_2+\d_1+2\g^2=0~~,~~\cr
+2\d_3(8\a_1+2\a_3)+\d_1+2\g^2=0~~.~~
\eea
With this, $V=\pa^{\un a}V_{\un a}=0$ and all longitudinal projections are eliminated.   To fully reduce $V_{\un a}$, the contraction of $D^\a$ on (\ref{old32Veom}) is performed.  In this calculation the only superprojector that remains is $\P_{1/2}^T$ and it is non-zero under this operation, thus, it must be removed $\a_4=-\frac 13\a_1$.  There are now two solutions according to (\ref{choice}):
\bea
&1.&\a_3=-4\a_1~~,~~\d_1=-2\g^2~~,~~\d_2=+\frac 18\g^2~~,~~
\d_3=+\frac 3{16}\g^2\a_1{}^{-1}~~,~~\\
&2.&\a_3=-\frac 43\a_1~~,~~\d_1=0~~,~~\d_2=-\frac 18\g^2~~,~~\d_3=-\frac 3{16}\g^2~~.~~
\eea
These are the solutions previously found in \cite{Buchbinder2002a}.

\subsubsection{\bf Massive Extension of Old-Minimal Supergravity}
\label{Massive Extension of Old-Minimal Supergravity}

The action for old-minimal linearized supergravity contains $V_a$ and a chiral compensator $\S$:
\bea
\cs_{\rm Old-Min}[V_a,\S]=\int d^8z\Big\{V^{\un a}\Box(
-\frac 13\P^L_0
+\frac 12\P^T_{3/2})V_{\un a}
-i(\S-\Bar\S)\pa^{\un a}V_{\un a}
-3\S\Bar\S\Big\}~~.~~
\eea
It would be nice to return to the this action when $m=0$.  So the auxiliary action should be proportional to $m$ or some other auxiliary superfields.  It is impossible to get the correct dynamics unless the chiral superfield is replaced with a potential.  The simplest potential is a real scalar superfield, $\S=-\frac 14\Bar D^2P$.  The mass terms for this theory are:
\bea
\cs_m[V_a, P]=\int d^8z\Big\{-\frac 12m^2V^{\un a}V_{\un a}+\frac 12\d m^2 P^2
\Big\}~~,~~
\eea
where $\d$ is a real coefficient.  The equations of motion are:
\bea
{\d\over\d V^{\un a}}\Big(\cs_{\rm Old-Min}[V_a,P]+\cs_m[V_a,P]\Big)
=\Box\Big[\P_{3/2}^T
-\frac 23\P_0^L\Big]V_{\un a}
\label{oldmin32Veom}
-m^2V_{\un a}
+i\pa_{\un a}(\S-\Bar \S)
=0~~,~~
\eea
\bea
\label{oldmin32Peom}
{\d\over\d V}\Big(\cs_{\rm Old-Min}[V_a, P]+\cs_m[P]\Big)=
+\frac i4\Bar D^2\pa^{\un a}V_{\un a}
-\frac i4 D^2\pa^{\un a}V_{\un a}
-\frac 3{16}\{D^2,\Bar D^2\}P
+\d{\rm m}^2P=0~~.~~
\eea
The equation of motion for $P$ implies that $D^\a\Bar D^2D_\a P=0$ and therefore $\{D^2,\Bar D^2\}P=16\Box P$.  Substituting (\ref{oldmin32Peom}) into the real part of the contraction of $D^2\pa^{\un a}$ on the equation of motion of $V_{\un a}$ implies that  $P=0$ if $\d=\frac 92$.  The vanishing of $P$ sets $\P^L_0V_{\un a}=0$ and equation (\ref{oldmin32Veom}) then implies $\pa^{\un a}V_{\un a}=0$ and $D^\a V_{\un a}=0$.  The full action is:
\bea
\nonumber
\cs_{Y=3/2}[V_a,P]=\int d^8z\Big\{V^{\un a}\Box(
-\frac 13\P^L_0
+\frac 12\P^T_{3/2})V_{\un a}
-i(\S-\Bar\S)\pa^{\un a}V_{\un a}
\eea
\bea
-3\S\Bar\S
-\frac 12m^2V^{\un a}V_{\un a}+\frac 94 m^2 P^2
\Big\}~~,~~
\eea
which is exactly the action presented in \cite{Gregoire2004}.  This action represents a massive extension of linearized old-minimal supergravity and is directly associated to the dimensional reduction of the five-dimensional action of \cite{Linch2003}.
\subsubsection{\bf Massive Extension of New-Minimal Supergravity}
\label{Massive Extension of New-Minimal Supergravity}

With the knowledge of how to extend old-minimal supergravity, it should be possible to create a massive extension of new-minimal supergravity.  New-minimal supergravity is described by the vector valued superfield $V_{\un a}$ and a real linear compensating superfield $\cu$.  Since $\cu$ has mass dimension $1$, it is not possible to write a mass term for this superfield.  Introducing a chiral spinor  potential for $\cu$:
\bea
\cu=D^\a \chi_\a+\Bar D_{\dot\a}\Bar \chi^{\dot\a}~~,~~
\eea
allows a mass term for the compensator.  The action for linearized new-minimal supergravity is:
\bea
\cs_{\rm New-Min}[V_{\un a}, \cu]=\int d^8z\Big\{V^{\un a}\Box(-\P^T_{1/2}+\frac 12\P^T_{3/2})V_{\un a}+\frac 12\cu [D_\a,\Bar D_{\dot\a}]V^{\un a}+\frac 32\cu^2\Big\}~~.~~
\eea
The only possible mass terms are:
\bea
\cs_m[V_{\un a}, \chi_\a]=
-\frac 12m^2\int d^8z V^{\un a}V_{\un a}
+\frac 12\g m^2\int d^6z
 \chi^\a\chi_\a
+\frac 12\g^* m^2\int d^6\bar z
\bar \chi_{\dot\a}\bar\chi^{\dot\a}~~.~~
\eea
The equations of motion are:
\bea
\label{sweet}
{\d\over\d V^{\un a}}\Big(\cs_{\rm New-Min}[V_a,\cu]+\cs_m[V_a,\chi_\a]\Big)=\Box\Big[
\P_{3/2}^T
-2\P_{1/2}^T
\Big]V_{\un a}
-{\rm m}^2V_{\un a}
+\frac 12[D_\a,\Bar D_{\dot\a}]\cu
=0~,~\,
\eea
\bea
\label{tasty}
{\d\over\d \chi^\a}\Big(\cs_{\rm New-Min}[V_a,\cu]+\cs_m[V_a,\chi_\a]\Big)=
\frac 18\Bar D^2D_\a[D_\b,\Bar D_{\dot\b}]V^{\un b}
+\frac 34\Bar D^2D_\a\cu
+m^2\g\chi_\a
=0~~.~~
\eea
Because $\cu$ is linear, (\ref{sweet}) implies that $V_{\un a}$ is linear, $D^2V_{\un a}=0$.  It does not seem possible to use the normal algorithm of substitution followed by elimination of the auxiliary superfield in this case.  This inability to use the standard procedure in superspace is a hint as to why constructing models with superspin greater than $\frac 32$ is so hard.  Although the standard procedure does not work, it is possible to show that $\Bar D^{\dot\a}V_{\un a}\propto \chi_\a$, both of these quantities will vanish on-shell, so this is not a contradictory statement.  To prove this proportionality, first contract $\Bar D^{\dot\a}$ on (\ref{sweet}) and use the following identities:
\bea
\label{goodvibes}
\Bar D^2 D_\b [D_\a, \Bar D_{\dot\a}]V^{\un a}
=2i\Bar D^2 D^\a\pa_{(\a}{}^{\dot\a}V_{\b)\dot\a}~~,~~
\eea
\bea
\Box\Bar D^{\dot\a}\P_{1/2}^TV_{\un a}=
-\frac i8\Bar D^2D^\d\pa_{(\a}{}^{\dot\b} V_{\d)\dot\b}
=-\frac 1{16}\Bar D^2 D_\b [D_\a, \Bar D_{\dot\a}]V^{\un a}~~,~~
\eea
to arrive at:
\bea
+\frac 18\Bar D^2 D_\b [D_\a, \Bar D_{\dot\a}]V^{\un a}
+\frac 34\Bar D^2D_\a\cu
- m^2 \Bar D^{\dot\a}V_{\un a}
=0~~.~~
\eea
Substituting the first two terms with (\ref{tasty}) leads to:
\bea
\label{thebomb}
\g\chi_\a
+ \Bar D^{\dot\a}V_{\un a}=0~~.~~
\eea
Substituting for $\cu$ in (\ref{tasty}) by plugging (\ref{thebomb}) back in yields:
\bea
+\frac 18\Bar D^2D_\a[D_\b,\Bar D_{\dot\b}]V^{\un b}
-\frac 34\frac 1\g\Bar D^2D_\a[D^\b\Bar D^{\dot\b}
-\frac \g{\g^*}\Bar D^{\dot\b}D^\b]V_{\un b}
+m^2\g\chi_\a
=0~~.~~
\eea
This means that $\chi_\a$ will vanish if $\g$ is real and $\g=6$.  Equation (\ref{thebomb}) implies that 
$V_{\un a}$ is irreducible when $\chi_\a$ vanishes.  This means that $\P^T_{3/2}V_{\un a}=V_{\un a}$ and the Klein-Gordon equation is obtained from (\ref{sweet}).  The full action is:
\bea
\nonumber
\cs_{Y=3/2}[V_{\un a}, \chi_\a]=
\int d^8z\Big\{V^{\un a}\Box(-\P^T_{1/2}+\frac 12\P^T_{3/2})V_{\un a}+\frac 12\cu [D_\a,\Bar D_{\dot\a}]V^{\un a}+\frac 32\cu^2
\eea
\bea
\label{newminmass}
-\frac 12m^2
V^{\un a}V_{\un a}\Big\}
+3 m^2\int d^6z
 \chi^\a\chi_\a
+3 m^2\int d^6\bar z \bar\chi_{\dot\a}\bar\chi^{\dot\a})~~.~~
\eea
This result is remarkable for three reasons. First, this action is a completely new result.  It has never appeared in the literature.
Second, the existence of this action gives more support to the idea that it is possible to take the kinetic terms for a known massless theory and simply add mass terms to obtain the massive theory.  This had only recently been seen for old-minimal and until now had not been corroborated for new-minimal. This insight will guide the construction of the higher superspin massive actions.  Third, by reversing the dimensional reduction procedure, it is possible to perform dimensional oxidation and create a new massless five-dimensional theory.  This theory will be dual to the theory that is related to old-minimal.  Instead of containing a five-dimensional gauge 1-form, it will contain a five-dimensional gauge 2-form.  The dimensional oxidation will be discussed in section \ref{Dimensional Oxidation from New-Minimal}.

\subsubsection{\bf Massive Extension of ``New"-New-Minimal Supergravity}
\label{Massive Extension of ``New"-New-Minimal Supergravity}

With the success of writing massive versions of old and new minimal it is only natural to look for a massive extension of the ``new"-new-minimal theory first described in \cite{Buchbinder2002a}.  This theory has the same superfield content as new-minimal.  The distinction between the two occurs in the gauge transformation of the compensator:
\bea
\d U=\frac i{12}(D^\a\Bar D^2L_\a-\Bar D_{\dot\a}D^2\Bar L^{\dot\a})~~.~~
\eea
The action for ``new"-new-minimal linearized supergravity is:
\bea
\cs_{\n^2\rm-Min}[V_{\un a}, U]=\int d^8z\Big\{
V^{\un a}\Box(\frac 12\P^T_{3/2}+\frac 13\P^L_{1/2})V_{\un a}
+U\pa_{\un a}V^{\un a}
+\frac 32U^2
\Big\}~~.~~
\eea
As in the case of new-minimal, a chiral spinor superfield potential is used:
\bea
U=iD^\a\l_\a-i\Bar D_{\dot\a}\bar\l^{\dot\a}~~,~~
\eea
which allows the following mass terms to be added to the action:
\bea
\cs_m[V_{\un a}, \l_\a]
=-\frac 12 m^2\int d^8zV^{\un a}V_{\un a}
+\frac 12\d m^2\int d^6z \l^\a\l_\a
+\frac 12\d^* m^2\int d^6\bar z \bar\l_{\dot\a}\bar\l^{\dot\a}
~~.~~
\eea
The equations of motion are:
\bea
\label{eomVnnmass}
{\d\over\d V^{\un a}}\Big(\cs_{\n^2\rm-Min}[V_a,U]+\cs_m[V_a,\l_\a]\Big)=2\Box(\frac 12\P^T_{3/2}+\frac 13\P^L_{1/2})V_{\un a}
-\pa_{\un a}U
- m^2V_{\un a}=0~~,~~
\eea
\bea
\label{eomlamnnmass}
{\d\over\d \l^\a}\Big(\cs_{\n^2\rm-Min}[V_a,U]+\cs_m[V_a,\l_\a]\Big)=+\frac i4\Bar D^2D_\a\pa_{\un b}V^{\un b}
+\frac 34i\Bar D^2D_\a U
+\d m^2 \l_\a=0~~.~~
\eea
It is possible to derive an identity between $\l_\a$ and $\Bar D^{\dot\a}V_{\un a}$ as in the new-minimal case.  This can be proven by using the identity:
\bea
\Box\Bar D^{\dot\a}\P^L_{1/2}V_{\un a}
=\frac i8\Bar D^2D_\g \pa_{\un b}V^{\un b}~~,~~
\eea
in the contraction of $\Bar D^{\dot\a}$ on (\ref{eomVnnmass}).  This leads to:
\bea
-\frac 13\d m^2 \l_\a
- m^2\Bar D^{\dot\a}V_{\un a}
=0~~,~~
\eea
which means that if $\l_\a$ vanishes, $V_{\un a}$ is irreducible.  Substituting this result into (\ref{eomlamnnmass}) yields:
\bea
+\frac i4\Bar D^2D_\a\pa_{\un b}V^{\un b}
+\frac 1\d\frac 94\Bar D^2D_\a 
( D^\b\Bar D^{\dot\b}
+\frac \d{\d^*} \Bar D^{\dot\b} D^\b) V_{\un b}
+\d m^2 \l_\a=0~~.~~
\eea
Thus, if $\d=18$, $\l_\a$ vanishes and the full action is:
\bea
\label{newnewminmass}
\cs_{Y=3/2}[V_{\un a}, \l_\a]=\int d^8z\Big\{
V^{\un a}\Box(\frac 12\P^T_{3/2}+\frac 13\P^L_{1/2})V_{\un a}
+U\pa_{\un a}V^{\un a}
+\frac 32U^2\cr
-\frac 12 m^2V^{\un a}V_{\un a}\Big\}
+9 m^2\int d^6z \l^\a\l_\a
+9 m^2\int d^6\bar z \bar\l_{\dot\a}\bar\l^{\dot\a}
~~.~~
\eea
This action is a new result and will be oxidized to produce a five-dimensional massless action in the next chapter.

This concludes the discussion of massive half-integer superspin models.  Three new superspin-$\frac 32$ models have been constructed. The action (\ref{psudomass}) corresponds to a gauge fixed massless theory when $m=0$ and also contains a real scalar auxiliary superfield.  This theory stands in contrast to the original two theories published in \cite{Buchbinder2002a}, which have real pseudo scalar superfields.  The two actions (\ref{newminmass}) and (\ref{newnewminmass}) represent massive extensions of linearized supergravity theories.  When $m=0$ these theories correspond to non-gauge fixed massless theories.  This distinction makes these theories more interesting phenomenologically, since they can be related to five-dimensional massless theories.

This section shows that it is possible to write massive extensions of massless theories in superspace.  There are three subtleties that complicate the construction of theories with superspin greater than $\frac 32$.  The first subtlety is in the choice of the potential of the compensating superfield of the massless theory, e.g. $\S$$=-\frac 14$$\Bar D^2$$P$.  In higher superspin theories the choice of potential is not always obvious.  Second, direct substitution of the auxiliary superfield equation of motion did not work in the final two examples.  This makes the procedure for solving the equations of motion more of an art than a  algorithm.  Third, the number of auxiliary fields in non-supersymmetric theories is not equal between massless and massive theories with spin higher than 2.  This means that for theories with superspin greater than $\frac 32$, new auxiliary multiplets must be introduced.  The considerations in this section shed no light on this problem.  These three points make it extremely difficult to construct higher superspin theories.  I have attempted to construct a superspin-$\frac 52$ theory using the longitudinal formulation of massless $\frac 52$.  I used a spinor potential for the longitudinal compensator.  It seems that this spinor superfield potential must be coupled through a Dirac mass term, i.e. $m\psi^\a\u_\a$, to an auxiliary spinor superfield.  It seems that this auxiliary superfield must be further coupled to a chiral auxiliary superfield.  I also attempted to construct a massless five-dimensional theory that could be reduced to four dimensions superspin-$\frac 52$, hoping to use gauge invariance as a guide.  In both of these attempts, the proliferation of superfields and complication of the equations of motion have proven to be an insurmountable challenge for publication in one graduate career.
\section{Massive Integer Superspin Actions}
\label{Massive Integer Superspin Actions}

Massive integer superspin models have spinors as the highest spin component fields.  This means that the superfields that are used to describe these models are also spinor valued.  The equations of motion associated with spinor superfields are relatively complicated compared to their bosonic counterparts.  It seems that there are some subspaces in spinor valued superfield equations that are not as intuitive.  This point will be discussed at the end of this section when a new version of superspin-$1$ is presented.  This section first reviews the massive gravitino or Rarita-Schwinger field in order to show that the method of auxiliary fields also works for spinors.   After this brief review, the known superspin-$1$ model is presented and a new version is derived.
\subsection{Massive Gravitino}
\label{Massive Gravitino}
A massive spin-$\frac 32$ representation is described with the irreducible fermion field $\Psi_{\a\b\dot\a}$.  This field is symmetric in $\a\b$, so the functional variation is:
\bea
{{\d\Psi^{\a\b\dot\a}(x)}\over{\d\Psi^{\g\d\dot\b}(y)}}=
\frac 12\d_{(\g}{}^\a\d_{\d)}{}^\b\d_{\dot\b}{}^{\dot\a}\d^4(x-y)
~~.~~
\eea
For higher spin fermions, the massive representations are governed by the Rarita-Schwinger equation and vanishing divergence:
\bea
\label{RSconstrsaints}
-\frac i2\pa_{(\b}{}^{\dot\b}\Bar\Psi_{\a)\dot\a\dot\b}
+m\Psi_{\a\b\dot\a}=0
~~,~~\pa^{\un a}\Psi_{\a\b\dot\a}=0~~.~~
\eea
An action containing only $\Psi_{\un a\b}$ is insufficient to produce these equations and an auxiliary spinor field must be introduced, $\chi_\a$.  The general action is:
\bea
\nonumber
\cs[\Psi_{\un a\b},\chi_\a]=\int d^4x\Big\{
-i\Psi^{\un a\g}\pa_\g{}^{\dot\g}\Bar\Psi_{\un a\dot\g}
+\frac 12m\Psi^{\a\b\dot\g}\Psi_{\a\b\dot\g}
+\frac 12m\Bar \Psi_{\dot\a\dot\b\g}\Bar\Psi^{\dot\a\dot\b\g}
\eea
\bea
+\Psi^{\a\b\dot\g}\pa_{\a\dot\g}\chi_\b
+\Bar\Psi_{\dot\a\dot\b\g}\pa^{\dot\a\g}\bar\chi^{\dot\b}
+\a i\chi^\a\pa_{\un a}\bar\chi^{\dot\a}
+\b m\chi^\a\chi_\a
+\b^*m\bar\chi_{\dot\a}\bar\chi^{\dot\a}
\Big\}~~,~~
\eea
where $\a$ and $\b$ are respectively real and complex coefficients.  The coefficients on the first three terms are fixed to produce (\ref{RSconstrsaints}) when $\chi_\a=0$, and the coefficient of the fourth and fifth terms has been absorbed into the normalization of $\chi_\a$.  The equations of motion are:
\bea
\label{eomrsphys}
{\d\over \d\Psi^{\un a\b}}\cs[\Psi_{\un a\b},\chi_\a]=
-\frac i2\pa_{(\b}{}^{\dot\b}\Bar\Psi_{\a)\dot\a\dot\b}
+m\Psi_{\a\b\dot\a}
+\frac 12\pa_{\dot\a(\a}\chi_{\b)}=0
~~,~~
\eea
\bea
\label{eomrsaux}
{\d\over \d\chi^\a}\cs[\Psi_{\un a\b},\chi_\a]=
-\pa^{\un b}\Psi_{\un b\a}
+i\a \pa_{\un a}\bar\chi^{\dot\a}
+2\b m\chi_\a=0
~~.~~
\eea
After plugging (\ref{eomrsaux}) into the divergence of (\ref{eomrsphys}), $\chi_\a$ can be set to zero if $\a=3$ and $\b=-3$.  When $\chi_\a=0$, $\Psi_{\un a\b}$ has no divergence and the constraints (\ref{RSconstrsaints}) are satisfied.  The final action takes the form:
\bea
\nonumber
\cs_{Rarita-Schwinger}=\int d^4x\Big\{
-i\Psi^{\a\b\dot\a}\pa_\a{}^{\dot\g}\Bar\Psi_{\b\dot\a\dot\g}
+\frac 12m\Psi^{\a\b\dot\g}\Psi_{\a\b\dot\g}
+\frac 12m\Bar \Psi_{\dot\a\dot\b\g}\Bar\Psi^{\dot\a\dot\b\g}
\eea
\bea
+\Psi^{\a\b\dot\g}\pa_{\a\dot\g}\chi_\b
+\Bar\Psi_{\dot\a\dot\b\g}\pa^{\dot\a\g}\bar\chi^{\dot\b}
-3i\chi^\a\pa_{\un a}\bar\chi^{\dot\a}
+3m\chi^\a\chi_\a
+3m\bar\chi_{\dot\a}\bar\chi^{\dot\a}
\Big\}~~.~~
\eea

\subsection{Superspin-1 Actions}
\label{Superspin-1 Actions}

In this section, the known superspin-1 action is presented and analyzed.  A new version is then derived that uses a real scalar auxiliary superfield instead of the chiral scalar auxiliary superfield.  The theory of \cite{Buchbinder2002b} is described by a physical spinor superfield $V_\a$ and a chiral auxiliary superfield $\Phi$.  
The following equations are required for $V_\a$ to be an irreducible superspin-$1$ representation:
\bea
\label{y1const}
\Bar D^2 V_\a=0~~~D^\a V_\a=0~~~~i\pa_{\un a}\Bar V^{\dot\a}+mV_\a=0
\eea
The action from \cite{Buchbinder2002b} is:
\bea
\nonumber
\cs_{\rm Y=1}[V_\a, \Phi]=\int d^8z\Big\{-V^\a\Bar D_{\dot\a} D_\a \Bar V^{\dot\a}
+m(V^\a V_\a+\Bar V_{\dot\a}\Bar V^{\dot\a})
+\g V^\a \Bar D^2 V_\a
\eea
\bea
\label{y1oldaction}
+\g^*\Bar V_{\dot\a}D^2\Bar V^{\dot\a}-\frac 12\Phi\Bar\Phi
-\frac 12\Phi D^\a V_\a-\frac 12\Bar\Phi \Bar D_{\dot\a}\Bar V^{\dot\a}
\Big\}
-\frac m4\int d^6z \Phi^2-\frac m4\int d^6\bar z\Bar\Phi^2~~,~~
\eea
where $\g$ is a complex coefficient.  The equations of motion are:
\bea
\label{y1veom}
{\d\over \d V^\a}\Big(\cs_{\rm Y=1}[V_\a, \Phi]\Big)=-\Bar D_{\dot\a} D_\a \Bar V^{\dot\a}+2mV_\a
+2\g\Bar D^2 V_\a+\frac 12D_\a \Phi=0~~,~~
\eea
\bea
\label{y1phieom}
{\d\over \d \Phi}\Big(\cs_{\rm Y=1}[V_\a, \Phi]\Big)
=+\frac 18\Bar D^2 D^\a V_\a
+\frac 18\Bar D^2\Bar \Phi
-\frac m2\Phi=0~~.~~
\eea
The first constraint in (\ref{y1const}) follows directly from (\ref{y1veom}) independent of how $\Phi$ behaves.  Contracting the operator $\frac 1{16}\Bar D^2D^\a$ on (\ref{y1veom}) and substituting (\ref{y1phieom}) proves that $\Phi$ is auxiliary:
\bea
0=\frac 14\Bar D^2[-\frac 18D^2\Phi+\frac m2\Bar \Phi]
-\frac m8\Bar D^2\Bar \Phi
+\frac {m^2}2\Phi+\frac 12\Box\Phi
=+\frac{m^2}2\Phi~~.~~
\eea
This means that $\Bar D^2 D^\a V_\a=0$.  
The second constraint in (\ref{y1const}) is obtained by contracting $D^\a$ on (\ref{y1veom}):
\bea
0=-D^\a\Bar D_{\dot\a} D_\a \Bar V^{\dot\a}+2mD^\a V_\a
=+\frac 12\{D^2,\Bar D_{\dot\a}\}\Bar V^{\dot\a}+2mD^\a V_\a
=+2mD^\a V_\a~~.~~
\eea
Now the equation of motion (\ref{y1veom}) becomes:
\bea
+2i\pa_{\un a} \Bar V^{\dot\a}+2mV_\a=0~~,~~
\eea
which is the final constraint in (\ref{y1const}).  Thus, the action
(\ref{y1oldaction}) describes a superspin-$1$ irreducible representation of the Poincar\'e superalgebra.  When $m=0$, this action is the massless gravitino multiplet that uses a chiral compensating superfield\cite{Ogievetsky1977}.  There is another version of the massless gravitino multiplet that uses a real scalar compensating superfield, $H$\cite{Fradkin1979, deWit1979, Gates1980}.  It is natural to ask if this theory has a massive extension.  The kinetic terms are given by:
\bea
\nonumber
S_{(1,\frac 32)}^{\|}[V_\a, H]=\int d^8z\Big\{
D^\a\Bar V^{\dot\a}\Bar D_{\dot\a}V_\a
-\frac 14\Bar D^{\dot\a}V^\a\Bar D_{\dot\a}V_\a
-\frac 14 D_\a \Bar V_{\dot\a} D^\a \Bar V^{\dot\a}
\eea
\bea
-\frac 1{16}HD^\a\Bar D^2D_\a H
-\frac 14\Bar D^2D^\a H V_\a-\frac 14D^2\Bar D_{\dot\a}H\Bar V^{\dot\a}
\Big\}~~,~~
\eea
and the only possible mass terms are:
\bea
\cs_m[V_\a,H]=\int d^8z\Big\{
mV^\a V_\a
+m\Bar V_{\dot\a}\Bar V^{\dot\a}
+\a m^2 H^2
+ mH\Big(\b D^\a V_\a 
+\b^* \Bar D_{\dot\a} V^{\dot\a}\Big)
\Big\}~~.~~
\eea
$\a$ and $\b$ are respectively real and imaginary coefficients, and the coefficient on the first and second terms is fixed according to (\ref{y1const}).  The equations of motion are:
\bea
\nonumber
{\d\over\d V^\a}\Big(\cs_{(1,\frac 32)}^{\|}[V_\a, H]+\cs_m[V_\a,H]\Big)
=-\Bar D_{\dot\a}D_\a\Bar V^{\dot\a}
+\frac 12\Bar D^2V_\a
+2mV_\a
\eea
\bea
\label{y1newVeom}
-\frac 14\Bar D^2D_\a H
-\b m D_\a H=0~~,~~
\eea
\bea
\nonumber
{\d\over\d H}\Big(\cs_{(1,\frac 32)}^{\|}[V_\a, H]+\cs_m[V_\a,H]\Big)=
-\frac 18D^\a\Bar D^2D_\a H
+\frac 14D^\a\Bar D^2 V_\a
+\frac 14\Bar D_{\dot\a} D^2 \Bar V^{\dot\a}
\eea
\bea
\label{y1newHeom}
+\b mD^\a V_\a 
+\b^* m\Bar D_{\dot\a} V^{\dot\a}
+2\a m^2H=0~~.~~
\eea
This system of equations is considerably more complicated than any of the previous examples.  One reason for this added complexity is the massive coupling between $H$ and $V_\a$, i.e. the terms proportional to $\b$.  A few quick observations can simplify the analysis.  First, multiplying (\ref{y1newVeom}) and (\ref{y1newHeom}) by $\Bar D^2$ yields:
\bea
\Bar D^2V_\a=\frac \b2  \Bar D^2D_\a H~~,~~
\eea
\bea
 \Bar D^2D^\a V_\a =-2\frac \a\b m\Bar D^2H~~.~~
\eea
Plugging these relations into the contraction of $D^\a$ on (\ref{y1newVeom}) leads to:
\bea
mD^\a V_\a = -\frac 18(\b+\b^*-1)D^\a\Bar D^2D_\a H
+\frac \b2\Big(1+{\a\over |\b|^2}\Big)mD^2H~~.~~
\eea
Substituting these three results into (\ref{y1newHeom}) gives:
\bea
0=-\frac 18(1-\b-\b^*)^2D^\a\Bar D^2D_\a H
+\frac 12m\Big(1+{\a\over |\b|^2}\Big)[\b^2D^2
+(\b^*)^2\Bar D^2]H
+2\a m^2H
\eea
proving that $H$ is auxiliary if $\b+\b^*=1$ and $\a=-|\b|^2$.  The final action takes the form:
\bea
\nonumber
\cs_{Y=1}[V_\a, H]=\int d^8z\Big\{
D^\a\Bar V^{\dot\a}\Bar D_{\dot\a}V_\a
+\frac 14V^\a\Bar D^2V_\a
+\frac 14 \Bar V_{\dot\a} D^2 \Bar V^{\dot\a}
-\frac 14V^\a\Bar D^2D_\a H 
-\frac 14\Bar V_{\dot\a}D^2\Bar D^{\dot\a}H
\eea
\bea
\label{ss1final}
-\frac 1{16}HD^\a\Bar D^2D_\a H
+mV^\a V_\a
+m\Bar V_{\dot\a}\Bar V^{\dot\a}
-|\b|^2 m^2 H^2
+ mH\Big(\b D^\a V_\a 
+\b^* \Bar D_{\dot\a} V^{\dot\a}\Big)
\Big\}~~,~~
\eea
where $\b+\b^*=1$.  This result is interesting because the imaginary part of $\b$ has not been determined.  Also, the superfield redefinition $V_\a\rightarrow V_\a+\frac \d m\Bar D^2V_\a$ can be used to change the coefficients of the second through fifth terms.  Thus, the general theory is parameterized by the imaginary part of $\b$ and the complex coefficient of the superfield redefinition, $\d$.

\section{Massive Discussion}
\label{Massive Discussion}

This concludes the chapter on higher superspin free massive irreducible representations of the Poincar\'e superalgebra.  Four new actions have been presented.  The action (\ref{ss1final}) is a new version of superspin-$1$ which has $V_\a$ as the physical superfield.  This model is new because the chiral scalar compensator, $\Phi$, has been replaced with a real scalar compensator, $H$.  The model is parameterized by one real coefficient and one complex coefficient.  Three new versions of the original superspin-$\frac 32$ models presented in \cite{Buchbinder2002a} were found.  The first is given in (\ref{Newresult1}).  This model is different than the models of \cite{Buchbinder2002a} because it contains a real scalar auxiliary superfield as opposed to a real pseudo scalar superfield.  The two new theories that are the most interesting are the massive extensions of new-minimal and ``new"-new-minimal supergravity.  The massive extension of new-minimal supergravity is given in (\ref{newminmass}) and contains a chiral spinor auxiliary superfield.  The ``new"-new-minimal massive extension is given in (\ref{newnewminmass}) and also contains a chiral spinor auxiliary superfield.  The difference between these two theories lies solely in the choice of superprojectors used in the kinetic terms for $V_{\un a}$.  Interestingly, the superspin-$\frac 32$ theories contain no continuous parameters unlike the superspin-$1$ models.

The massive extensions of the supergravities or $(2,3/2)$ multiplets are interesting in the sense that these theories can be related to massless theories in one dimension higher.  The next chapter is devoted to exploring this connection in light of the two new theories (\ref{newminmass}) and (\ref{newnewminmass}).  It will be shown that there are two new versions of linearized five-dimensional supergravity.  These two theories could have a profound affect on the phenomenology of extra dimensions.

\newpage
\chapter{Dimensional Reduction and Oxidation in N=1 Superspace}
\label{Dimensional Reduction and Oxidation in N=1 Superspace}

This chapter begins with a discussion of the geometric approach to gauge theory for non supersymmetric theories, including gravity.  Then, using the same general ideas, discusses super geometry.  Super geometry gives a natural interpretation of the superfield-strengths that are found in supersymmetric gauge theories and supergravity.  The differential geometry used in this introduction can be used in any dimension and for any extension of supersymmetry.  Therefore, the five-dimensional superfield-strengths that will be found in the technical part of this chapter can be understood from a super geometric point of view.  For this reason it is necessary to have some understanding of how to create gauge theories from differential geometry.

In non-supersymmetric particle physics it is easy to construct an action for a gauge theory.  The gauge group is chosen, a covariant derivative is constructed and kinetic terms are determined using gauge invariance.  In superspace this procedure is not as straightforward for two reasons.  First, there are more differential operators.  The spacetime gauge covariant derivative may be easy to understand, but how the spinor derivatives are made into gauge covariant objects is not obvious.  Second, although it may be relatively simple to construct linearized theories in superspace by embedding component fields into superfields, the non-linear versions can not be easily constructed in this fashion.  A consistent method for producing actions for gauge theories comes from differential geometry.

This chapter discusses the differential geometry of non-supersymmetric gauge theory and gravity.  Using the same methods, supersymmetric gauge theory and supergravity are then discussed.  These methods give a general understanding of the superfield-strengths that will be derived in the technical part of this chapter on five-dimensional linearized supergravity.  In particular, the new superfield-strengths associated with five dimensions will stand out.  The new superfield-strengths can be understood as originating from the fifth direction of the covariant vector derivative $\nabla_5$ and the extra covariant spinor derivative $\nabla_\a^-$.  The latter being necessary for supersymmetry in five dimensions, since the spinors are four component complex objects, or Dirac spinors.

This chapter is organized into three sections.  The first section \ref{Spacetime Geometry}, discusses spacetime geometry for non-Abelian gauge theory and gravity.  The second section \ref{Superspace Geometry}, works through supersymmetric non-Abelian theories and old-minimal supergravity.  The third section \ref{Five-dimensional Supergravities}, describes the relation of the massive extensions of four-dimensional supergravities from chapter \ref{Massive Superspin Actions} to five-dimensional supergravities.  The five-dimensional theories related to new-minimal and ``New"-new-minimal supergravity are new results and may have interesting implications for phenomenology.

\section{Spacetime Geometry}
\label{Spacetime Geometry}
This section is a review of spacetime geometry.  Non-Abelian gauge theory and gravity are discussed.  In general, the algebra of some set of linearly independent differential operators is considered.  A good understanding of spacetime geometry is necessary in order to work with superspace geometry, which is significantly more complicated.

\subsection{Non-Abelian Gauge Theory}
\label{Non-Abelian Gauge Theory}

The general covariant derivative is a combination of the normal partial derivative and a gauge transformation:
\bea
\label{covder}
\de_a = \pa_a +A_a^it^i~~,~~
\eea
where $A_a$ is the well known vector potential.  $t^i$ are the generators of the gauge group, and the coupling constant has been absorbed into the definition of $A_a$ for now.  Information about geometry comes from examining parallel transport about a closed curve.  This gets mapped into the commutator of two covariant derivatives $[\de_a,\de_b]$.  If $\de_a$ and $t^i$ are the only operators in this theory, then this commutator must be proportional to a covariant derivative or a gauge transformation:
\bea
[\de_a,\de_b]=T_{ab}{}^c\de_c+F_{ab}{}^it^i~~,~~
\eea
$T_{ab}{}^c$ is the torsion and $F_{ab}{}^i$ is the field-strength.  Expanding the covariant derivative using (\ref{covder}) shows that the torsion is zero and the field-strength can be expressed in terms of the vector potential:
\bea
F_{ab}{}^k=\pa_{[a}A_{b]}^k
+A_a^iA_b^jf^{ijk}~~,~~
\eea
where the structure constants are defined by $[t^i, t^j]=f^{ijk}t^k$.  
Actions are constructed by squaring the field-strengths and using the covariant derivative in place of the flat derivative:
\bea
S_{YM}=\int d^4x\Big\{-\frac 14F_{ab}{}^iF^{abi}
-\frac i2\psi^\a \de_{\un a}\bar \psi^{\dot\a}
+\de^a\phi \de_a \bar\phi
\Big\}~~.~~
\eea
The covariant derivatives, under commutation, satisfy a lie algebra and therefore obey the Jacobi identity:
\bea
[\de_a,[\de_b,\de_c]]+[\de_b,[\de_c,\de_a]]+[\de_c,[\de_a,\de_b]]=0~~.~~
\eea
The Jacobi identity yields the Bianchi identity for the field-strength:
\bea
[\de_a,F_{bc}^it^i]+[\de_b,F_{ca}^it^i]+[\de_c,F_{ab}^it^i]
=\frac 12\de_{[a}F_{bc]}{}^it^i
=0~~.~~
\eea
Although the Bianchi identities play a small role in the understanding of non-supersymmetric gauge theory, they will be in-disposable in superspace.
\subsection{Gravity}
\label{Gravity}

Gravity is a more complicated gauge theory because the momentum and Lorentz rotation generators are gauged.  Since translations occur in the curved spacetime and rotations occur in the tangent space to the spacetime, the spacetime manifold and its tangent space must be distinguished.  Roman letters from the middle of the alphabet are reserved for the manifold and are called curved indices.  Roman letters from the beginning of the alphabet are reserved for the tangent space and are called flat indices.  To incorporate spinors and supersymmetry it is necessary to work in the vierbein formalism, where the metric is replaced by the frame fields $e_a{}^m$:
\bea
g_{mn}(x)=e_m{}^a(x)\h_{ab}e_n{}^b(x)~~.~~
\eea
The spacetime dependence of the frame fields will be suppressed throughout the rest of this document.  The flat(curved) metric to raise and lower flat(curved) indices.  $e_m{}^a$ is also used to convert between flat and curved tensors.  Sticking to the analog of non-Abelian gauge theory, the covariant derivative is a sum of a translation and Lorentz rotation:
\bea
\de_a=e_a{}^m\pa_m+\frac 12\o_{ab}{}^c\cm_c{}^b~~.~~
\eea
Thus, the vierbein is the gauge field for the diffeomorphism group and the spin connection, $\o_{ab}{}^c$, is the gauge field for the tangent space Lorentz rotations.  Expanding the commutator of covariant derivatives over these two generators leads to:
\bea
[\de_a,\de_b]=T_{ab}{}^c\de_c+R_{abc}{}^d\cm_d{}^c~~,~~
\eea
$T_{ab}{}^c$ is the torsion tensor and $R_{abc}{}^d$ is the Riemann curvature tensor.  So the curvature tensor is the field-strength corresponding to gauging Lorentz transformations in the tangent space.  Expanding the covariant derivatives in the commutator yields expressions for the torsion and curvature in terms of the potentials:
\bea
\label{tors}
T_{ab}{}^c=c_{ab}{}^c
-\o_{[ab]}{}^c~~,~~
\eea
\bea
\label{anhol}
c_{ab}{}^c=e_{[a}{}^m(\pa_me_{b]}{}^n)e_n{}^c~~,~~
\eea
\bea
\label{curv}
R_{ab}{}^{ef}=\frac 12e_{[a}{}^m(\pa_m\o_{b]}{}^{ef})
-\frac 12c_{ab}{}^c\o_c{}^{ef}
-\frac 12\o_{[a}{}^{ec}\o_{b]c}{}^f~~.~~
\eea
Note that $\cm_a{}^b$ only acts on flat indices and has the vector space representation: $(\cm_a{}^b)_c{}^d=\h_{ca}\h^{bd}-\d_a{}^d\d_c{}^b$.  The anholonomity $c_{abc}$ is introduced to simplify the notation.  Non-Abelian gauge theory is only described by the gauge vector fields and gravity should only involve the metric or vierbein.  It is necessary to remove the spin connection so it does not become a propagating degree of freedom.  This can be accomplished by setting the torsion to zero.  This constraint relates the spin connection directly to derivatives of the vierbein.  This can be proven by subtracting and adding two cyclic permutations of (\ref{tors}):
\bea
0=c_{abc}-\o_{[ab]c}
-c_{bca}+\o_{[bc]a}
+c_{cab}-\o_{[ca]b}~~,~~\cr
\rightarrow ~~\o_{abc}(e)=\frac 12(+c_{abc}
-c_{bca}+c_{cab})~~.~~
\eea
Setting the torsion to zero is called a conventional constraint.  Conventional constraints involve removing redundant degrees of freedom from a geometry.  Noting the mass dimension of the curvature tensor, it can be used in an action:
\bea
S_{Grav-Matt}=\int d^4x e^{-1}\Big\{ \h^{ab}R_{acb}{}^c(e)
-\frac i2\psi^\a \de_{\un a}\bar \psi^{\dot\a}
+\de^a\phi \de_a \bar\phi
\Big\}~~.~~
\eea
This action is good, but is not general enough to add gauge fields.  This minimal coupling is not correct since the gauge symmetry is local.  The gauge symmetry should be added into the definition of the covariant derivative from the beginning:
\bea
\de_a=e_a{}^m\pa_m+\frac 12\o_{ab}{}^c\cm_b{}^c+A_a^it^i
~~.~~
\eea
The commutator becomes:
\bea
[\de_a,\de_b]=T_{ab}{}^c\de_c+R_{abc}{}^d\cm_d{}^c
+F_{ab}^it^i~~.~~
\eea
The addition of the non-Abelian gauge group does not affect the torsion and curvature, since the generators, $t^i$ have no action on the other gravity potentials.  The field-strength is:
\bea
\label{YMcoupGrav}
F_{ab}{}^i = e_{[a}{}^m\pa_m A_{b]}^i
-c_{ab}{}^cA_c{}^i+A_a{}^jA_b{}^kf^{jki}~~.~~
\eea
So the coupling of gravity to Yang-Mills should take the form:
\bea
S_{G-YM}=\int d^4x e^{-1}g^{ac}g^{bd}F_{ab}{}^iF_{cb}{}^i
\eea
A good review of these concepts is given in \cite{Gates1}.  What is most important from this derivation is the fact that the torsion constraint must be imposed in order to remove the independence of the spin connection.  Constraints are fundamental to the geometric description of gauge theories.  In superspace, constraints are used to distinguish between different types of supergravities.  Thus, the difference between old and new minimal supergravity can be stated concisely by listing both sets of constraints.  In the literature the constraints are often the only description of the supergravity that is given, making these papers inaccessible to those who do not understand the geometric description of gauge theories.
\section{Superspace Geometry}
\label{Superspace Geometry}
The same procedure can be followed for constructing geometries in superspace.  In superspace things are more complicated.  There are three covariant derivatives $\de_a,~\de_\a,~\Bar\de_{\dot\a}$.  This leads to three times as many potentials as in normal spacetime.  The potentials are also superfields, which means that these super geometries contain a large amount of field content.  The fundamental role of constraints to strip away these excess degrees of freedom is a major tool in constructing super geometries.

There are two ways of looking at super geometry.  They are both consistent, and depending on the problem they both can give insight into the solution of the geometry.  The first perspective follows the previous section.  Commutators of covariant derivatives are expanded in terms of the potentials and the superfield-strengths are analyze.  The second perspective involves analyzing the superspace Bianchi identities.  This procedure gives valid information after constraints have been set because the identities are no longer algebraic identities.  It turns out that in superspace, the Bianchi identities of constrained algebras can not be solved using the potentials defined in the covariant derivatives.  The potentials must be replaced by derivatives of prepotentials.  This singular property makes super geometry a rich and complicated mathematical structure.
\subsection{Supersymmetric Non-Abelian Gauge Theory}
\label{Supersymmetric Non-Abelian Gauge Theory}

To construct the non-Abelian supersymmetric gauge theory, all supersymmetric covariant derivatives are extended to gauge covariant derivatives:
\bea
\de_\a=D_\a -i\G_\a^it^i~~,~~\\
\Bar \de_{\dot\a}=\Bar D_{\dot\a}-i\Bar\G_{\dot\a}^it^i~~,~~\\
\de_a=\pa_a -i\G_a^it^i~~,~~
\eea
here the conjugate spinor derivative is consistent because $(t^i)^*=-t^i$.  The super vector superfield gauge potential $\G_A^i$ has been used so that there is no confusion with the usual gauge potential $A_a^i$.  The super vector notation means: $A=\{\,a,\,\a,\,\dot\a\,\}$.  This notation is extremely convenient.  Supervector notation does not specify the spinor representation and, therefore, can be applied to any dimension.  In super vector notation the covariant derivatives look like the non-supersymmetric covariant derivatives:
\bea
\de_A = D_A-i\G_A^it^i~~,~~
\eea
with $D_a:=\pa_a$ and $\G_{\dot\a}^i=\Bar \G_{\dot\a}^i$.  In order to talk about super Lie algebras, it is necessary to define graded symmetrization of two super vectors objects:
\bea
X_{[A}Y_{B)}=X_AY_B-(-)^{\e(A)\e(B)}X_BY_A~~,~~
\eea
where $\e(A)=0$ for bosons and $\e(A)=1$ for fermions.  For spinor indices this is symmetric and for vector-vector or vector-spinor this is antisymmetric.  In supervector notation, the graded commutator of two covariant derivatives is:
\bea
\de_{[A}\de_{B)}=[\de_A,\de_B\}=T_{AB}{}^C\de_C-iF_{AB}^it^i
~~.~~
\eea 
In the spacetime non-Abelian gauge theory case, the torsion was identically zero, something different occurs in superspace.  Plugging in the covariant derivatives and using the anholonimity for the simple covariant derivatives:
\bea
[D_A, D_B\}=C_{AB}{}^CD_C~~,~~
\eea
leads to:
\bea
T_{AB}{}^C=C_{AB}{}^C~~,~~
\eea
\bea
\label{YMFS}
F_{AB}=D_{[A}\G_{B)}-i\G_{[A}\G_{B)}-C_{AB}{}^C\G_C~~,~~
\eea
where the simplifying notation $F_{AB}=F_{AB}^it^i$ and $\G_A=\G_A^it^i$ has been used.  This result is not so obvious and only looks simple because of the notation.  In particular, super non-Abelian gauge theory picks up the torsion of simple superspace.  Also, the field-strength looks strangely similar to the spacetime non-Abelian field-strength coupled to gravity (\ref{YMcoupGrav}).  

In the mathematical introduction, see (\ref{simplemax}), it was shown that a vector field can be modeled with a real scalar superfield.  It seems plausible to look for a consistent way to constrain the superfield-strength to reduce the degrees of freedom contained in $\G_A$, to some real scalar superfield.  Since the anholonimity has one nonzero entry $C_{\a\dot\a}{}^c=-2i(\s^c)_{\a\dot\a}$ the $F_{\a\dot\a}$ piece in (\ref{YMFS})  will contain $\G_{\un a}$ algebraically.  Setting the conventional constraint $F_{\a\dot\b}=0$ allows for $\G_{\un a}$ to be solved in terms of the spinor superpotential:
\bea
\label{ymconvcons}
0=F_{\a\dot\b}=D_\a\G_{\dot\b}
+\Bar D_{\dot\b}\G_\a
-i\G_\a\G_{\dot\b}-i\G_{\dot\b}\G_\a
+2i\G_{\a\dot\b}~~.~~
\eea
This constraint establishes an algebraic relation between $\G_a$ and $\G_\a$.  The same situation occurred in spacetime gravity when the spin connection was written in terms of the frame fields.  This is a conventional constraint and removes a vector superfield worth of degrees of freedom.  The algebra can be constrained even more by requiring that the notion of a chiral superfield still makes sense on arbitrary representations of the gauge group.  Constraints of this nature are called representation preserving constraints:
\bea
 \de_\a \Bar \F = 0~~\Rightarrow
[ \de_\a,\de_\b\}\Bar \F =  F_{\a\b}\Bar \F=0~~.~~
\eea
This constraint leads to:
\bea
0=F_{\a\b}=D_{(\a}\G_{\b)}-i\G_{(\a}\G_{\b)}~~,~~
\eea
which can be solved if $\G_\a=\G_\a^it^i=ie^{-\O}D_\a e^{\O}$.  $\O$ is shorthand for $\O^it^i$.  The proof is a little tricky and it should be noted that $D_\a$ only acts on $\O^i$:
\bea
F_{\a\b}
=iD_{(\a}\Big[e^{-\O}D_{\b)} e^{\O}\Big]
+ie^{-\O}\Big[D_{(\a} e^{\O}\Big]e^{-\O}\Big[D_{\b)} e^{\O}\Big]
~~~~~~~~~~~~~~\cr
=i\Big[D_{(\a}e^{-\O}\Big]\Big[D_{\b)} e^{\O}\Big]
-ie^{-\O}e^{\O}\Big[D_{(\a} e^{-\O}\Big]\Big[D_{\b)} e^{\O}\Big]
=0~~.~~
\eea
This is solution that represents all of the super potentials in terms of the super pre-potential $\O$, which is a complex scalar superfield.  This means that this theory still contains one extra vector field which can be removed by further constraining the pre-potential to be either real or imaginary.  Taking another look at the form of $\G_a$ from (\ref{ymconvcons}) and using the solution for $\G_\a$ reveals:
\bea
\nonumber
-2i\G_{\un a}=D_\a\G_{\dot\b}+\Bar D_{\dot\b}\G_\a
-i\G_\a\G_{\dot\b}-i\G_{\dot\b}\G_\a\\ \cr
=ie^{\Bar\O}D_\a\Bar D_{\dot\a}e^{-\Bar\O}
+ie^{-\Bar\O}D_\a(e^\O e^{\Bar\O})\Bar D_{\dot\a}e^{-\Bar\O}
+c.c.
\eea
Note that $(\G_\a)^*=-\Bar\G_{\dot\a}^it^i=-\G_{\dot\a}$, and $(\O)^*=-\Bar\O^it^i=-\Bar\O$.  From here it is obvious that the gauge choice $\O=-\Bar\O$ is a pure gauge configuration.  No such simplification occurs for the gauge choice $\O=\Bar\O$.  This gauge choice means that the pre-potential is real, $\O^j=\Bar \O^j=V^j$.  

To recapitulate, two superfield-strengths have been constrained to vanish.  $F_{\un a}=0$ is a conventional constraint.  This constraint relates $\G_{\un a}$ to $\G_\a$.  $F_{\a\b}=0$ is a representation preserving constraint, allowing the definition of gauge covariant chiral superfields.  Removing this superfield-strength allows the connection $\G_\a^i$ to be written in terms of a complex superfield $\O^i$, the pre-potential.  Finally, the real part of $\O^i$ represents a pure gauge configuration.  This allows the gauge fixing such that  $\O^i$ is purely imaginary, leaving only one real scalar gauge pre-potential superfield $V$.  At this point, the values of the connections can be substituted into the remaining superfield-strengths and the algebra can be completely determine.  Instead of completing this analysis, it is more instructive to look at this algebra from the perspective of the Bianchi identities to see how the superfield-strengths are determined.

Once the constraints $F_{\a\b}=F_{\un a}=0$ have been set, the Bianchi identities are no longer algebraic identities.  They imply constraints on the remaining superfield-strengths.  The super Bianchi identities are:
\bea
(-)^{AC}[[\de_A,\de_B\},\de_C\} +(-)^{BA}[[\de_B,\de_C\},\de_A\} 
+(-)^{CB}[[\de_C,\de_A\},\de_B\}=0~~,~~
\eea
and once constraints have been set it is important that these equations can be satisfied without setting all superfield-strengths to zero.  The Bianchi identity with undotted spinor indices vanishes  identically.  Taking two undotted and one dotted index gives:
\bea
0=[[\de_\a,\de_\b\},\Bar\de_{\dot\g}\} 
+[[\Bar \de_{\dot\g},\de_{(\a}\},\de_{\b)}\}
=0-2i[\de_{\dot\g(\b}, \de_{\a)}\}
~~,~~
\eea
\bea
\Rightarrow F_{(\a\b)\dot\g}=0~~.~~
\eea
So only the part antisymmetric in $\a\b$ of $F_{\un a\b}$ remains.  This part of $F_{\un a\b}$ is usually written as:
\bea
F_{\un a\b}=\e_{\a\b}i\Bar W_{\dot\a}~~.~~
\eea
The Bianchi identity with two undotted and one vector index leads to:
\bea
0=[[\de_\a,\de_\b\},\de_{\un c}\} 
+[[\de_{\un c},\de_{(\a}\},\de_{\b)}\}
=0-i[\Bar W_{\dot\g},\e_{\g(\b}\de_{\a)}\}
=+i\e_{\g(\b}\de_{\a)} \Bar W_{\dot\g}~~,~~
\eea
which means that $W_\a$ must be a gauge covariant chiral superfield.  This is the same type of superfield-strength that was discussed after (\ref{simplemax}) for supersymmetric Maxwell theory.  Finally, the Bianchi identity with one of each type of index yields:
\bea
2iF_{\a\dot\b\un c}-i\e_{\dot\g\dot\b}\de_\a W_\g
-i\e_{\g\a}\Bar \de_{\dot\b}\Bar W_{\dot\g}=0~~.~~
\eea
Tracing this equation on all spinor indices and recalling the symmetry of the vector-vector superfield-strength leads to:
\bea
\label{BI1}
\de^\a W_\a -\Bar\de_{\dot\a}\Bar W^{\dot\a}=0~~,~~
\eea
which then implies:
\bea
\label{u1fs}
F_{\un a\un b}=-\frac 14\e_{\a\b}\Bar \de_{(\dot\a}\Bar W_{\dot\b)}
-\frac 14\e_{\dot\a\dot\b}\de_{(\a}W_{\b)}~~.~~
\eea
Thus, the two constraints $F_{\a\b}=0$ and $F_{\un a}=0$, have completely constrained all superfield-strengths in terms of $W_\a$.  
Further, the superfield-strength $W_\a$ can be solved in terms of the prepotential $V^i$.  The full algebra can now be written:
\bea
\label{ymalg}
[\de_\a, \de_\b\}=0~~,~~
[\de_\a, \Bar \de_{\dot\a}\}=-2i\de_{\un a}~~,~~\cr
[\de_{\un a}, \de_\b\}=\e_{\a\b}\Bar W_{\dot\a}~~,~~
[\de_{\un a}, \de_{\un b}\}=-iF_{\un a\un b}~~.~~
\eea
If too many constraints had been set, this process would have led to equations of motion or all superfield-strengths would have vanished.  If too few constraints had been set, there would have been more then one independent superfield-strength.  Outstanding problems in higher dimensional supergravity revolve around finding the appropriate constraints and their solutions in terms of prepotentials.  

Before examining the $U(1)$ case, the general super Yang-Mills action can be written.  By dimensional analysis the only possibility is $W^\a W_\a$ under a chiral measure.  Coupling to supersymmetric covariantly chiral matter $\Phi_\pm$ leads to:
\bea
\cs_{YM}=\int d^6z W^\a W_\a
+\int d^8z [\Phi_+\Bar \Phi_++\Phi_-\Bar \Phi_-]
+\int d^6zW(\Phi_+,\Phi_-)+{\rm c.c.}~~.~~
\eea
The function $W(\Phi_+,\Phi_-)$ is the famous superpotential that determines all of the wonderful phases in SQCD.  It is protected from perturbative corrections since it is under the chiral measure.  Finally, the component action is obtained by replacing the measures with the simple superspace density projector using covariant derivatives: $d^4\q=\frac 1{16}\de^2\Bar \de^2$ and $d^2\q=-\frac 14\de^2$ taking care to use the algebra (\ref{ymalg}) when pushing derivatives around.  

After linearizing and specializing to $U(1)$ many of the results are simpler.  First, the solution for $\G_\a$ simplifies to:
\bea
\G_\a t = ie^{-Vt}D_\a e^{Vt}=iD_\a Vt~~.~~
\eea
Plugging this into the solution for $\G_{\un a}$ gives:
\bea
-2i\G_{\un a}
=-i[D_\a, \Bar D_{\dot\a}]V~~,~~
\eea
and the non-zero superfield-strengths can be calculated easily:
\bea
F_{\un a\b}=\pa_{\un a}\G_\b-D_\a \G_{\un a}
=-\frac 12\e_{\a\b}D^2\Bar D_{\dot\a}V~~.~~
\eea
So $W_{\a}\sim i\Bar D^2D_\a V$ which is what was derived chapter \ref{Mathematical Background}.  Plugging this solution into the definitions of the covariant derivatives yields:
\bea
\de_\a = D_\a + D_\a Vt~~,~~\Bar \de_{\dot\a}=\Bar D_{\dot\a} -\Bar D_{\dot\a}Vt~~,~~\cr
\de_{\un a} = \pa_{\un a}-\frac i2[D_\a, \Bar D_{\dot\a}]Vt~~.~~
\eea
These derivatives satisfy the algebra (\ref{ymalg}).  The  component action for the kinetic term is:
\bea
\int d^6z W^\a W_\a
=-\frac 14\int d^4x (2\de^2 W^\b| W_\b|+\frac 12\de_{(\a} W_{\b)}|\de^{(\a} W^{\b)}|
+\de^\a W_\a|\de^\b W_\b|)~~.~~
\eea
Two observations are necessary in order to define components.  First, the lowest component of $F_{ab}$ is the usual component field-strength and equation (\ref{u1fs}) implies:
\bea
\frac 12\de_{(\a}W_{\b)}|=\e^{\dot\a\dot\b}F_{\un a\un b}|=:-2f_{\a\b}~~.~~
\eea
Second, taking $\de_\b$ on (\ref{BI1}) gives the relation:
\bea
-\frac 12\de^2 W_\b+2i\de_{\b\dot\a}\Bar W^{\dot\a}=0
\eea
So the component $\de^2 W_\a|$ is just a covariant derivative on the lowest component of $\Bar W_{\dot\a}|=:\frac 12\bar\l_{\dot\a}$, which is called the gaugino.  The auxiliary field is defined by $D^\a W_\a|=d$.  With these component definitions, the final form of the component action is:
\bea
=\int d^4x (-\frac i2\l^\b\de_{\un b} \Bar \l^{\dot\b} -\frac 12F^{ab}F_{ab}
+d^2)~~,~~
\eea
where the following identities have been used for
the component field-strength:
\bea
-\frac 12F^{ab}F_{ab}=f^{\a\b}f_{\a\b}+\bar f^{\dot\a\dot\b}\bar f_{\dot\a\dot\b}\cr
\int d^4xf^{\a\b}f_{\a\b}=\int d^4x\bar f^{\dot\a\dot\b}\bar f_{\dot\a\dot\b}
\eea

In this section, supersymmetric non-Abelian gauge theory has been constructed from a geometric standpoint.  The main point is that there are two ways to look at  the geometry.  The first is based on the potentials.  Superfield-strengths are constrained and the potentials are solved in terms of prepotentials.  The second view is based on the consistency of the Bianchi identities.  The constraints remove entire superfield-strengths and set non-trivial differential constraints on the non-zero superfield-strengths.  All of this structure weighs heavily on the form of the component action, since the algebra of covariant derivatives is used when projecting the superfield action into components.  In supergravity, the relationship between the algebra and component action is even more intertwined.  This is because the supergravity density projector is not just the simple replacement of the measure with covariant derivatives.  But in normal gravity the measure is replaced by $d^4x\sqrt g$ so a similar complication is to be expected in supergravity.

\subsection{Old-Minimal Supergravity}
\label{Old-Minimal Supergravity}

In this section, old-minimal supergravity is constructed following the same reasoning as in the non-Abelian gauge theory case.  This time the Lorentz generators are gauged along with super diffeomorphisms.  The covariant derivatives take the form:
\bea
\label{sugracovder}
\de_A=E_A{}^MD_M+\frac 12\o_{A\g}{}^\d\cm_\d{}^\g=E_A+\o_A
~~.~~
\eea
The commutator is expanded using the diffeomorphism and Lorentz rotation generators:
\bea
[\de_A,\de_B\}=T_{AB}{}^C\de_C
+\frac 12 R_{AB\g}{}^\d\cm_\d{}^\g~~.~~
\eea
The torsion and curvatures can be written in terms of the potentials.  The torsion is all that is necessary to set all of the constraints:
\bea
T_{AB}{}^C=C_{AB}{}^C-\o_{[AB)}{}^C~~,~~~~~~~~~~\cr
[E_A,E_B\}=C_{AB}{}^CE_C~~,~~
C_{AB}{}^C=E_{[A}(E_{B)}{}^M)E_M{}^C~~,~~
\eea
The super vector notation is deceiving when used with the super spin connection.  As can be seen in (\ref{sugracovder}), $\o_{AB}{}^C$ has the same index structure of $\cm_{\a\b}$ on the last two indices, not the full super vector structure.  In ordinary gravity a constraint was set to remove the spin connection as a degree of freedom.  The super spin connection can also be completely removed.  The following torsion contain the spin connection algebraically:
\bea
\label{conspin}
T_{\a\b}{}^\g=C_{\a\b}{}^\g-\o_{(\a\b)}{}^\g~~,~~\cr
T_{\a\un b}{}^\g=C_{\a\un b}{}^\g+\o_{\un b\a}{}^\g~~,~~\cr
T_{\a\dot\b}{}^{\dot\g}=C_{\a\dot\b}{}^{\dot\g}-\o_{\a\dot\b}{}^{\dot\g}~~,~~\cr
T_{\a\un b}{}^{\un c}=C_{\a\un b}{}^{\un c}-\o_{\a\un b}{}^{\un c}~~,~~\cr
T_{[ab]}{}^c=C_{[ab]}{}^c-\o_{[ab]}{}^c~~.~~
\eea
Setting these torsion to zero completely determines the super spin connection in terms of the super anholonomity.   The algebra must be constrained further, because there are still two many degrees of freedom in the frame superfields, $E_A$.  In particular, there are two many fields that could be component vierbeins.  A closer look at the super anholonomity equation:
\bea
[E_\a, \Bar E_{\dot\a}\}=C_{\a\dot\a}{}^\b E_\b
+C_{\a\dot\a}{}^{\dot\b}\Bar E_{\dot\b}
+C_{\a\dot\a}{}^{\un b}E_{\un b}~~,~~
\eea
reveals that $E_{\un b}$ can be solved for in terms of $E_\a$ if $C_{\a\dot\a}{}^{\un d}$ is a constant.  The only Lorentz covariant constant with this index structure is $\d_\a{}^\b\d_{\dot\a}{}^{\dot\b}$.  Thus, the final conventional constraint is:
\bea
\label{conveir}
T_{\a\dot\a}{}^{\un b}=-2i\d_\a{}^\b\d_{\dot\a}{}^{\dot\b}
=C_{\a\dot\a}{}^{\un b}~~.~~
\eea
The constraints (\ref{conspin})=0 and (\ref{conveir}) are the conventional constraints of supergravity.  They leave only the fields $E_\a^M$ as unconstrained degrees of freedom.  

The final step is to impose the representation preserving constraints.  The requirement that covariantly chiral superfields exist means:
\bea
\de_\a\Bar \Phi=0~~,~~
0=[\de_\a, \de_\b\}\Bar\Phi=T_{\a\b}{}^{\un c}\de_{\un c}\Phi
+T_{\a\b}{}^{\dot\g}\Bar\de_{\dot\g}\Phi~~,~~\cr
\Rightarrow~~T_{\a\b}{}^{\un c}=0~~,~~T_{\a\b}{}^{\dot\g}=0~~.~~
\eea
With these last two constraints the spin-spin anholonomity equation reads:
\bea
[E_\a,E_\b\}=C_{\a\b}{}^\g E_\g
\eea
This means that $E_\a$ form a closed set under graded commutation.  For this reason, a result from differential geometry, Frobenius's theorem, can be used to write the solution to this equation.  If there exists some set of $q$ vector fields $V_i$ on some $p$-dimensional manifold such that $[V_i, V_j]=C_{ij}{}^kV_k$, then Frobenius's theorem allows these vector fields to be written in terms of the flat derivative $\pa_m$:
\bea
V_i=A_i{}^je^U\pa_je^{-U}~~~~
U=U^m\pa_m
\eea
This can be applied directly to superspace equations.  The manifold is the supermanifold with partial derivatives $\pa_\mu$ and $\pa_m$, and the vector fields are $E_\a$.  So $q$ runs over the spinor index and $p$ runs over the full set of indices.  Thus, Frobenius's theorem implies:
\bea
E_\a=A_\a{}^\mu e^UD_\mu e^{-U}~~,~~
U=U^m\pa_m+U^\mu D_\mu+U^{\dot\mu}\Bar D_{\dot\mu}
\eea
This solution looks much like the non-Abelian gauge theory solution found in the previous section.  Here $U$ is a super diffeomorphism gauge transformation like $\O$ for non-Abelian gauge theory.  $A_\a{}^\mu$ must have non-zero determinant and is usually factored into a scalar $\Psi$ and unimodular matrix $N_\a{}^\mu$:
\bea
E_\a=\Psi N_\a{}^\mu e^UD_\mu e^{-U}~~,~~
\eea
It turns out there is a gauge where $N_\a{}^\mu=\d_\a{}^\mu,$ and $U^{\dot\a}=U^{\dot\a}=0$, leaving only $\Psi$ and $U^a$.  These superfields are the prepotentials and represent the pre-geometry of supergravity.  Finding this pre-geometry is one of the main problems in higher dimensional supergravity.

This derivation actually generates conformal supergravity.  To get Einstein supergravity, the algebra must be constrained further:
\bea
T_{\a\un b}{}^{\un c}=0~~.~~
\eea
This entire super torsion was not previously removed, just $T_{\a(\b}{}^{\dot\g}{}_{\g)\dot\g}=0$.  This final constraint effectively replaces the scalar $\Psi$ with a covariantly chiral scalar $\varphi$.  Thus, the pre-geometry of Einstein supergravity is represented by a vector superfield and a chiral scalar superfield.  The off-shell structure of supergravity is completely determined by this pre-geometry.

With the full set of constraints the algebra becomes:
\bea
\nonumber
\{\de_\a, \Bar\de_{\dot\a} \}=-2i\de_{\un a}~~,~~
\{ \de_\a, \de_\b\}=-4\Bar R M_{\a\b}~~,~~
\eea
\bea
\nonumber
[\de_\a, \de_{\un b}]=i\e_{\a\b}(\Bar R\Bar \de_{\dot\b}+G^\g{}_{\dot\b}\de_\g
-\de^\g G^\d{}_{\dot\b}\cm_{\g\d}
+2\Bar W_{\dot\b}{}^{\dot\g\dot\d}\Bar\cm_{\dot\g\dot\d})
+i\Bar \de_{\dot\b}\Bar R \cm_{\a\b}~~,~~
\eea
\bea
\nonumber
[\de_{\un a},\de_{\un b}]=\e_{\dot\a\dot\b}\Big(-iG_\b{}^{\dot\g}\de_{\a\dot\g}
+\frac 12\de_\a R\de_\b+\frac 12\de_\a G_\b{}^{\dot\g}\Bar \de_{\dot\g}
+W_{\a\b}{}^\g\de_\g~~,~~
\eea
\bea
+\frac 14(\Bar \de^2-8R)\Bar R\cm_{\a\b}
+\de_\a W_\b{}^{\g\d}\cm_{\g\d}
-\frac 12\de_\a\Bar \de^{\dot\g}D_\b{}^{\dot\d}\Bar\cm_{\dot\g\dot\d}\Big)
+{\rm c.c.}~~.~~
\eea
The closure of this algebra under the Bianchi identities implies:
\bea
~~~G_{\un a}=\Bar G_{\un a}~~,~~W_{\a\b\g}=3!W_{(\a\b\g)}~~,~~\cr
\Bar \de_{\dot\a}R=0~~,~~\Bar \de_{\dot\a}W_{\a\b\g}=0~~,~~\cr
\Bar \de^{\dot\a}G_{\un a}=\de_\a R~~,~~\cr
\de^\a W_{\a\b\g}=\frac i2\de_{(\a}{}^{\dot\a}G_{\b)\dot\a}~~,~~
\eea
Since the entirety of the technical part of this chapter deals with linearized theory, it is instructive to discuss the linearization of this algebra.  Linearized supergravity is written in terms of the vector superfield $H_{\un a}$ and the chiral compensator $\s$ and the flat covariant derivatives.  These are the linearized prepotentials in a particular gauge.  The superfield-strengths that are used to write the algebra can be written in terms of the prepotentials as:
\bea
R=-\frac 14\Bar D^2 \Bar \s+\frac i6\Bar D^2\pa^a H_a
~~,~~~~~~~~~~~~~~~~~~~~~~~~~~~~~\cr
G_{\un a}=i\pa_{\un a}(\bar\s-\s)
+\frac 18D^\b\Bar D^2 D_\b H_{\un a}
-\frac 1{24}[D_\a, \Bar D_{\dot\a}][D_\b, \Bar D_{\dot\b}]H^{\un b}
+\pa_{\un a}\pa_{\un b}H^{\un b}~~,~~\cr
W_{\a\b\g}=\frac i{3!8}\Bar D^2\pa_{(\a}{}^{\dot\g}D_\b H_{\g)\dot\g}
~~,~~~~~~~~~~~~~~~~~~~~~~~~~~~~~
\eea
These equations solve the Bianchi identities.  The supergravity action is simply the integration of the super determinant of the frame superfields over the superspace:
\bea
\int d^8z E^{-1}\cl~~.~~
\eea
With $\cl=1$, after some gauge fixing and other various field redefinitions this action can be written in terms of the prepotentials as:
\bea
\int d^8z (\varphi e^{-2i H}\bar\varphi)(1\cdot e^{2i\overleftarrow{H}})\Hat E^{-\frac 13}~~.~~
\eea 
In this expression $H=H^a\pa_a$ and:
\bea
1\cdot e^{2i\overleftarrow{H}}=1+2iH^a\overleftarrow\pa_a
+\frac 12(2i)^2(H^a\overleftarrow\pa_aH^b)\overleftarrow\pa_b
+\cdots~~.~~
\eea
When linearized this action becomes:
\bea
\label{linsugra}
\cs_{SUGRA}=\int d^8z\Big\{\frac 18H^aD^\b\Bar D^2D_\b H_a
-3\s\bar\s+\frac 1{48}([D_\a,\Bar D_{\dot\a}]H^{\un a})^2\cr
-(\pa_a H^a)^2+2i(\s-\bar\s)\pa_a H^a
\Big\}
\eea
The equations of motion with respect to $H_{\un a}$ and $\s$ are:
\bea
G_{\un a}=0~~,~~R=0~~.~~
\eea 
Using these equations of motion the final Bianchi identity becomes:
\bea
D^\a W_{\a\b\g}=0
\eea
This is the correct equation for propagating helicity-$2$ and helicity-$\frac 32$ component fields.  The rest of this chapter is devoted to describing linearized supergravity in five dimensions using four-dimensional $N=1$ superspace.  Many of the superfield-strengths and Bianchi identities encountered in the next sections will bare a striking resemblance to the four-dimensional objects discussed in this section.

\section{Five-dimensional Supergravities}
\label{Five-dimensional Supergravities}

Consistent dimensional reductions from properly constrained five-dimensional superalgebras are not known.  A significant amount of information about higher dimensional theories can be gained by working at the linearized level using four-dimensional covariant derivatives.  In particular, completely off-shell descriptions of higher dimensional supergravities are not known.  Thus, writing higher dimensional theories on the supersymmetric 3-brane, gives insight into the off-shell structure of those theories.  Further, theories that are written covariantly with respect to the full higher dimensional supersymmetry or even Lorentz invariance can still be used for phenomenological purposes (i.e. brane-world scenarios\cite{Buchbinder2003}).

Although this section deals with five-dimensional supersymmetric systems, five-dimensional supersymmetry and five-dimensional Lorentz invariance will be discarded.  That is, only the four-dimensional subset of the five-dimensional supersymmetry and Lorentz invariance will remain covariant.  The analysis begins in section \ref{Dimensional Reduction to Old-Minimal} with the dimensional reduction of the known five-dimensional supergravity into a massive four-dimensional theory, which is shown to be the massive extension of old-minimal supergravity.  The final two chapters reverse the method of dimensional reduction to obtain two new versions of five-dimensional supergravity.  Obtaining higher dimensional theories from lower dimensional theories is sometimes called oxidization.  The new theory associated with new-minimal supergravity is particularly interesting because it contains a physical 2-form gauge potential as opposed to a gauge 1-form, which makes it the leading candidate for dimensional reductions of string and M-theory.

\subsection{Dimensional Reduction to Old-Minimal}
\label{Dimensional Reduction to Old-Minimal}

The five-dimensional supergravity model from \cite{Linch2003} is governed by the following action:
\bea
\nonumber
\cs_{\rm 5DOld-Min}[V_a,P,\Psi_\a,T]=\int dx_5\cs_{\rm Old-Min}[V_a,P]
-\frac 12\int d^9z\Big\{
\Big[\Bar T(\S-i\pa_{\un a}V^{\un a})+c.c.\Big]\\\cr
-\frac 12\Big[D^\a\Psi_\a
+\Bar D_{\dot\a}\Bar\Psi^{\dot\a}
-\pa_5 P\Big]^2
+\Big[\pa_5 V_{\un a}-(\Bar D_{\dot\a}\Psi_\a
-D_\a\Bar\Psi_{\dot\a})\Big]^2
\Big\}~~,~~
\eea
where $\S=-\frac 14\Bar D^2P$.  This action is invariant under the gauge transformations:
\bea
\d V_{\un a}=\Bar D_{\dot\a}L_\a-D_\a\Bar L_{\dot\a}&,&
\d P = D^\a L_\a +\Bar D_{\dot\a}\Bar L^{\dot\a}~~,~~\\\cr
\d \Psi = \pa_5 L_\a-\frac 14D_\a\O&,&
\d T= \pa_5\O~~.~~
\eea
Taking $\pa_5=m$, disregarding the extra integration and changing the minus sign on the $T$ terms gives:
\bea
\nonumber
\cs_{Y=3/2}[V_a,P,\Psi_\a,T]=\cs_{\rm Old-Min}[V_a,P]
-\frac 12\int d^8z\Big\{
\Big[-\Bar T(\S-i\pa_{\un a}V^{\un a})+c.c.\Big]\\\cr
\label{oldminmasga}
-\frac 12\Big[D^\a\Psi_\a
+\Bar D_{\dot\a}\Bar\Psi^{\dot\a}
-m P\Big]^2
+\Big[m V_{\un a}-(\Bar D_{\dot\a}\Psi_\a
-D_\a\Bar\Psi_{\dot\a})\Big]^2
\Big\}~~.~~
\eea
The extra minus sign is necessary to keep gauge invariance since $\pa_5$ picks up a minus sign upon integration by parts, where as, $m$ acts as a constant.  It is possible to put extra minus signs on the last two squares in the action instead of the $T$ terms, but this theory would lead to the wrong sign in the Klein-Gordon equation, i.e. $(\Box+m^2)V_a=0~$.  The action (\ref{oldminmasga}) with $\Psi_\a=0$ and $T=0$ is the massive extension of old-minimal supergravity discussed in chapter \ref{Massive Extension of Old-Minimal Supergravity}.  In fact (\ref{oldminmasga}) is exactly the same action.  This can be seen in two ways.  The gauge transformations for $\Psi_\a$ and $T$ are now algebraic.  This means that there is a gauge where $\Psi_\a=0$ and $T=0$.  Alternatively, $\Psi_\a$ and $T$ can be eliminated by performing superfield redefinitions.  This is a much more interesting viewpoint, because it is an operational method that can be reversed.  The reversal of this procedure, dimensional oxidation, will be used in the following sections.
First, $T$ is removed by shifting $\Psi_\a$:
\bea
\Psi_\a \rightarrow \Psi_\a-\frac 1{4m}D_\a T~~,~~
\eea
and then $\Psi_\a$ is removed by shifting $V_a$ and $P$:
\bea
V_{\un a}\rightarrow V_{\un a}+\frac 1m(\Bar D_{\dot\a}\Psi_\a
-D_\a\Bar\Psi_{\dot\a})~~,~~\cr
P\rightarrow P+\frac 1mD^\a\Psi_\a
+\frac 1m\Bar D_{\dot\a}\Bar\Psi^{\dot\a}~~.~~
\eea
These superfield redefinitions are the same form as the gauge transformations of the massless theories.  This means that the dimensional oxidation procedure should begin by shifting all superfields by their massless gauge transformations.  The second step is to shift the newly introduced superfield by another superfield.  This second step will require a little ingenuity.
\subsection{Dimensional Oxidation from New-Minimal}
\label{Dimensional Oxidation from New-Minimal}

Two steps are necessary to perform dimensional oxidation from the massive extension of new-minimal.  First, the the superfields $V_{\un a}$ and $\chi_\a$ are shifted according to their massless gauge invariance:
\bea
V_{\un a}\rightarrow V_{\un a}
-\frac 1m\Bar D_{\dot\a}\Psi_\a
+\frac 1m D_\a\Bar \Psi_{\dot\a}~~,~~\\
\chi_\a\rightarrow\chi_\a
-\frac 1{4m}\Bar D^2\Psi_\a
-\frac im \Bar D^2D_\a V~~.~~
\eea
The second term in the shift of $\chi_\a$ accounts for the added gauge invariance which occurs from using $\chi_\a$ instead of $\cu$.  With this shift the action takes the form:
\bea
\label{partial5dnewmin}
\cs_{Y=3/2}[V_{\un a}, \chi_\a, \Psi_\a, V]=
\cs_{\rm New-Min}[V_{\un a}, \chi_\a]
-\frac 12\int d^8z\Big[mV^{\un a}
-(\Bar D^{\dot\a}\Psi^\a
-D^\a\Bar\Psi^{\dot\a})\Big]^2\cr
+3\Big\{\int d^6z\Big(m\chi^\a
-i\Bar D^2 D^\a V
-\frac 14\Bar D^2\Psi^\a\Big)^2+c.c\Big\}
\eea
This action is invariant under $\d \Psi_\a=mL_\a$ and $\d V=mL$.  The second step is to shift the new fields $\Psi_\a$ and $V$.  This shift corresponds to a new gauge freedom.  This means that the next new superfield compensates for this gauge freedom.  To find this new superfield, general terms are added to the gauge transformations; $\d\Psi_\a=mL_\a+\L_\a$ and $\d V=mL+\tilde\O$.  The variation of (\ref{partial5dnewmin}) with respect to $\L_\a$ and $\tilde\O$ is:
\bea
\label{newmassgauge}
\nonumber
\d_{\tilde\O,\L}\cs_{Y=3/2}[V_{\un a}, \chi_\a, \Psi_\a, V]
=\int d^8z\Big\{
\Psi^\a(-\frac 12\Bar D^2\L_\a
+\Bar D^{\dot\a}D_\a\Bar\L_{\dot\a}
-6i\Bar D^2D_\a\tilde\O)
-\Bar D_{\dot\a}V^{\un a}(m\L_\a)
\\
\cr
+\chi^\a(6m\L_\a+24imD_\a\tilde\O)
+\frac 12V(6iD^\a\Bar D^2\L_\a
-6i\Bar D_{\dot\a}D^2\Bar \L^{\dot\a}
-48D^\a\Bar D^2D_\a\tilde\O)
+c.c.\Big\}~~,~~
\eea
A solution to this variation can be found by taking $\L_\a=D_\a\tilde\L$ and then decomposing the two scalar gauge parameters $\tilde\L$ and $\tilde\O$ into chiral, anti-chiral and linear pieces:
\bea
\tilde\L=\L_c+\L_a+\L~~,~~\Bar D_{\dot\a}\L_c=0~~,~~
D_\a \L_a=0~~,~~D^2\L=\Bar D^2\L=0~~,~~\cr
\tilde\O=\O+\Bar \O+\O_L~~,~~\Bar D_{\dot\a}\O=0
~~,~~D^2\O_L=0~~,~~\O_L=\Bar\O_L~~.~~
\eea
Plugging these back into (\ref{newmassgauge}) leads to the following set of constraints:
\bea
0=D^\a\Bar D^2D_\a(i(\L-\Bar\L)-8\O_L)
~~,~~
0=\Bar D^2D_\a(i(\L-\Bar\L)-12\O_L)~~,~~\cr
0=D_\a\Bar D^2\Bar\L_c~~.~~~~~~~~~~~~
~~~~~~~~~~~~~~~~~~~~
\eea
The First two constraints imply that $\L$ is real and that $\O_L=0$.  The third constraint means that the chiral part of $\L$ vanishes, $\L_c=0$.  So the gauge parameters satisfy $\L_\a=D_\a \L$ with $\L=\Bar \L$ and $D^2 \L=0$ and $\tilde\O=\O+\Bar \O$ with $\Bar D_\a\O=0$.  With this solution the left over variation of the action is:
\bea
\label{leftovers}
\d_{\O\L}\cs_{Y=3/2}[V_{\un a}, \chi_\a, \Psi_\a, V]
=\int d^8z\Big\{
(6\chi^\a-\Bar D_{\dot\a}V^{\un a})(m\L_\a)
+c.c.\Big\}~~,~~
\eea
which can be canceled easily by adding a real linear superfield $X$ with gauge transformation $\d X=m\L$.  The massive gauge invariant action takes the form:
\bea
\cs_{Y=3/2}[V_{\un a}, \chi_\a, \Psi_\a, X]=\cs_{Y=3/2}[V_{\un a}, \chi_\a, \Psi_\a]
+\Big([D_\a, \Bar D_{\dot\a}]V^{\un a}+6\cu\Big)X~~.~~
\eea
The final five-dimensional action is obtained from this action by replacing $m$ with $\pa_5$ and getting the minus signs correct to regain gauge invariance.  The final answer is:
\bea
\label{5dnewminaction}
\cs_{\rm 5D~New-Min}=
\int dx_5\cs_{\rm New-Min}[V_{\un a}, \chi_\a]
+3c\Big\{\int d^7z\Big[\pa_5\chi^\a
-i\Bar D^2 D^\a V
-\frac 14\Bar D^2\Psi^\a\Big]^2+c.c.\Big\}
\cr
+c\int d^9z\Big\{-\frac 12\Big[\pa_5V^{\un a}
-(\Bar D^{\dot\a}\Psi^\a
-D^\a\Bar\Psi^{\dot\a})\Big]^2
-\Big([D_\a, \Bar D_{\dot\a}]V^{\un a}+6\cu\Big)X\Big\}~~,~~~
\eea
where $c$ is a real constant that will be fixed by analyzing the theory on-shell.  This action is invariant under:
\bea
\d \chi_\a = \frac 14\Bar D^2 L_\a+i\Bar D^2D_\a L
~~,~~L=\Bar L~~,~~\cr
\d \Psi_\a=\pa_5 L_\a+D_\a\L~~,~~\L=\Bar \L~~,~~D^2\L=0~~,~~\cr
\d X=\pa_5\L~~,~~X=\Bar X~~,~~D^2X=0~~,~~\cr
\d V=\pa_5 L+\O~~,~~\Bar D_{\dot\a}\O=0~~.~~
\eea
There are two ways to prove that this theory really does represent linearized five-dimensional supergravity.  One way is to calculate the component action and eliminate all auxiliary fields.  Another way is to work solely in superspace, by analyzing the superfield-strengths on-shell and showing that the only propagating degrees of freedom are those of five-dimensional supergravity.  This method was applied to the five-dimensional extension of old minimal supergravity in \cite{Gates2003} and will be used in this thesis.

The first step in proving that (\ref{5dnewminaction}) is five-dimensional supergravity is to enumerate all possible superfield-strengths:
\bea
W_{\a\b\g}=\frac i{3!8}\Bar D^2D_{(\a}\pa_\b{}^{\dot\a}V_{\g)\dot\a}~~,~~
\eea
\bea
G_{\un a}:=(-2\P^T_{1/2}+\P^T_{3/2})\Box V_{\un a}
+\frac 12[D_\a, \Bar D_{\dot\a}]\cu~~,~~
\eea
\bea
G^{\prime}_{\un a}:=
\pa_5^2V_{\un a}
-\pa_5\Bar D_{\dot\a}\Psi_\a
+\pa_5D_\a\Bar \Psi_{\dot\a}
-[D_\a, \Bar D_{\dot\a}]X~~,~~
\eea
\bea
T_\a:=+\frac 18\Bar D^2D_\a[D_\b,\Bar D_{\dot\b}]V^{\un b}
+\frac 34\Bar D^2D_\a\cu~~,~~
\eea
\bea
T_\a^{\prime}:=-6\pa_5^2\chi_\a
+6i\pa_5\Bar D^2D_\a V
+\frac 32\pa_5\Bar D^2\Psi_\a
-\frac 32\Bar D^2D_\a X~~,~~
\eea
\bea
\l_\a:=6\pa_5\chi_\a
-6i\Bar D^2D_\a V
-\frac 12\Bar D^2\Psi_\a
+\pa_5\Bar D^{\dot\a}V_{\un a}
+\Bar D^{\dot\a}D_\a\Bar \Psi_{\dot\a}~~,~~
\eea
\bea
\l:=-48D^\a\Bar D^2D_\a V
-24i\pa_5D^\a\chi_\a
+24i\pa_5\Bar D_{\dot\a}\Bar\chi^{\dot\a}
+6iD^\a\Bar D^2\Psi_\a
-6i\Bar D_{\dot\a}D^2\Bar\Psi^{\dot\a}~~,~~
\eea
\bea
F_{\a\b}:=\frac 32\Bar D^2D_{(\a}\Psi_{\b)}
+\frac 12D_{(\a}\Bar D^2\Psi_{\b)}
-D_{(\a}\Bar D^{\dot\a}D_{\b)}\Bar \Psi_{\dot\a}\cr
+4i\pa_5\pa_{(\a}{}^{\dot\a}V_{\b)\dot\a}
+\pa_5[D_{(\a}, \Bar D^{\dot\a}]V_{\b)\dot\a}~~,~~
\eea
\bea
F_{\a\b}^{\prime}:=
-\frac 18\Bar D^2D_{(\a}\Psi_{\b)}
-\frac 18D_{(\a}\Bar D^2\Psi_{\b)}
+\frac 14D_{(\a}\Bar D^{\dot\a}D_{\b)}\Bar \Psi_{\dot\a}\cr
-\frac i2\pa_5\pa_{(\a}{}^{\dot\a}V_{\b)\dot\a}
+\pa_5D_{(\a}\chi_{\b)}
-iD_{(\a}\Bar D^2D_{\b)}V~~.~~
\eea
This is quite a long list, but there are about the same number of superfield-strengths here as for the old-minimal theory in \cite{Gates2003}.  The second step is to list several Bianchi identities.  These identities are relations between various derivatives of the superfield-strengths.  It is assumed that these relations would be obtained by the dimensional reduction of a manifestly supersymmetric five-dimensional superalgebra.  A partial list is:
\bea
\label{BIDivG}
\pa^{\un a}G_{\un a}=0~~,~~
\eea
\bea
D_{(\a}\l_{\b)}-\frac 12F_{\a\b}-6F_{\a\b}^{\prime}=0~~,~~
\eea
\bea
\Bar D^{\dot\a}G_{\un a}-T_\a=0~~,~~
\eea
\bea
\Bar D^{\dot\a}G_{\un a}^{\prime}-\pa_5\l_\a-T_\a^{\prime}=0~~,~~
\eea
\bea
\Bar D^2 F_{\a\b}+\Bar D^2 D_{(\a}\l_{\b)}=0~~,~~
\eea
\bea
+\frac {4i}3D^\a F_{\a\b}
+8i\Bar D^{\dot\a}D_\a\Bar \l_{\dot\a}
-2iD^2\l_\a
+4iD_\a\Bar D^{\dot\a}\Bar \l_{\dot\a}
+D_\a \l=0~~,~~
\eea
\bea
\label{BIcurlG}
-i\pa_{\dot\a}{}^\b(\frac 14F_{\a\b}+2F_{\a\b}^\prime)
+i\pa_\a{}^{\dot\b}(+\frac 14\Bar F_{\dot\a\dot\b}
+2\Bar F_{\dot\a\dot\b}^\prime)
-2\pa_5G_{\un a}=0~~,~~
\eea
\bea
\label{BIeomf}
D^\b\Bar D_{\dot\a}(F_{\a\b}+12F_{\a\b}^\prime)
+4i\pa_{\un a}D^\b\l_\b
+\Bar D_{\dot\a}D^2\l_\a
+2D^2\Bar D_{\dot\a}\l_\a
-c.c.=0~~,~~
\eea
\bea
\label{BIeomg}
\pa_5(\frac 14F_{\a\b}+2F^\prime_{\a\b})
+\frac 13D_{(\a}T^\prime_{\b)}
-\frac 14[D_{(\a},\Bar D^{\dot\g}]G^\prime_{\b)\dot\g}=0
~~.~~
\eea
These Bianchi identities in conjunction with the equations of motion will give the proper dimensionally reduced component equations of motion for five-dimensional linearized supergravity with a gauge 2-form.  The equations of motion are simply:
\bea
\label{EOMV5DNEwmin}
{\d\over\d V^{\un a}}\Big(\cs_{\rm 5D~New-Min}\Big)=G_{\un a}+cG_{\un a}^{\prime}=0~~,~~
\eea
\bea
{\d\over\d \chi^\a}\Big(\cs_{\rm 5D~New-Min}\Big)=T_\a+cT_\a^\prime=0~~,~~
\eea
\bea
{\d\over\d \Psi^\a}\Big(\cs_{\rm 5D~New-Min}\Big)=\l_\a=0~~,~~
\eea
\bea
{\d\over\d V}\Big(\cs_{\rm 5D~New-Min}\Big)=\l=0~~,~~
\eea
\bea
{\d\over\d X}\Big(\cs_{\rm 5D~New-Min}\Big)\Rightarrow~~T_\a=0~~.~~
\eea
On-shell, the only non-zero linearly independent superfield-strengths are $W_{\a\b\g}$, $G_{\un a}$ and $F_{\a\b}$ which are now irreducible four-dimensional transverse linear representations:
\bea
D^\a G_{\un a}=\Bar D^2F_{\a\b}=D^\a F_{\a\b}=0~~.~~
\eea
From \cite{Gates2003} these superfield-strengths constrained in this manner are enough to describe the graviton and gravitino.  The only question that remains is to show that the 3-form field-strength obeys the appropriate equations.  The appropriate equations are the dimensionally reduced 3-form Bianchi identity and equation of motion.  The Bianchi identity for a five-dimensional 3-form, $\pa_{[A}B_{BCD]}$, in $SL(2,C)$ notation breaks into two relations:  
\bea
\label{3formBI1}
\pa_{[a}B_{bcd]}=0~~~~\rightarrow 
~~~~\pa^{\un a}B_{\un a}=0~~,~~
\eea\bea
\label{3formBI2}
\pa_{[a}B_{bc]5}-2\pa_5B_{abc}=0~~~~\rightarrow
~~~~i\pa_\d{}^{\dot\b}\bar b_{\dot\d\dot\b}
-i\pa_{\dot\d}{}^\b b_{\d\b}+\pa_5 B_{\un d}=0~~,~~
\eea
where $B^a:=\frac 1{3!}\e^{abcd}B_{bcd}$ and $B_{\un a\un b5}=2\e_{\a\b}\bar b_{\dot\a\dot\b}+2\e_{\dot\a\dot\b}b_{\a\b}$.  
These Bianchi identities are mimicked in (\ref{BIDivG}) and (\ref{BIcurlG}).  The equation of motion for a five-dimensional 3-form, $\pa^AG_{ABC}=0$, also breaks into two equations in $SL(2,C)$ notation:
\bea
\label{3formEM1}
\pa^{a}B_{ab5}=0~~~~\rightarrow
~~~~i\pa_\d{}^{\dot\b}\bar b_{\dot\d\dot\b}
+i\pa_{\dot\d}{}^\b b_{\d\b}=0~~,~~
\eea
\bea
\label{3formEM2}
\pa^aB_{abc}+\pa_5B_{bc5}=0~~~~\rightarrow
~~~~2\pa_5b_{\a\b}-\frac i2\pa_{(\a}{}^{\dot\g}B_{\b)\dot\g}~~.~~
\eea
These equations of motion follow from the identities (\ref{BIeomf}) and (\ref{BIeomg}) when the theory is taken on-shell.  This allows the unknown coefficient $c$ to be fixed.  Defining the component 2-form as:
\bea
b_{\a\b}:=(\frac 14F_{\a\b}+2F^\prime_{\a\b})|~~,~~
\eea
and comparing (\ref{BIeomf}) and (\ref{BIeomg}) to (\ref{3formEM1}) and (\ref{3formEM2}) leads to the the component definition of $B_{\un a}$:
\bea
B_{\un a}:=-2G_{\un a}|~~,~~B_{\un a}:=2G_{\un a}^\prime~~.~~
\eea
The equation of motion for $V_{\un a}$, (\ref{EOMV5DNEwmin}), relates these two superfield-strengths by $G_{\un a}=-cG_{\un a}^\prime$ therefore $c=1$.  From this point some more Bianchi identities are needed to prove that $W_{\a\b\g}$ propagates, but for an initial analysis these results can be taken as proof of the consistency of this theory.  The full list of on-shell components can be taken as:
\bea
C_{\a\b\g\d}:=D_{(\a}W_{\b\g\d)}|~~,~~
\eea
\bea
\label{nmweylscraps}
C_{\a\b\g\dot\a}:=\frac 12[D_{(\a}, \Bar D_{\dot\a}]F_{\b\g)}|~~,~~
C_{\a\b\dot\a\dot\b}:=\frac 12[D_{(\a}, \Bar D_{(\dot\a}G_{\b)\dot\b)}|~~,~~
\eea
\bea
f^{(+)}_{\a\b\g}:=W_{\a\b\g}|~~,~~
\eea
\bea
\label{nmgravcurlscrap}
f^{(-)}_{\a\b\dot\a}:=D_{(\a}G_{\b)\dot\a}|~~,~~
\bar f^{(-)}_{\a\b\g}:=D_{(\a}F_{\b\g)}|~~,~~
\bar f^{(+)}_{\a\b\dot\a}:=\Bar D_{\dot\a}F_{\a\b}|~~,~~
\eea
\bea
B_{\un a}:=-2G_{\un a}|~~,~~
b_{\a\b}:=(\frac 14F_{\a\b}+2F^\prime_{\a\b})|~~.~~
\eea
The higher components of $G_{\un a}$ and $F_{\a\b}$ are the dimensionally reduced parts of the five-dimensional Weyl tensor (\ref{nmweylscraps}) and the curl of the five-dimensional gravitino (\ref{nmgravcurlscrap}).  The dimensional reduction using $SL(2,C)$ notation of these component objects is discussed thoroughly in \cite{Gates2003} and will not be reproduced in this document.

This concludes the presentation of the five-dimensional version of new-minimal supergravity.  The full action is:
\bea
\nonumber
\cs_{\rm 5D~New-Min}=
\int d^9z\Big\{V^{\un a}\Box(-\P^T_{1/2}+\frac 12\P^T_{3/2})V_{\un a}+\frac 12\cu [D_\a,\Bar D_{\dot\a}]V^{\un a}+\frac 32\cu^2\\\cr
\nonumber
-\frac 12\Big[\pa_5V^{\un a}
-(\Bar D^{\dot\a}\Psi^\a
-D^\a\Bar\Psi^{\dot\a})\Big]^2
-\Big([D_\a, \Bar D_{\dot\a}]V^{\un a}+6\cu\Big)X\Big\}
\\\cr
\label{5dnewminfinaction}
+3\Big\{\int d^7z\Big[\pa_5\chi^\a
-i\Bar D^2 D^\a V
-\frac 14\Bar D^2\Psi^\a\Big]^2+c.c.\Big\}~~.~
\eea
The main difference between this theory and the five-dimensional version of old-minimal is the form of the radion multiplet.  The radion multiplet contains the $g_{55}$ component of the metric, which holds information about the size of the extra dimension and therefore is of phenomenological importance.  In the old minimal theory the radion multiplet is the chiral scalar superfield $T$.  In new-minimal it is a real linear superfield $X$.  It seems that the radion multiplet mimics the form of the compensator used in the base four-dimensional supergravity.  To check this logic the next section discusses the five-dimensional extension of ``New"new-minimal supergravity.

\subsection{Dimensional Oxidation from New-New-Minimal}
\label{Dimensional Oxidation from New-New-Minimal}
The superfields of the massive new-new-minimal theory, $V_{\un a}$ and $\l_\a$ are shifted according to their massless gauge transformations:
\bea
V_{\un a}\rightarrow V_{\un a}
-\frac 1m\Bar D_{\dot\a}\Psi_\a
+\frac 1m D_\a\Bar \Psi_{\dot\a}~~,~~
\eea
\bea
\l_\a\rightarrow\l_\a
-\frac 1{12m}\Bar D^2\Psi_\a
-\frac 1m \Bar D^2D_\a V~~.~~
\eea
These field redefinitions lead to the following massive gauge invariant action:
\bea
\nonumber
\cs_{Y=3/2}[V_{\un a}, \l_\a, \Psi_\a, V]=
\cs_{\rm \n^2-Min}[V_{\un a}, \l_\a]
-\frac 12\int d^8z\Big[mV^{\un a}
-(\Bar D^{\dot\a}\Psi^\a
-D^\a\Bar\Psi^{\dot\a})\Big]^2
\eea
\bea
\label{partial5dnewmin}
+9\Big\{\int d^6z\Big(m\l^\a
-\Bar D^2 D^\a V
-\frac 1{12}\Bar D^2\Psi^\a\Big)^2+c.c\Big\}~~.~~
\eea
It is necessary to add more gauge invariance for the new superfields, $\d\Psi_\a=mL_\a+\L_\a$ and $\d V=mL+\tilde\O$.  The gauge variation of (\ref{partial5dnewmin}) under the arbitrary $\L_\a$ and $\tilde \O $ transformations is:
\bea
\nonumber
\d_{\tilde\O\L} \cs_{Y=3/2}[V_{\un a}, \l_\a, \Psi_\a, V]=
\int d^8z\Big\{\Psi^\a\Big(\frac 12\Bar D^2\L_\a
+\Bar D^{\dot\a}D_\a\Bar\L_{\dot\a}
-6\Bar D^2D_\a\tilde\O\Big)
-\Bar D_{\dot\a}V^{\un a}(m\L_\a)
\eea
\bea
+\frac 12V\Big(72D^\a\Bar D^2D_\a \tilde\O
+6D^\a\Bar D^2\L_\a
+6\Bar D_{\dot\a}D^2\Bar\L^{\dot\a}
\Big)
+\l^\a\Big(6m\L_\a+72mD_\a\tilde\O
\Big)
+c.c.\Big\}~~,~~
\eea
The solution to the vanishing of this variation is almost the same as the new minimal case.  $\tilde\O$ is again purely chiral $\tilde\O=\O+\Bar\O$ where $\Bar D_{\dot\a}\O=0$.  $\L_\a$ is replaced by $\L_\a=iD_\a\L$ where $\L=\Bar\L$ and $D^2\L=0$.
The main difference between this solution and that for new-minimal is the peculiar, but necessary, factor of $i$ in $\L_\a$.  Under these transformations, the gauge variation of (\ref{partial5dnewmin}) only contains two terms:
\bea
\d_{\tilde\O\L} \cs_{Y=3/2}[V_{\un a}, \l_\a, \Psi_\a, V]=
\int d^8z\Big\{
-\Bar D_{\dot\a}V^{\un a}(miD_\a\L)
+\l^\a\Big(6miD_\a\L\Big)
+c.c.\Big\}~~.~~
\eea
These terms can be canceled by adding a real linear superfield, $Y$, with gauge transformation $\d Y=m\L$.  The expanded massive gauge invariant action is now:
\bea
\nonumber
\cs_{Y=3/2}[V_{\un a}, \l_\a, \Psi_\a, V]=
\cs_{\rm \n^2-Min}[V_{\un a}, \l_\a]
+\int d^8z\Big\{-\frac 12\Big[mV^{\un a}
-(\Bar D^{\dot\a}\Psi^\a
-D^\a\Bar\Psi^{\dot\a})\Big]^2
\eea
\bea
\label{partial5dnewmin}
+(2\pa_{\un a}V^{\un a}+6U)Y\Big\}
+9\Big\{\int d^6z\Big(m\l^\a
-\Bar D^2 D^\a V
-\frac 1{12}\Bar D^2\Psi^\a\Big)^2+c.c\Big\}~~.~~
\eea
The oxidation is completed by changing $m$ to $\pa_5$ and switching some minus signs.  The five-dimensional action is:
\bea
\nonumber
\cs_{\rm 5D\,\n^2\,Min}=
\int dx_5\cs_{\rm \n^2-Min}[V_{\un a}, \l_\a]
+c\int d^9z\Big\{-\frac 12\Big[\pa_5V^{\un a}
-(\Bar D^{\dot\a}\Psi^\a
-D^\a\Bar\Psi^{\dot\a})\Big]^2
\eea
\bea
\label{Result5DNNMIN}
-(2\pa_{\un a}V^{\un a}+6U)Y\Big\}
+9c\Big\{\int d^7z\Big(\pa_5\l^\a
-\Bar D^2 D^\a V
-\frac 1{12}\Bar D^2\Psi^\a\Big)^2+c.c\Big\}~~,~~
\eea
and is gauge invariant under the following transformations:
\bea
\d V_{\un a}=\Bar D_{\dot\a}L_\a-D_\a\Bar L_{\dot\a}~~,~~\cr
\d \l_\a=\frac 1{12}\Bar D^2L_\a+\Bar D^2D_\a L~~,~~\cr
\d \Psi_\a=\pa_5 L_\a+iD_\a\L~~,~~\L=\Bar \L~~,~~D^2\L=0
~~,~~\cr
\d Y=\pa_5\L~~,~~Y=\Bar Y~~,~~D^2Y=0~~,~~\cr
\d V = \pa_5 L+\O+\Bar\O~~~~\Bar D_{\dot\a}\O=0~~.~~
\eea
Knowledge of the superfield-strengths and their Bianchi identities is crucial, in order to fix the unknown coefficient $c$.  The full set of superfield-strengths is:
\bea
W_{\a\b\g}:=\frac i{3!8}\Bar D^2D_{(\a}\pa_\b{}^{\dot\a}V_{\g)\dot\a}~~,~~
\eea
\bea
G_{\un a}:=
2\Box(\frac 12\P^T_{3/2}+\frac 13\P^L_{1/2})V_{\un a}
-\pa_{\un a}U~~,~~
\eea
\bea
G_{\un a}^{\prime}:=
\pa_5^2V_{\un a}
-\pa_5(\Bar D_{\dot\a}\Psi_\a-D_\a\Bar\Psi_{\dot\a})
+2\pa_{\un a}Y~~,~~
\eea
\bea
T_\a:=+\frac i4\Bar D^2D_\a\pa_{\un b}V^{\un b}
+\frac 34i\Bar D^2D_\a U~~,~~
\eea
\bea
T_{\a}^{\prime}:=-18\pa_5^2\l_\a
+\frac 32\pa_5\Bar D^2\Psi_\a
-\frac 32i\Bar D^2D_\a Y
+18\pa_5\Bar D^2D_\a V~~,~~
\eea
\bea
\l_\a^\prime:=
+\frac 12\Bar D^2\Psi_\a
+\Bar D^{\dot\a}D_\a\Bar \Psi_{\dot\a}
+\pa_5\Bar D^{\dot\a}V_{\un a}
+6\pa_5\l_\a
-6\Bar D^2D_\a V
\eea
\bea
\l:=72D^\a\Bar D^2D_\a V-72\pa_5D^\a\l_\a
+6D^\a\Bar D^2\Psi_\a+c.c.
\eea
\bea
F_{\a\b}:=\frac 14\Bar D^2D_{(\a}\Psi_{\b)}
-\frac 14D_{(\a}\Bar D^2\Psi_{\b)}
-\frac 12D_{(\a}\Bar D^{\dot\a}D_{\b)}\Bar\Psi_{\dot\a}
+i\pa_5\pa_{(\a}{}^{\dot\a}V_{\b)\dot\a}~~,~~
\eea
\bea
F_{\a\b}^\prime:=\frac 12\Bar D^2D_{(\a}\Psi_{\b)}
-\pa_5\Bar D^{\dot\a}D_{(\a}V_{\b)\dot\a}
+6\pa_5D_{(\a}\l_{\b)}
-6D_{(\a}\Bar D^2D_{\b)}V~~.~~
\eea
These superfield-strengths obey this partial list of Bianchi identities:
\bea
F_{\a\b}-\frac 12F_{\a\b}^\prime
+\frac 12D_{(\a}\l_{\b)}^\prime=0~~,~~
\eea
\bea
\label{reduceF1}
D^\a F_{\a\b}
-\frac 1{24}D_\b\l
+\frac 14D^2\l_\b^\prime
+\frac 12D_\b\Bar D^{\dot\b}\bar\l_{\dot\b}^\prime
+\Bar D^{\dot\b}D_\b\bar\l_{\dot\b}^\prime=0~~,~~
\eea
\bea
\label{reduceF2}
\Bar D^2 F_{\a\b}+\frac 12\Bar D^2D_{(\a}\l_{\b)}^\prime=0~~,~~
\eea
\bea
\label{reduceG1}
\Bar D^{\dot\a} G_{\un a}^\prime
-\pa_5\l_\a^\prime-\frac 13T_\a^\prime=0~~,~~
\eea
\bea
\label{45}
\pa_5 F_{\a\b}-i\pa_{(\a}{}^{\dot\a}G_{\b)\dot\a}^\prime=0~~,~~
\eea
\bea
\label{RegBI1}
\pa_{\a}{}^{\dot\b}\Bar F_{\dot\b\dot\a}
+\pa_{\dot\a}{}^{\b} F_{\b\a}=0~~,~~
\eea
\bea
\label{44}
D^\a W_{\a\b\g}+\frac i2\pa_{(\b}{}^{\dot\a}G_{\g)\dot\a}=0~~,~~
\eea
\bea
\label{48}
-\frac 14\Bar D^2 D_{(\a}F_{\b\g)}+24\pa_5 W_{\a\b\g}=0~~.~~
\eea
The identities (\ref{45}) and (\ref{RegBI1}) are the dimensionally reduced Bianchi identities of a five-dimensional $2$-form field-strength\cite{Gates2003}.  This means that $F_{\b\a}$ and $G_{\un a}$ contain the component $2$-form field-strength.  The equations of motion can be written in terms of the superfield-strengths:
\bea
\label{EOMV5Dnewnewmin}
{\d\over\d V^{\un a}}\Big(\cs_{\rm 5D\,\n^2\,Min}\Big)=G_{\un a}+cG_{\un a}^{\prime}=0~~,~~
\eea
\bea
{\d\over\d \l^\a}\Big(\cs_{\rm 5D\,\n^2\,Min}\Big)=T_\a+cT_\a^\prime=0~~,~~
\eea
\bea
{\d\over\d \Psi^\a}\Big(\cs_{\rm 5D\,\n^2\,Min}\Big)=\l_\a^\prime=0~~,~~
\eea
\bea
{\d\over\d V}\Big(\cs_{\rm 5D\,\n^2\,Min}\Big)=\l=0~~,~~
\eea
\bea
{\d\over\d Y}\Big(\cs_{\rm 5D\,\n^2\,Min}\Big)\Rightarrow~~T_\a=0~~.~~
\eea
There are three non-zero linearly independent superfield-strengths on-shell $W_{\a\b\g}$, $F_{\a\b}$ and $G_{\un a}$.  The Bianchi identities (\ref{reduceF1}-\ref{reduceG1}) mean that both $F_{\a\b}$ and $G_{\un a}$ are transversal linear superfields on-shell.  The coefficient $c$ can be determined by taking some derivatives of the equation of motion for $V_{\un a}$, (\ref{EOMV5Dnewnewmin}), to obtain the five-dimensional D'Alembertian acting on $W_{\a\b\g}$:
\bea
{i\over 8\cdot 3!}\Bar D^2 D_{(\a}\pa_\b{}^{\dot\b}\Big(G_{\g)\dot\b}
+cG_{\g)\dot\b}^\prime\Big)
=\Big(\Box +c\pa_5^2\Big) W_{\a\b\g}~~,~~
\eea
where the Bianchi identities (\ref{45}), (\ref{44}), and (\ref{48}) have been used.  This proves that this theory cannot be be five-dimensionally Lorentz invariant unless $c=1$.  The final action with  no unknown coefficients is:
\bea
\nonumber
\cs_{\rm 5D\,\n^2\,Min}=
\int dx_5\cs_{\rm \n^2-Min}[V_{\un a}, \l_\a]
+c\int d^9z\Big\{-\frac 12\Big[\pa_5V^{\un a}
-(\Bar D^{\dot\a}\Psi^\a
-D^\a\Bar\Psi^{\dot\a})\Big]^2
\eea
\bea
\label{final5DNNMIN}
-(2\pa_{\un a}V^{\un a}+6U)Y\Big\}
+9c\Big\{\int d^7z\Big(\pa_5\l^\a
-\Bar D^2 D^\a V
-\frac 1{12}\Bar D^2\Psi^\a\Big)^2+c.c\Big\}~~,~~
\eea
This action differs from the old-minimal model in the same way as the new-minimal five-dimensional theory.  The radion multiplet is a real linear scalar superfield, $Y$.  Of course, this theory contains a five-dimensional gauge $1$-form just like the old-minimal.

\section{Five-dimensional Discussion}
\label{Five-dimensional Discussion}

Two new theories representing linearized five-dimensional supergravity were presented in this chapter.  The actions are given in equations (\ref{5dnewminfinaction}) and (\ref{final5DNNMIN}).  In both cases the off-shell superfield-strengths were enumerated and partial lists of Bianchi identities were given.  The unknown coefficient $c$ was fixed using purely superspace methods which rely on the knowledge of the superfield-strengths and their Bianchi identities.  This coefficient represents the fact that dimensionally reduced actions can break into separate gauge invariant pieces, and does not signify a lack of understanding.  

This chapter yields important information about the off-shell structure of higher dimensional supersymmetric theories.  The most obvious observation is that many of the superfield-strengths have the same mass dimension and index structure.  The primed superfield-strengths $G_{\un a}^\prime$ and $T_\a^\prime$ are good examples.  Both of these superfield-strengths are duplicates of the usual four-dimensional superfield-strengths.  This means that in higher dimensional theories there are multiple copies of auxiliary fields off-shell, even in Wess-Zumino gauge.  Another rather striking observation is the appearance of the low mass dimension superfield-strength $\l_\a$.  By dimensional analysis, there are only two places in which this superfield-strength can appear in a covariant derivative algebra.  It must appear as a super torsion with either of the following index structure:
\bea
T_{\tilde\a\tilde\b}{}^{\tilde\g}~~,~~T_{\tilde\a A}{}^B~~,~~
\eea
where $\tilde\a$ is a five-dimensional spinor index.  In four dimensions, both of these torsion are set to zero in Einstein supergravity.  In dimensions greater than four, the vanishing of both of these torsion will probably lead to an on-shell covariant derivative algebra.  

Perhaps that most interesting result of this thesis is the five-dimensional extension of new-minimal supergravity (\ref{5dnewminfinaction}).  This theory contains a component $3$-form field-strength which is required by supersymmetry in five dimensions.  The $3$-form is not simply coupled to supergravity, it is an essential part of the supergravity multiplet.  This theory is related to the low energy limits of superstring theory and M-theory, which both contain higher rank forms as part of their background supergravity multiplets.  

Two directions come to mind to push this analysis further.  It should be possible to connect this $N=1$ analysis to the usual covariant higher dimensional analysis.  The explicit dimensional reduction of a fully covariant five-dimensional super covariant derivative algebra should lead to the superfield-strengths and Bianchi identities presented in this chapter.  Alternatively, it may be possible to find the algebra associated with the extra five-dimensional supersymmetry that is not manifest in the $N=1$ formulations.  In this sense, the full five-dimensional off-shell theory can be constructed using methods analogous to component constructions of supersymmetric theories.

The second direction is to develop a five-dimensional  AdS supergravity background using $N=1$ superfields.  It seems likely that the dimensional oxidation method could be used to create this theory once a massive AdS theory is written in four dimensions.  A concrete understanding of $N=1$ AdS flat superspace would be necessary for such an endeavor.  While the covariant derivative algebra of $N=1$ AdS superspace is known, the Casimir operators that define the supersymmetric representation theory are not known.  There is a way to construct these operators based on the formulation given in \cite{Jarvis1979}.  The analogue of the D'Alembertian is:
\bea
C_2=-{2\over{|\mu|^2}}\mathbb P^{\un a} \mathbb P_{\un a}
-4\mathbb J^{\a\b}\mathbb J_{\a\b}
-4\Bar{\mathbb J}^{\dot\a\dot\b}\Bar{\mathbb J}_{\dot\a\dot\b}
+\frac 1{\bar\mu}\mathbb Q^\a\mathbb Q_\a
+\frac 1{\mu}\Bar{\mathbb Q}_{\dot\a}\Bar{\mathbb Q}^{\dot\a}
~~,~~
\eea
where the operators $ \mathbb P_{\un a}$, $\mathbb J_{\a\b}$, and $\mathbb Q_\a$ satisfy the AdS superalgebra:
\bea
[\mathbb Q_\a,\Bar {\mathbb Q}_{\dot\a}\}=2\mathbb P_{\un a}
~~,~~
\eea
\bea
[\mathbb Q_\a, \mathbb Q_\b\}=-4i\bar\mu\mathbb J_{\a\b}~~,~~
[\mathbb Q_\a,\mathbb P_{\un b}\}=-\e_{\a\b}\bar\mu\Bar{\mathbb Q}_{\dot\b}~~,~~
\eea
\bea
[\mathbb P_{\un a},\mathbb P_{\un b}\}=-2i|\mu|^2\Big(
\e_{\dot\a\dot\b}\mathbb J_{\a\b}+\e_{\a\b}\Bar {\mathbb J}_{\dot\a\dot\b}
\Big)~~,~~
\eea
\bea
[\mathbb J_{\a\b},\mathbb Q_\g\}=\frac i2\e_{\g(\a}\mathbb Q_{\b)}~~,~~
[\mathbb J_{\a\b},\mathbb P_{\un c}\}=\frac i2\e_{\g(\a}\mathbb P_{\b)\dot\g}~~,~~
\eea
\bea
[\mathbb J_{\a\b}, \mathbb J_{\g\d}\}=\frac i2\e_{\g(\a}\mathbb J_{\b)\d}
+\frac i2\e_{\d(\a}\mathbb J_{\b)\g}~~.~~
\eea
The parameter $\mu$ is related to the curvature of the spacetime.  
The analogue of the superspin operator is much more complicated:
\bea
C_4=+\S_{1\b}\S^\b{}_1\S_{1\d}\S^\d{}_1
+\S_{1\dot\b}\S^{\dot\b}{}_1\S_{1\d}\S^\d{}_1
+\S_{1\b}\S^\b{}_1\S_{1\dot\d}\S^{\dot\d}{}_1
+\S_{1\dot\b}\S^{\dot\b}{}_1\S_{1\dot\d}\S^{\dot\d}{}_1
\cr
-\S_{1\b}\S^\b{}_\g\S^\g{}_\d\S^\d{}_1
-\S_{1\dot\b}\S^{\dot\b}{}_\g\S^\g{}_\d\S^\d{}_1
-\S_{1\b}\S^\b{}_\g\S^\g{}_{\dot\d}\S^{\dot\d}{}_1
-\S_{1\dot\b}\S^{\dot\b}{}_\g\S^\g{}_{\dot\d}\S^{\dot\d}{}_1
\cr
-\S_{1\b}\S^\b{}_{\dot\g}\S^{\dot\g}{}_\d\S^\d{}_1
-\S_{1\dot\b}\S^{\dot\b}{}_{\dot\g}\S^{\dot\g}{}_\d\S^\d{}_1
-\S_{1\b}\S^\b{}_{\dot\g}\S^{\dot\g}{}_{\dot\d}\S^{\dot\d}{}_1
-\S_{1\dot\b}\S^{\dot\b}{}_{\dot\g}\S^{\dot\g}{}_{\dot\d}\S^{\dot\d}{}_1
\cr
+\Big\{
+\S^\a{}_1\S_{1\g}\S^\g{}_\d\S^\d{}_\a
+\S^\a{}_1\S_{1\g}\S^\g{}_{\dot\d}\S^{\dot\d}{}_\a
+\S^\a{}_1\S_{1\dot\g}\S^{\dot\g}{}_\d\S^\d{}_\a
+\S^\a{}_1\S_{1\dot\g}\S^{\dot\g}{}_{\dot\d}\S^{\dot\d}{}_\a\cr
+\S^\a{}_\b\S^\b{}_1\S^1{}_\d\S^\d{}_\a
+\S^\a{}_\b\S^\b{}_1\S^1{}_{\dot\d}\S^{\dot\d}{}_\a
+\S^\a{}_{\dot\b}\S^{\dot\b}{}_1\S^1{}_\d\S^\d{}_\a
+\S^\a{}_{\dot\b}\S^{\dot\b}{}_1\S^1{}_{\dot\d}\S^{\dot\d}{}_\a\cr
-\S^\a{}_\b\S^\b{}_\g\S^\g{}_\d\S^\d{}_\a
-\S^\a{}_\b\S^\b{}_\g\S^\g{}_{\dot\d}\S^{\dot\d}{}_\a
-\S^\a{}_\b\S^\b{}_{\dot\g}\S^{\dot\g}{}_\d\S^\d{}_\a
-\S^\a{}_\b\S^\b{}_{\dot\g}\S^{\dot\g}{}_{\dot\d}\S^{\dot\d}{}_\a\cr
-\S^\a{}_{\dot\b}\S^{\dot\b}{}_\g\S^\g{}_\d\S^\d{}_\a
-\S^\a{}_{\dot\b}\S^{\dot\b}{}_\g\S^\g{}_{\dot\d}\S^{\dot\d}{}_\a
-\S^\a{}_{\dot\b}\S^{\dot\b}{}_{\dot\g}\S^{\dot\g}{}_\d\S^\d{}_\a
-\S^\a{}_{\dot\b}\S^{\dot\b}{}_{\dot\g}\S^{\dot\g}{}_{\dot\d}\S^{\dot\d}{}_\a\cr
+\S^\a{}_{\dot\b}\S^{\dot\b}{}_{\dot\g}\S^{\dot\g}{}_1\S^1{}_\a
+\S^\a{}_{\dot\b}\S^{\dot\b}{}_\g\S^\g{}_1\S^1{}_\a
+\S^\a{}_\b\S^\b{}_{\dot\g}\S^{\dot\g}{}_1\S^1{}_\a
+\S^\a{}_\b\S^\b{}_\g\S^\g{}_1\S^1{}_\a\cr
-\S^\a{}_1\S_{1\g}\S^\g{}_1\S^1{}_\a
-\S^\a{}_1\S_{1\dot\g}\S^{\dot\g}{}_1\S^1{}_\a
+{(\a\rightarrow\dot\a)}\Big\}~~,~~
\eea
where the $\S$ objects are defined as:
\bea
\S_{\a\b}=-i2\mathbb J_{\a\b}~~,~~
\S_{\dot\a\dot\b}=-i2\Bar{\mathbb J}_{\dot\a\dot\b}~~,~~
\eea
\bea
\S_{\un a}=\frac i{\sqrt{\mu\bar\mu}}\mathbb P_{\un a}~~,~~
\eea
\bea
\S_{1\a}=\frac i{\sqrt{2\bar\mu}}\mathbb Q_\a~~,~~
\S_{1\dot\a}=-\frac 1{\sqrt{2\mu}}\Bar{\mathbb Q}_{\dot\a}~~.~~
\eea
After writing the Casimir operators in terms of supercovariant derivatives, the constraints that determine the irreducible representations of the AdS superalgebra on superfields can be determined.  Then superspace actions can be constructed that lead to these constraints on-shell.  There may also be a question pertaining to the superspace integration measure, since the spacetime is curved.  The density projector can be easily solved by using the ectoplasmic integration theorem \cite{Gates1998}\cite{Gates1997b} or superspace normal coordinates \cite{Grisaru1997} or both \cite{Gates1997a}.  Both ectoplasmic integration and normal coordinate expansions are state of the art superspace techniques that use only the super covariant derivative algebra as the basis for calculating.

\end{document}